\documentclass[acmtog, authorversion]{acmart}

\usepackage{booktabs} 
\usepackage{cancel}

\setcitestyle{square}

\usepackage[ruled]{algorithm2e} 

\SetAlFnt{\small}
\SetAlCapFnt{\small}
\SetAlCapNameFnt{\small}
\SetAlCapHSkip{0pt}
\IncMargin{-\parindent}

\setcopyright{acmcopyright}
\acmJournal{TOG}
\acmYear{2018}\acmVolume{37}\acmNumber{4}\acmArticle{78}\acmMonth{8} \acmDOI{10.1145/3197517.3201282}
\acmTotPages{13}

\acmISBN{123-4567-24-567/08/06}

\copyrightyear{2018}

\acmPrice{15.00}

\acmSubmissionID{0128}

\usepackage{gensymb}
\usepackage{float}
\usepackage{soul}
\usepackage{array}
\usepackage{multirow}
\usepackage{hhline}
\usepackage{setspace}
\usepackage{csquotes}
\usepackage{siunitx}
\usepackage{xr}
\usepackage{tcolorbox}

\usepackage{pbox}
\usepackage{amsmath}
\usepackage{xspace}
\usepackage{empheq}
\usepackage{soul}

\renewcommand{\paragraph}[1]{\par\medskip\noindent\textbf{#1}\hspace{1em}}

\newcommand{\diff}[1]{\ensuremath{\mathrm{d}#1}}
\newcommand{\Diff}[1]{\ensuremath{\nabla#1}}
\newcommand{\Mean}[1]{\ensuremath{\overline{#1}}}
\DeclareMathOperator{\variance}{Var}
\newcommand{\Var}[1]{\ensuremath{\variance(#1)}}

\newcommand{\dGamma}[3]{\ensuremath{\Gamma(#1;#2,#3)}}
\newcommand{\fGamma}[1]{\ensuremath{\gamma(#1)}}
\newcommand{\fGammaIncomplete}[1]{\ensuremath{\gamma_{p}(#1)}}

\newcommand{\imgformat}{png}

\usepackage{color}
\definecolor{gray}{rgb}{0.6,0.6,0.6}
\definecolor{cyan}{cmyk}{1,0,0,0}
\definecolor{blue}{rgb}{0,0,0.7}
\definecolor{red}{rgb}{0.5,0.0,0}
\definecolor{darkgreen}{rgb}{0,0.6,0}
\definecolor{darkergreen}{rgb}{0,0.3,0}
\definecolor{orange}{rgb}{1,0.5,0}
\definecolor{magenta}{cmyk}{0,1,0,0}
\definecolor{darkyellow}{cmyk}{0,0,0.75,0}
\definecolor{purple}{cmyk}{0.5,1,0,0}

\usepackage{ifpdf}
\usepackage{graphicx}
\ifpdf
  
\else
\fi

\DeclareSIUnit{\unitless}{unitless}

\newcommand{\toSuppMaterial}[1]{}

\newcommand{\new}[1]{#1}
\newcommand{\newad}[1]{\new{#1}}

\newcommand{\old}[1]{\textcolor{gray}{\st{#1}}}
\newcommand{\revised}[1]{\textcolor{red}{{#1}}}
\newcommand{\revisedSecond}[1]{\textcolor{red}{{#1}}}
\newcommand{\revisedThird}[1]{\textcolor{red}{{#1}}}

\renewcommand{\old}[1]{}
\renewcommand{\revised}[1]{#1}
\renewcommand{\revisedSecond}[1]{#1}
\renewcommand{\revisedThird}[1]{#1}

\newcommand{\mycomment}[3]{}
\newcommand{\adrianc}[1]{\mycomment{Adrian}{orange}{#1}}

\newcommand{\diego}[1]{\mycomment{Diego}{blue}{#1}}

\usepackage{tabularx}

\newcommand{\Fig}[1] {Figure~\ref{fig:#1}}
\newcommand{\Figs}[2] {Figures~\ref{fig:#1} and~\ref{fig:#2}}

\newcommand{\Tab}[1] {Table~\ref{table:#1}}
\newcommand{\Sec}[1] {Section~\ref{sec:#1}}
\newcommand{\Secs}[2] {Sections~\ref{sec:#1} and~\ref{sec:#2}}
\newcommand{\Eq}[1] {Equation~\eqref{eq:#1}}
\newcommand{\Eqs}[2] {Equations~\eqref{eq:#1} and \eqref{eq:#2}}
\newcommand{\Eqss}[3] {Equations~\eqref{eq:#1}, \eqref{eq:#2}, and \eqref{eq:#3}}
\newcommand{\Eqsss}[4] {Equations~\eqref{eq:#1}, \eqref{eq:#2}, \eqref{eq:#3}, and \eqref{eq:#4}}
\newcommand{\EqsRange}[2] {Equations~\eqref{eq:#1} to \eqref{eq:#2}}
\newcommand{\ParEq}[1] {[\Eq{#1}]}
\newcommand{\ParEqs}[2] {[\Eqs{#1}{#2}]}

\newcommand{\Supp}[1]{\Sec{#1} in the supplemental\xspace}
\newcommand{\Supps}[2]{\Secs{#1}{#2} in the supplemental\xspace}
\newcommand{\App}[1]{Appendix~\ref{app:#1}}
\newcommand{\suppimages}{../images}

\newcommand{\sTime}{\ensuremath{\mathfrak{t}}}

\newcommand{\point}[1]{\ensuremath{\mathbf{#1}}}
\newcommand{\px}{\point{x}}

\newcommand{\volume}[1]{\ensuremath{#1}}
\newcommand{\vV}{\diff{\volume{V}}}

\newcommand{\vect}[1]{\ensuremath{\omega_{#1}}}
\newcommand{\dir}{\vect{}}
\newcommand{\diro}{\vect{o}}
\newcommand{\diri}{\vect{i}}

\newcommand{\sMatrix}[1]{\mathbf{#1}}
\newcommand{\sVarMatrix}{\sMatrix{V}}

\newcommand{\Sphere}{\ensuremath{{\Omega}}}

\newcommand{\sDistance}{\ensuremath{t}}
\newcommand{\sMFP}{\ensuremath{\Mean{t}}}

\newcommand{\sLightSpeed}{\ensuremath{c}}
\newcommand{\sParticlesSpeed}{\ensuremath{v}}
\newcommand{\sPlank}{\ensuremath{h}}
\newcommand{\sFlux}{\ensuremath{\Phi}}
\newcommand{\sEmissionFlux}{\ensuremath{q}}

\newcommand{\sArvo}[1]{\ensuremath{\mathbf{#1}}}
\newcommand{\sArvoSource}{\sArvo{E}}
\newcommand{\sArvoStream}{\sArvo{S}}
\newcommand{\sArvoInscattering}{\sArvo{C}_\mathrm{in}}
\newcommand{\sArvoExtinction}{\sArvo{C}_\mathrm{ext}}

\newcommand{\sRad}{\ensuremath{L}}
\newcommand{\sRadI}{\ensuremath{S}}
\newcommand{\sRadICorr}{\ensuremath{\sRad_\sRadI}}

\newcommand{\sRadQiCorr}[1]{\ensuremath{{\sRad_{\sEmission_{#1}}}}}

\newcommand{\sTranmittance}{\ensuremath{T}}
\newcommand{\sTranmittanceCorr}{\ensuremath{\sTranmittance}}
\newcommand{\sTranmittanceCorrI}[1]{\ensuremath{\sTranmittanceCorr_{#1}}}

\newcommand{\sCoefficient}{\ensuremath{\mu}}
\newcommand{\sAbsorption}{\ensuremath{\sCoefficient_a}}
\newcommand{\sScattering}{\ensuremath{\sCoefficient_s}}
\newcommand{\sExtinction}{\ensuremath{\sCoefficient}}
\newcommand{\sExtinctionIdx}[1]{\ensuremath{\sCoefficient_{#1}}}
\newcommand{\sEmission}{\ensuremath{Q}}
\newcommand{\sOpticalDepth}{\ensuremath{\tau}}

\newcommand{\sCoefficientResolved}{\ensuremath{\Sigma}}
\newcommand{\sExtinctionResolved}{\ensuremath{\sCoefficientResolved}}
\newcommand{\sScatteringResolved}{\ensuremath{\sCoefficientResolved_s}}
\newcommand{\sExtinctionResolvedI}[1]{\ensuremath{\sCoefficientResolved_{#1}}}

\newcommand{\sAlbedoI}[1]{\ensuremath{\sAlbedo_{#1}}}
\newcommand{\sPFI}[1]{\ensuremath{f_{r,#1}}}
\DeclareMathOperator{\sScatteringOperator}{B}

\newcommand{\sBCSDF}{\ensuremath{f}}

\newcommand{\sPF}{\ensuremath{f_r}}

\newcommand{\sAlbedo}{\ensuremath{\Lambda}}

\newcommand{\sCrossSection}{\ensuremath{\sigma}}

\newcommand{\sConcentration}{\ensuremath{\mathcal{C}}}

\newcommand{\sParticle}{\ensuremath{k}}
\newcommand{\sParticles}{\ensuremath{\mathcal{P}}}

\newcommand{\sProb}[1]{\ensuremath{p\left(#1\right)}}
\newcommand{\sProbI}[2]{\ensuremath{p_{#1}\left(#2\right)}}

\newcommand{\sMGFProb}[1]{\ensuremath{M(#1)}}

\newcommand{\sProbConditionalI}[3]{\ensuremath{\sProbI{#1}{#2;#3}}}

\newcommand{\sProbExtinction}[1]{\ensuremath{\sProbI{\sOpticalDepth}{#1}}}
\newcommand{\sProbExtinctionConditional}[2]{\ensuremath{\sProbConditionalI{\sOpticalDepth}{#1}{#2}}}
\newcommand{\sProbConc}[1]{\ensuremath{\sProbI{\sConcentration}{#1}}}
\newcommand{\sProbConcConditional}[2]{\ensuremath{\sProbConditionalI{\sConcentration}{#1}{#2}}}
\newcommand{\sProbLight}[1]{\ensuremath{\sProbI{\sRad}{#1}}}

\newcommand{\sAttenuation}{\ensuremath{\mathcal{T}}}
\newcommand{\sBeamFront}{\ensuremath{\mathcal{R}}}
\newcommand{\sRay}{\ensuremath{\mathfrak{r}}}
\newcommand{\sCorr}[1]{\ensuremath{\eta_{#1}}}
\newcommand{\sAngle}{\ensuremath{\theta}}

\usepackage{datenumber}
\usepackage{calc}

\newcounter{datetoday}
\newcounter{diffyears}
\newcounter{diffmonths}
\newcounter{diffdays}

\newcommand{\difftoday}[3]{%
      \setmydatenumber{datetoday}{\the\year}{\the\month}{\the\day}%
      \setmydatenumber{diffdays}{#1}{#2}{#3}%
      \addtocounter{diffdays}{-\thedatetoday}%
      \ifnum\value{diffdays}>0
        \def\diffbefore{}%
        \def\diffafter{left}%
      \else
        \def\diffbefore{}%
        \def\diffafter{ago}%
        \setcounter{diffdays}{-\value{diffdays}}%
      \fi
      \setcounter{diffyears}{\value{diffdays}/365}%
      \setcounter{diffdays}{\value{diffdays}-365*\value{diffyears}}%
      \setcounter{diffmonths}{\value{diffdays}/30}%
      \setcounter{diffdays}{\value{diffdays}-30*\value{diffmonths}}%
      \diffbefore
      \ifnum\value{diffyears}=0
      \else
        \ifnum\value{diffyears}>1
            \thediffyears\space years,
        \else
            \thediffyears\space year,
        \fi
      \fi
      \ifnum\value{diffmonths}=0
      \else
        \ifnum\value{diffmonths}>1
            \thediffmonths\space months
        \else
            \thediffmonths\space month
        \fi
      \fi
      \ifnum\value{diffdays}=0
      \else
        \ifnum\value{diffdays}>1
            \thediffdays\space days
        \else
            \thediffdays\space day
        \fi
      \fi
      \diffafter
}

\begin{document}
\title{A Radiative Transfer Framework for Spatially-Correlated Materials}

\author{Adrian Jarabo}
\affiliation{%
  \institution{Universidad de Zaragoza -- I3A}
}

\author{Carlos Aliaga}
\affiliation{%
  \institution{Universidad de Zaragoza -- I3A, and Desilico Labs}
}

\author{Diego Gutierrez}
\affiliation{%
  \institution{Universidad de Zaragoza -- I3A}
}

\renewcommand\shortauthors{Jarabo et al.}



\begin{teaserfigure}
 \includegraphics[width=\textwidth]{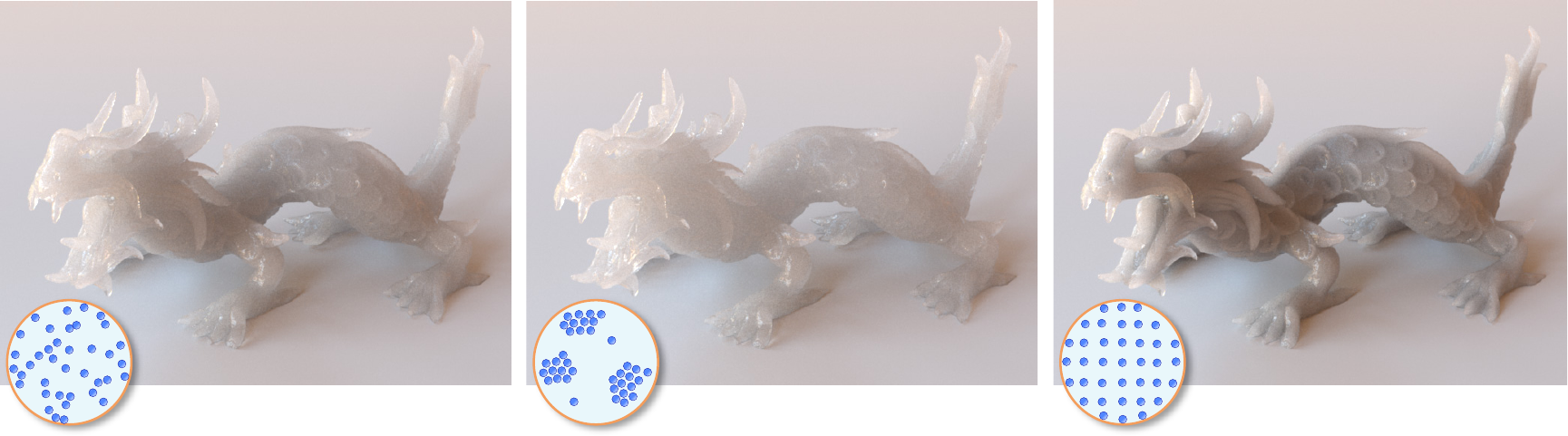} \\
 \caption{\new{Spatial correlation in media leads to non-exponential light transport, which significantly affects appearance. The image shows volumetric renderings of translucent dragons made of materials with the same density per unit differential volume $\Mean{\sExtinction}=10$ (isotropic, albedo $\sAlbedo=.8$), but different correlations. Left: using classic light transport, where material particles are assumed to be uncorrelated. Middle and right: positive and negative correlation, respectively, using our novel framework for spatially-correlated materials. The insets show illustrative views of scatterer correlation for each dragon. }}
   \label{fig:isotropic_dragons}
\end{teaserfigure}

\begin{abstract}
\adrianc
{\small
\color{blue}{(Last updated: \today)}
\color{darkgreen}{[Submission: January 23, 2018]}
\color{red}{(\difftoday{2018}{01}{23})}
\newline
}
\diego{check all: }We introduce a \old{novel} non-exponential radiative framework that takes into account the local spatial correlation of scattering particles in a medium. Most previous works in graphics have ignored this, assuming uncorrelated media with a uniform, random local distribution of particles. However, positive and negative correlation lead to slower- and faster-than-exponential attenuation respectively, which cannot be predicted by the Beer-Lambert law. As our results show, this has a major effect on extinction, and thus appearance. 
From recent advances in neutron transport, we first introduce our Extended Generalized Boltzmann Equation, and develop a general framework for light transport in correlated media. We lift the limitations of the original formulation, including an analysis of the boundary conditions, and present a model suitable for computer graphics, based on optical properties of the media and statistical distributions of scatterers. In addition, we present an analytic expression for transmittance in the case of positive correlation, and show how to incorporate it efficiently into a Monte Carlo renderer.  
We show results with a wide range of both positive and negative correlation, and demonstrate the differences compared to classic light transport.

\end{abstract}

%
%


%
%


\maketitle

\renewcommand{\Supp}[1]{\Sec{#1}}
\renewcommand{\Supps}[2]{\Secs{#1}{#2}}
\renewcommand{\imgformat}{jpg}


\section{Introduction}
\label{sec:intro}
Volumetric appearances are ubiquitous in the real world, from translucent organic materials to clouds, smoke, or densely packed granular media.
Voxel-based representations with anistropic scattering functions~\cite{Jakob2010Radiative,Heitz2015SGGX} have been widely used in recent years to represent the appearance of complex geometries such as trees~\cite{Neyret1998,Loubet2017}, cloth and hair~\cite{Aliaga2017,Schroder2011Volumetric,Zhao2011Building,Khungurn2015Matching}, or particles' aggregates~\cite{Moon2007Discrete,Meng2015Granular,Muller2016Efficient}. 

%
Many translucent objects and participating media present a strong spatial correlation between scatterers\footnote{Following other works' terminology, through the paper we use the term ``scatterers'' for all particles in the media, including perfect absorbers.}~\cite{Lovejoy1995,Knyazikhin1998,Coquard2006}, where scatterers' densities are non-uniform within a differential volume. The aerosol of clouds, for instance, tends to form clusters, resulting in areas with very different optical thicknesses~\cite{Marshak1998}. 
%
As a result, the probability of a photon interacting with a scatterer inside each differential volume is also non-uniform, which in turn has a great effect in the final appearance, as Figure~\ref{fig:isotropic_dragons} shows. 

\begin{figure} [t]
 \centering
 \includegraphics[width=0.30\columnwidth]{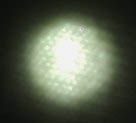}
 \includegraphics[width=0.30\columnwidth]{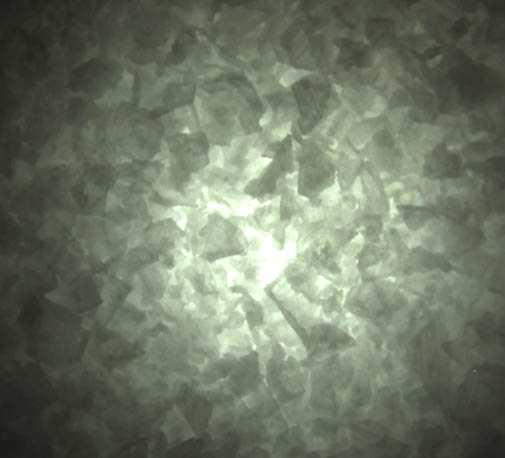}
\hspace{.1cm}
\raisebox{1.3cm}{\begin{tabular}{l c} 
 Media & $R^2$ \\
\hline
Milk$^*$ & 0.99 \\
Black Fabric & 0.87 \\
White Fabric & 0.76 \\
Maldon Salt & 0.76 \\
Sugar & 0.69 \\
\hline
\end{tabular}
}
\caption{\new{Left: Photographs of spatially-correlated media (white fabric, and maldon salt), lit from behind using a mobile flash. Right: We evaluate the transmittance of different media, by fitting measurements at different optical thickness to the exponential decay predicted by the classic Beer-Lambert law. As expected, a diluted liquid such as milk (marked with an asterisk) shows a very close fit to the exponential decay (measured using the $R^2$ metric); however, transmittance in locally correlated media cannot be modeled using classic radiative transfer. \revised{Details on the experiment can be found in \Supp{app_figure2}.}}}
\label{fig:capture}
\end{figure}

\new{Most previous works in graphics have assumed an uncorrelated distribution of scatterers, \old{ignored this effect,}considering only spatial correlation at a macroscopic scale as heterogeneous media. This results in the well-known exponential transmittance predicted by the Beer-Lambert law. However, in the presence of correlation at differential-volume scale, the predictions of the Radiative Transfer Equation (RTE)~\cite{Chandrasekhar1960} break, and therefore attenuation is no longer exponential (see \Fig{capture}):
In such cases, negatively correlated media leads to \emph{faster-than-exponential} transmittance, whereas positive correlation leads to \emph{slower-than-exponential} transmittance~\cite{Davis1999Horizontal}. 
\newad{Works rendering granular aggregates observed such non-exponential transmittance; however, they either formulate it in an uncorrelated radiative (exponential) framework~\cite{Meng2015Granular}, precalculate the full light transport explicitly~\cite{Moon2007Discrete}, or combine both approaches~\cite{Muller2016Efficient}. }
}


%

\new{
In this work we introduce a theoretical framework for  \revised{simulating light transport in spatially-correlated media}, which accounts for the local structure of scatterers. }
%
\old{Our framework builds upon the well-established radiative theory, generalizing the RTE to spatially-correlated media. We leverage recent advances in \emph{non-classical transport} in the neutron transport field; from the Generalized Boltzmann Equation (GBE)~, we extend it and lift its main limitations, leading to a general framework suitable for computer graphics.}
\revised{Our framework builds upon the well-established radiative theory, and leverages recent advances in \emph{non-classical transport} in the neutron transport field: We extend the Generalized Boltzmann Equation (GBE)~\cite{Larsen2007Generalized}, which generalizes the RTE to correlated media, and lift its main limitations, leading to a general framework suitable for computer graphics.}
In addition, we present an analytic expression of transmittance for positive correlation, leading to  a compact representation of directionally-dependent spatial correlation based on a gamma distribution of scatterers. We also present efficient sampling techniques, enabling the use of our model within any existing volumetric renderer.
Our framework is able to accurately simulate light transport inside correlated media. We show results with a wide range of correlations, both negative and positive, and demonstrate the differences with classic (uncorrelated) light transport. Our model is general and intuitive, and can be seen as complementary in the spatial domain to angular anisotropy in media~\protect\cite{Jakob2010Radiative,Heitz2015SGGX}. It might also be useful in other areas such as volumetric level of detail, or accelerating light transport using similarity theory.

\paragraph{Overview }
The technical sections of the paper are organized as follows: We first present a general background of radiative transport in \revised{uncorrelated media}, a brief summary of the effect of spatial correlation on extinction, and the Generalized Boltzmann Equation (\Sec{RTCM}). 
Unfortunately, \old{this equation is not suitable for rendering, due to its strong simplifying assumptions}\revised{the original formulation of the GBE presents some simplifying assumptions valid for neutron transport in reactors, but which limit its applicability in rendering}. In \Sec{ourgbe} we present our Extended GBE, which lifts the limitations of the original GBE to support more general media, and include a thorough analysis of its boundary conditions. 
Finally, in \Sec{ggbe_rendering} we propose \old{a formulation for appearance of}\revised{an appearance model for} positively correlated media based on local optical parameters, which is intuitive and easy to manipulate, and which can be plugged directly into our Extended GBE. 

%

%
%
%
%
%
%
%


\section{Related Work}
\label{sec:rw}

\paragraph{Volumetric light transport}
Simulating light transport in participating media has a long history in computer graphics (see e.g. \cite{Gutierrez2008}). Existing methods aim to efficiently solve the RTE~\cite{Chandrasekhar1960} by means of path tracing~\cite{Lafortune1996Media,Veach1997Thesis}, photon mapping~\cite{Jensen2001}, photon beams~\cite{Jarosz2011}, or a combination of these techniques~\cite{Krivanek2014}. Our framework is independent of the particular algorithm used for rendering.
%
%
Jakob et al.~\shortcite{Jakob2010Radiative} extended the RTE to account for directional (angular) anisotropy. %
Later, Heitz et al.~\shortcite{Heitz2015SGGX} further extended this model with the SGGX microflakes distribution. While these works focus on the local angular dependence of scattering and extinction, they still assume that the scatterers are uncorrelated, distributed uniformly in the spatial domain. Our work is orthogonal to these approaches, focusing on the effects of spatial correlation. 

\paragraph{Volumetric representation of appearance}
Volumetric representations of explicit geometry have been successfully used to approximate complex appearances.
%
\new{Meng et al.~\shortcite{Meng2015Granular} used a classical radiative approximation of light transport in particulate media for efficient rendering.}
%
Fiber-level cloth appearance models based either on micro-CT geometry~\cite{Zhao2011Building,Zhao2012Structure} or procedural modeling~\protect\cite{Schroder2011Volumetric}, have used volumetric anisotropic representations for rendering high-detailed garments~\cite{Aliaga2017}, similar in quality to explicit fiber representations~\protect\cite{Khungurn2015Matching}.
Zhao et al.~\shortcite{Zhao2016} presented an optimization-based approach to downsample volumetric appearance representation by altering the rendering parameters (scattering and phase function) to match the desired appearance. 
All these works make again the assumption of perfect decorrelation of the scatterers in the medium. Our theoretical framework departs from this assumption. 

\paragraph{Correlated volumetric media} 
Correlated volumetric media have been studied in computational transport in fields such as nuclear engineering~\cite{Levermore1986Linear,Larsen2011Generalized,Camminady2017}, atmospheric sciences~\cite{Newman1995Systematic,Davis1999Horizontal,Davis2004Photon}, or thermal propagation~\cite{Coquard2006,Bellet2009,Taine2010}, leading to \textit{non-classical} transport theories~\cite{Larsen2007Generalized,Frank2010Generalized}. 
\new{Non-classical transport has been however largely unexplored in graphics: The first work modeling non-exponential flights in graphics is the work of Moon et al.~\shortcite{Moon2007Discrete}, which precomputed transport functions of granular materials as a set of homogeneous shells. However, they required precomputing all light transport operators rather than attempting to express such non-exponential flights into a new radiative transport theory, and did not take into account the effect of correlation at boundaries. M\"{u}ller et al.~\shortcite{Muller2016Efficient} later used a similar approach in combination with other volumetric estimators for rendering heterogeneous discrete media.
%
Concurrently to us, Wrenninge et al.~\shortcite{Wrenninge2017} used non-exponential flights for increased artist control on volumetric light transport, but omitted the underlying theory, and did not relate their model with the physical process of extinction.  
More formally, d'Eon analyzed rigorously the effect of isotropic non-Poissonian extinction on the diffusion (multiple scattering) regime~\shortcite{dEon2014diffusion,dEon2016diffusion}, and discussed the connections between graphics and non-classical transport, including the limitations of such theories to be used in rendering~\cite{dEon2014digipro,dEon2016book}. We generalize these works, offering a non-classic transport theory suitable for rendering, and introduce an intuitive formulation for rendering spatially-correlated media based on local optical parameters. }
%
%
%

%


\section{Radiative Transport in Correlated Media}
\label{sec:RTCM}
In this section, we first introduce light transport in participating media as modeled by the Radiative Transfer Equation (RTE)~\cite{Chandrasekhar1960} (\Sec{background}). We then describe the notion of spatial correlation in media, and its effect on light extinction (\Sec{correlation}), as well as the Generalized Boltzmann Equation (GBE), first proposed by Larsen~\shortcite{Larsen2007Generalized} in the context of neutron transport (\Sec{larsengbe}). 

\begin{figure}
  \def\svgwidth{.99\columnwidth}
\footnotesize
  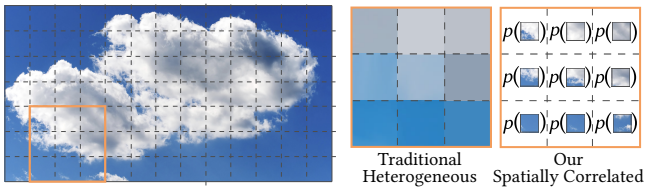
\caption{Difference between heterogeneous and spatially-correlated media, from a computer graphics perspective. Heterogeneous media assume heterogeneity at macroscopic level, but local homogeneity at each differential volume $\vV(\px)$ (see "Traditional Heterogeneous", represented as solid colors per voxel). \new{Instead, our model for spatially-correlated media takes into account the uneven distribution of scatterers within each $\vV(\px)$ (represented on the right as probabilities of extinction $\sProb{\cdot}$).}}
\label{fig:heterogenous_vs_correlated}
\end{figure}

It is important to first clarify the difference between spatially-correlated media, and heterogeneous media, as commonly used in computer graphics. As illustrated in \Fig{heterogenous_vs_correlated}, heterogeneous media assume \textit{local} homogeneity at each differential volume $\vV(\px)$. In contrast, correlated media take into account the average effect of uneven scatterer distributions within each $\vV(\px)$. Therefore, a medium can be \emph{statistically homogeneous}, meaning that its statistical moments are invariant over all the volume, but spatially correlated~\cite{Kostinski2001Extinction}. 

\subsection{Background: The Radiative Transport Equation}
\label{sec:background}
In its integro-differential form, the Radiative Transfer Equation (RTE) models \old{the directional derivative of flux $\sRad$ at point $\px$}\revised{the amount of radiance $\sRad$ at point $\px$ in direction $\diro$} as:
\begin{equation}
\diro\cdot\Diff{\sRad(\px,\diro)} + \sExtinction\sRad(\px,\diro) = \sRadI(\px,\diro) + \sEmission(\px,\diro),
\label{eq:rte}
\end{equation}
where $\sEmission(\diro)$ is the volume source term, and $\sRadI$ is the in-scattered radiance:
\begin{equation}
\sRadI(\px,\diro) = \sScattering\int_\Sphere \sRad(\px,\diri)\sPF(\px,\diri,\diro)\,\diff{\diri},
\label{eq:inscattering}
\end{equation}
which is the directional integral over the sphere $\Sphere$ of the light scattered towards $\diro$, modeled using the phase function $\sPF$; $\diri$ represents the incoming direction of light. Note that we have omitted the spatial dependency of all terms in \Eqs{rte}{inscattering} for simplicity.
Finally, $\sExtinction=\sAbsorption+\sScattering$ [\si{\per\meter}] is the extinction coefficient, with $\sAbsorption$ and $\sScattering$ the absorption and scattering coefficients respectively. These terms model the probability of a beam of light to be attenuated either by absorption or scattering, and are defined as the product of the number of scatterers per unit volume $\sConcentration$ [\si{\per\cubic\meter}], and the scatterers' cross section $\sCrossSection$ [\si{\square\meter}], assuming that the scatterers are uniformly distributed in the differential volume (\Fig{isotropic_dragons}, left) (see~\cite{Arvo1993} for a detailed derivation). Jakob et al.~\shortcite{Jakob2010Radiative} later generalized the RTE to model directionally anisotropic media, by taking into account the angular (directional) dependence of the scatterers's cross section in media. 

\begin{figure}
\centering
  \def\svgwidth{.99\columnwidth}
  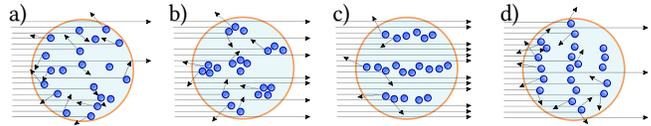
\caption{\new{Illustration of the effect of spatial correlation in a medium for different hypothetical distributions of scatterers.} (a) A random distribution of scatterers within a differential volume $\vV(\px)$. (b) In the presence of scatterer correlation, the probability of interaction changes within $\vV(\px)$. (c, d) This correlation might further exhibit directional behavior, leading to very different interaction probabilities according to the degree of alignment with the propagation of light.
}
\label{fig:correlation}
\end{figure}

\subsection{\new{Effect of Spatial Correlation on Extinction}}
\label{sec:correlation}
When light propagates through a participating medium, it scatters as a function of the distribution of the scatterers. When this distribution is random and uniformly distributed, extinction becomes a Poissonian process, and the exponential Beer-Lambert law accurately describes the attenuation of light (see~\cite{Gallavotti1972} for a rigorous derivation). However, the distribution of scatterers in many media exhibit different forms and degrees of spatial correlation (e.g. clouds~\cite{Lovejoy1995,Davis2004Photon}, textiles~\cite{Coquard2006}, porous materials~\cite{Bellet2009,Taine2010}, or granular aggregates~\cite{Meng2015Granular}). This affects light transport, as Figure~\ref{fig:correlation} illustrates; as a consequence, attenuation is no longer exponential, and light extinction becomes non-Poissonian.

\begin{figure}[t]
\includegraphics[width=\columnwidth]{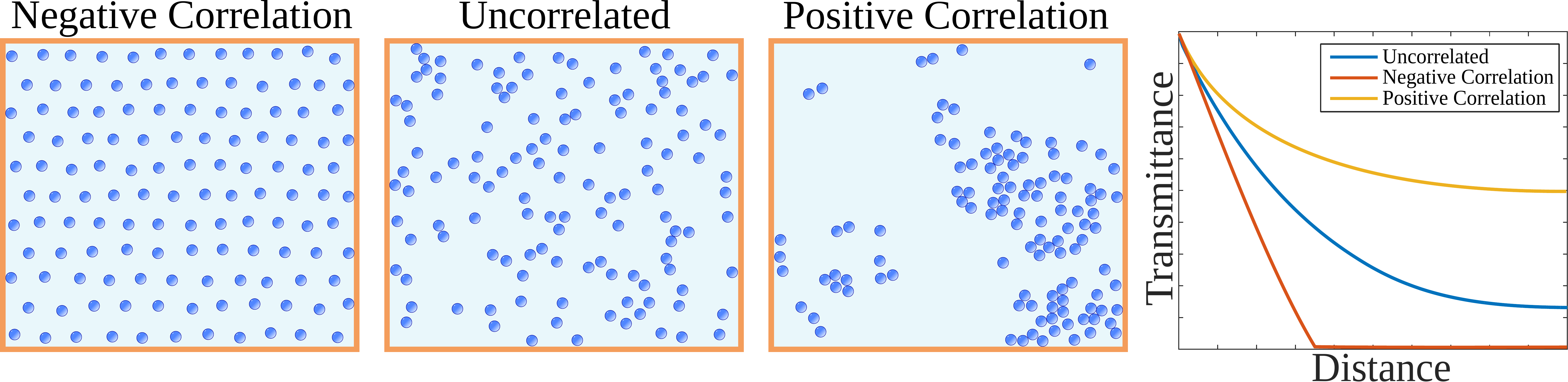}
\caption{\new{Different types of scatterer correlation, and their effect on transmittance. From left to right, the first three figures depict negative correlation, no correlation (extinction is a Poissonian process), and positive correlation. The plot on the right shows extinction\revised{, averaged for several procedural realizations of the media (see \Supp{app_simulations}) and ray directions}: While uncorrelated media results in the classic exponential extinction, negative and positive correlation lead to faster- and slower-than-exponential extinction, respectively.}}
\label{fig:pos_neg}
\end{figure}

\begin{figure}[t]
  \def\svgwidth{.9\columnwidth}
\footnotesize
  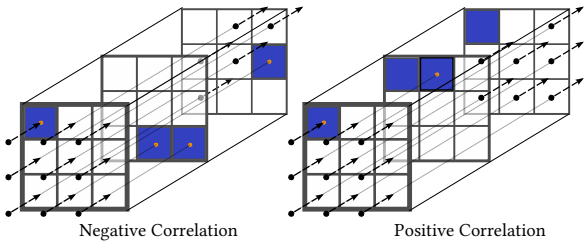
\caption{\new{Intuitive explanation for non-exponential transmittance in negatively- (left) and positively-correlated (right) media. Solid squares represent perfect absorbers. Although both media have the same number of absorbers, shadowing (or overlapping) of such scatterers (positive correlation) results in less (sub-exponential) extinction. Figure inspired from~\cite{Kostinski2002}.}}
\label{fig:kostinskis}
\end{figure}

\new{Negative correlation occurs when the distribution of scatterers is more uniform than Poisson (as in electrostatic repulsion), and leads to super-exponential (faster) extinction. Clustered scatterers, on the other hand, yield positive correlation, which leads to sub-exponential (slower) extinction; this is illustrated in \Fig{pos_neg}. }
\new{The reason for such non-exponential transmittance can be further visualized intuitively in \Fig{kostinskis}: In negatively-correlated media, absorbers are less likely to "shadow" one another; as a result, more light becomes extinct. Positively-correlated media presents the opposite case, with many absorbers shadowing others; this creates empty regions which in turn lead to more light passing through.}




\new{More formally, in uncorrelated media (Poissonian process) the extinction probability after traveling a distance $t$ from the previous scattering (or emission\footnote{We will refer only to scattering events from now on for simplicity.}) event is $\sProb{t}=\sExtinction\,\exp(-\sExtinction\,t)$, as predicted by the Beer-Lambert law. Thus, defining the differential probability of extinction $\sExtinctionResolved(t)$ [\si{\per\meter}] as \cite{Larsen2011Generalized} 
\begin{equation}
\sExtinctionResolved(t) = \frac{\sProb{t}}{1-\int_0^t\sProb{s}\diff{s}},
\label{eq:extinctionresolved}
\end{equation}
\old{where the denominator describes transmittance}\revised{where the denominator is the physical definition of transmittance $\sTranmittanceCorr(t)$}, we obtain $\sExtinctionResolved(t) = \sExtinction$, the extinction coefficient of the media (uniform for each differential volume, and independent of the distance $t$).} 

\new{Let us now define a simple positively correlated (clustered) medium, composed of regions with a high density of scatterers (extinction coefficient $\sExtinctionIdx{1}$), and regions with low density ($\sExtinctionIdx{2}$). The probability of light extinction\footnote{The probability of extinction $\sProb{t}$ is also termed in the literature as ``path length distribution'', ``free path distribution'', or ``chord length distribution''.} after traveling a distance $t$ is given by
%
\begin{equation}
\sProb{t} = \sExtinctionIdx{1}\,\sProbExtinction{\sExtinctionIdx{1}}\,e^{-\sExtinctionIdx{1}\,t} + \sExtinctionIdx{2}\,\sProbExtinction{\sExtinctionIdx{2}}\,e^{-\sExtinctionIdx{2}\,t},
\label{eq:prob_two_extinction}
\end{equation}
with $\sProbExtinction{\sExtinctionIdx{1}}$ and $\sProbExtinction{\sExtinctionIdx{2}}$ the probability of traversing a region with extinction coefficients $\sExtinctionIdx{1}$ and $\sExtinctionIdx{2}$ respectively, where $\sProbExtinction{\sExtinctionIdx{1}}+\sProbExtinction{\sExtinctionIdx{2}} = 1$.
From this simple example, we can see that $\sProb{t}$ is no longer exponential, and thus extinction is no longer a Poissonian process with a constant $\sExtinctionResolved(t) = \sExtinction$. Instead, plugging  \Eq{prob_two_extinction} into \Eq{extinctionresolved} leads to a function dependent on $t$.
In other words, spatial correlation introduces a \emph{memory effect}~\cite{Kostinski2002}, where the differential probability of extinction depends on the traveled distance $t$ since the previous scattering event. This has a significant effect in the final volumetric appearance of the medium, as shown in \Fig{isotropic_dragons} and throughout this paper.}

\subsection{The Generalized Boltzmann Equation}
\label{sec:larsengbe}
\new{Since $\sExtinctionResolved(t)$ is a function of $t$ in the presence of correlation, \old{flux $\sRad(\px,\diro)$ now becomes $\sRad(\px,\diro,t)$,}\revised{we need to introduce the $t$-dependent flux $\sRad(\px,\diro,t)$ [\si{\watt \per\square\meter\per\steradian\per\meter}]}, the flux at $\px$ after traveling a distance $t$ from the last scattering event. \revised{It relates with classic flux $\sRad(\px,\diro)$ [\si{\watt \per\square\meter\per\steradian}] as $\sRad(\px,\diro)= \int_0^\infty \sRad(\px,\diro,t) \diff{t}$, and} in turn introduces an additional derivative term in \Eq{rte}, resulting in the Generalized Boltzmann Equation~\cite{Larsen2007Generalized,Larsen2011Generalized}:} 
\begin{multline}
\frac{d}{dt}\sRad(\px,\diro,t) + \diro\cdot\Diff{\sRad(\px,\diro,t)} + \sExtinctionResolved(t)\sRad(\px,\diro,t) = 0, \\
\sRad(\px,\diro,0) = \underbrace{\int_0^\infty\sScatteringResolved(t)\int_\Sphere \sRad(\px,\diri,t)\sPF(\diri,\diro)\,\diff{\diri}\,\diff{t}}_{\text{Inscattering $\sRadI(\px,\diro)$}} + \sEmission(\px,\diro),
\label{eq:larsengrte}
\end{multline}
where \new{$\sScatteringResolved(t)=\sAlbedo\,\sExtinctionResolved(t)$} is the probability of a photon being scattered after traveling a distance $t$ \revised{(see \Supp{app_gbe} for the full derivation)}, and $\sAlbedo$ represents albedo. \revised{The second line of the equation represents the value for $t=0$, in which light is scattered or emitted. Thus, after each scattering event the memory effect for the extinction is reset to zero.}

As expected, by removing the $t$-dependency as $\sExtinctionResolved(t) = \sExtinctionResolved$, and integrating \Eq{larsengrte}, we obtain the classic RTE \ParEq{rte} \new{(see \Supp{app_gbe2rte})}. 
\new{Moreover, \Eq{larsengrte} can also support directionally anisotropic media \cite{Jakob2010Radiative} by formulating $\sExtinctionResolved$ as a function of $\diro$ \cite{Vasques2014anisotropic}}.

\section{Our Extended GBE}
\label{sec:ourgbe}

\subsection{Limitations of the GBE}
\label{sec:larsenslimitations}
\new{Unfortunately, \Eq{larsengrte} relies on a set of simplifying assumptions, which \revised{limit its applicability in rendering applications}. In particular~\cite{Larsen2007Generalized,Larsen2011Generalized}:\footnote{\revised{Larsen and Vasques also assume a monoenergetic system; for simplicity, we assume also a single wavelength, although removing this limitation is straight forward. }}}
%
\new{\begin{enumerate}
\item The medium is statistically homogeneous, and infinite; no system boundaries exist. 
 \item The phase function $\sPF(\diri,\diro)$ and albedo $\sAlbedo$ are independent of $t$. For example, in a mixture of two types of scatterers with different phase function or albedo, this assumes that both types have the same structure. 
\item The source term $\sEmission(\px,\diro)$ is correlated with the scatterers in the volume. This assumption does not hold in most cases\revisedSecond{, as illustrated in \Fig{correlation_source}.}
 \end{enumerate}
}
In order to make the GBE useful for rendering, we need to extend it beyond these limiting assumptions. \old{Removing the first one is straightforward; } We describe this in the rest of this section, introducing our novel Extended GBE.

\subsection{Extending the GBE}
\label{sec:extendinggbe}
\new{To lift the \old{second and third}\revised{first and second} limitations of the standard GBE, we first reformulate $\sExtinctionResolved$, $\sAlbedo$, and $\sPF$ as functions of the spatial position and the traveled distance, as $\sExtinctionResolved(\px,t)$, $\sAlbedo(\px,t)$, and $\sPF(\px,\diri,\diro,t)$ respectively. 
This means that, depending on the traveled distance $t$, light will be scattered differently, according to the different spatial correlation of the scatterers. Note that some previous works in graphics~\cite{Frisvad2007,Sadeghi2012} have included a mixture of scatterer sizes in the medium, but not spatial correlation.}
\new{Defining the directional scattering operator $\sScatteringOperator(\px,\diri,\diro,t)=\sAlbedo(\px,t)\,\sExtinctionResolved(\px,t)\sPF(\px,\diri,\diro,t)$ for compactness, \Eq{larsengrte} becomes
\begin{multline}
\frac{d}{dt}\sRad(\px,\diro,t) + \diro\cdot\Diff{\sRad(\px,\diro,t)} + \sExtinctionResolved(t)\sRad(\px,\diro,t) = 0, \\
\sRad(\px,\diro,0) = \int_0^\infty\int_\Sphere \sRad(\px,\diri,t)\sScatteringOperator(\px,\diri,\diro,t)\,\diff{\diri}\,\diff{t} + \sEmission(\px,\diro),
\label{eq:grte1}
\end{multline}
where we assume an isotropic formulation to avoid cluttering.
}

\begin{figure}
  \def\svgwidth{.8\columnwidth}
\footnotesize
  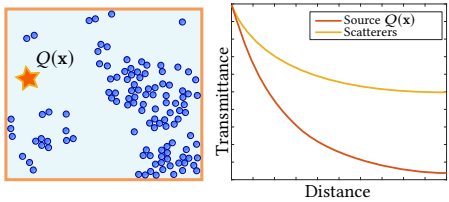
\caption{\new{Left: an uncorrelated light source $\sEmission(\px)$ in a positively correlated medium. The differential probability of extinction $\sExtinctionResolved(\px,t)$ is therefore different for each, which significantly modifies light transport, as shown in the right\revised{, where transmittance is numerically computed from several procedurally generated media with identical positive correlation}. This difference is not captured in Larsen's original GBE~\cite{Larsen2011Generalized}.}}
\label{fig:correlation_source}
\end{figure}

\new{The \old{fourth}\revised{third} assumption, on the other hand, requires a more significant change of \Eq{larsengrte}. In Larsen's original formulation of the GBE, since $\sRad(\px,\diro,0)$ includes both scattering $\sRadI(\px,\diro)$ \textit{and} light emitted by sources $\sEmission(\px,\diro)$, both terms implicitly share the same differential probability of extinction $\sExtinctionResolved(\px,t)$. 
%
However, this would only be true if they present the exact same correlation (e.g. the scattering and the emissive particles are the same); in the general case, $\sExtinctionResolved(\px,t)$ is different for $S$ and $Q$. 
Moreover, different sources $\sEmission$ might correlate differently with the medium, leading to different $\sExtinctionResolved(\px,t)$ per source. 
\Fig{correlation_source} shows how different  $\sExtinctionResolved(\px,t)$ for scatterers and sources significantly affects light transport. \revisedSecond{This different correlation between sources and scatterers is in fact very important for rendering realistic scenes, since as we shown later in \Sec{boundary_conditions} reflection at media boundaries act as uncorrelated sources.}
}

\new{
Taking all this into account, we can express radiance $\sRad(\px,\diro,t)$ as
\begin{equation}
\sRad(\px,\diro,t) = \sRadICorr(\px,\diro,t) + \sum_j \sRadQiCorr{j}(\px,\diro,t),
\end{equation}
where $\sRadICorr(\px,\diro,t)$ is the \textit{scattered} radiance reaching $\px$ after traveling a distance $t$ since the last scattering event, and $\sRadQiCorr{j}(\px,\diro,t)$ is the \textit{unscattered} radiance directly emitted by source $\sEmission_j$, which has traveled a distance $t$ since emission. We can then transform \Eq{grte1} into our Extended GBE as 
\begin{tcolorbox}[colback=white,colframe=black,boxrule=0.2mm]
\new{
\begin{align}
\label{eq:grteours}
\frac{d}{dt}\sRad(\px,\diro,t) & + \diro\cdot\Diff{\sRad(\px,\diro,t)} + \sExtinctionResolvedI{\sRadI}(\px,t)\,\sRadICorr(\px,\diro,t) \nonumber \\ & + \sum_j \sExtinctionResolvedI{\sEmission_j}(\px,t)\,\sRadQiCorr{j}(\px,\diro,t) = 0, 
\end{align}
}
\end{tcolorbox}
%
\noindent where $\sExtinctionResolvedI{\sRadI}(\px,t)$ and $\sExtinctionResolvedI{\sEmission_j}(\px,t)$ are the differential extinction probabilities for the scattered photons and the (unscattered) photons emitted by light source $\sEmission_j$, respectively. \old{The initial conditions of this partial differential equation are}\revisedThird{Then, for $t=0$ we have}
\begin{align}
\label{eq:grteoursinitial}
\sRadICorr(\px,\diro,0) = \int_0^\infty&\int_\Sphere\Big(\sScatteringOperator_{\sRadI}(\px,\diri,\diro,t)\,\sRadICorr(\px,\diri,t)  \\ &+ \sum_j \sScatteringOperator_{\sEmission_j}(\px,\diri,\diro,t)\,\sRadQiCorr{j}(\px,\diri,t)\Big) \,\diff{\diri}\,\diff{t}, \nonumber \\
\sRadQiCorr{j}(\px,\diro,0) & = \sEmission_j(\px,\diro),
\label{eq:grteoursinitial_source}
\end{align}
where $\sScatteringOperator_{\sRadI}(\px,\diri,\diro,t) = \sAlbedoI{\sRadI}(\px,t)\,\sExtinctionResolvedI{\sRadI}(\px,t)\sPFI{\sRadI}(\px,\diri,\diro,t)$ is the scattering operator for scattered photons (thus representing a \textit{multiple scattering} operator), and $\sScatteringOperator_{\sEmission_j}(\px,\diri,\diro,t)$ is the scattering operator for photons emitted by light source $\sEmission_j$ (\textit{single scattering} operator). 
Note how, interestingly, \Eq{grteoursinitial} makes the convenient separation between multiple and single scattering explicit. 
}
\new{Similar to $\sExtinctionResolved(\px,t)$, the phase function and albedo terms might also be different, depending on the correlation between sources and the scatterers. It is easy to verify that when the sources and scatterers are equally correlated with the rest of the medium, the Extended GBE in \Eq{grteours} simplifies to \Eq{larsengrte} (see \Supp{app_ggbe2gbe}). 
}

\paragraph{Integral form of the Extended GBE}
\new{
In order to get an integral formulation of our Extended GBE usable in a general Monte Carlo renderer, we solve \Eq{grteours} for the incoming radiance at point $\px$ as (see \Supp{app_integralggbe} for the full derivation): 
%
%
%
%
\begin{tcolorbox}[colback=white,colframe=black,boxrule=0.2mm]
\new{
\begin{align}
\label{eq:integral_grte}
\sRad(\px,\diro) = \int_0^\infty & \sTranmittanceCorrI{\sRadI}(\px,\px_t) \sRadI(\px_t,\diro) \\ 
+ & \sum_j \sTranmittanceCorrI{\sEmission_j}(\px,\px_t)  \,\sEmission_{j}(\px_t,\diro) \,\diff{t}, \nonumber
\end{align}
}
\end{tcolorbox}
\noindent where $\px_t = \px - \diro\,t$.
%
The terms $\sTranmittanceCorrI{\sRadI}(\px,\px_t)=e^{-\int_{0}^{t}\sExtinctionResolvedI{\sRadI}(\px,s)\diff{s}}$ and $\sTranmittanceCorrI{\sEmission_j}(\px,\px_{t})=e^{-\int_{0}^{t}\sExtinctionResolvedI{\sEmission_j}(\px,s)\diff{s}}$ represent transmittance between $\px$ and $\px_t$ for the scattered and emitted radiance, respectively. 
Last, $\sRadI(\px_t,\diro) $ is
\begin{align}
\nonumber
\sRadI(\px_t,\diro) & = \int_0^\infty\int_\Sphere \Big( \sScatteringOperator_{\sRadI}(\px_{t'},\diri,\diro,t')\sRadI(\px_{t'},\diri)\sTranmittanceCorrI{\sRadI}(\px_t,\px_{t'})\\   
& + \sum_j \sScatteringOperator_{\sEmission_j}(\px_{t'},\diri,\diro,t') \sEmission_{j}(\px_{t'},\diro) \sTranmittanceCorrI{\sEmission_j}(\px_t,\px_{t'}) \Big) \,\diff{\diri}\,\diff{t'}, 
\label{eq:integral_inscattering}
\end{align}
where $\px_{t'}=\px_t+\diri\,t'$. Next, we describe boundary conditions, and how they affect light transport.
}
\begin{figure}
  \def\svgwidth{.6\columnwidth}
\footnotesize
  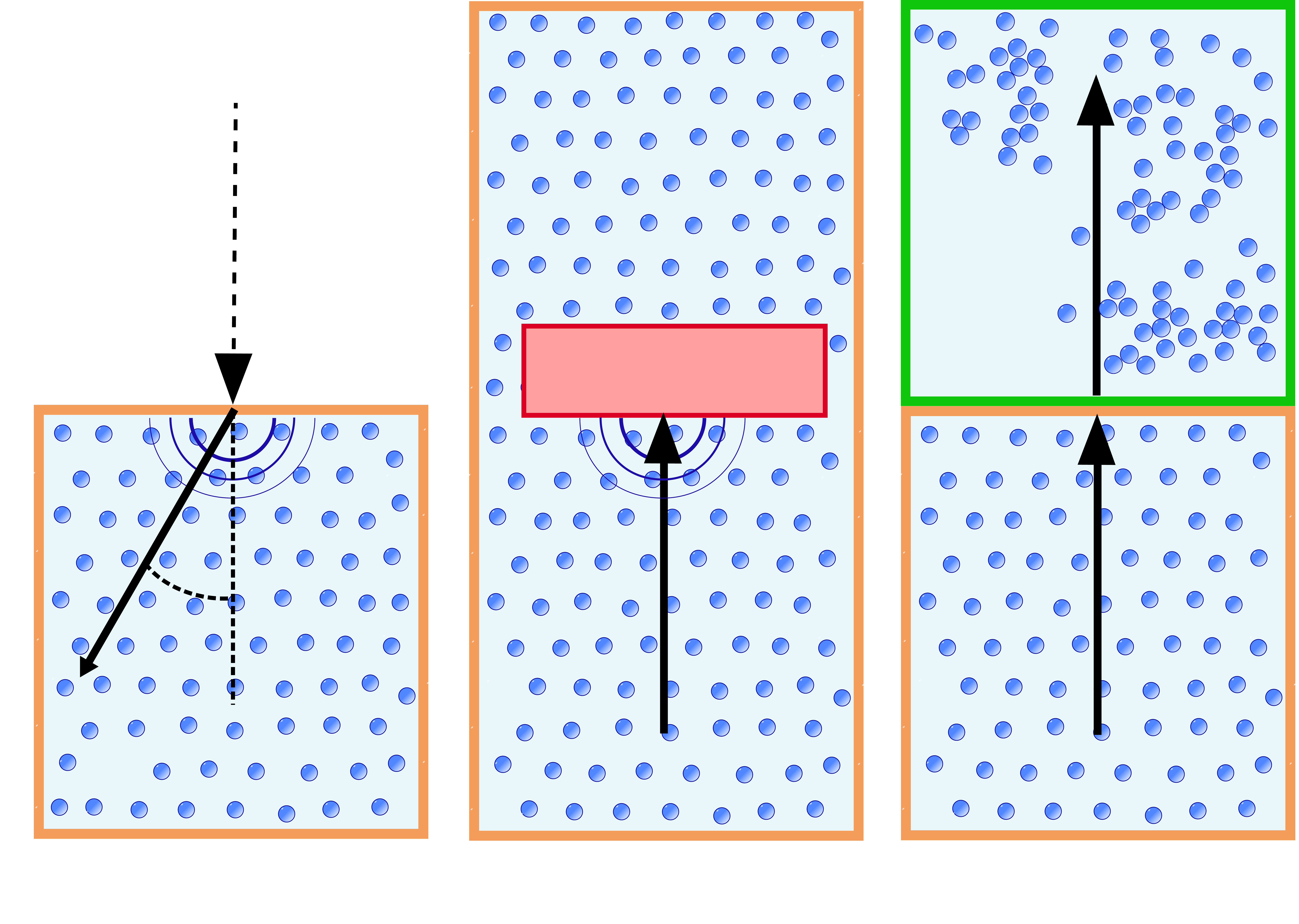
\caption{\new{Schematic example of the different boundary conditions: a) light entering a medium (\emph{Vacuum to Medium}); b) light being reflected from a boundary back into the medium (\emph{Medium to Surface}); and c) light crossing the interface between two different media (\emph{Medium to Medium}). 
Refer to the text for details. }}
\label{fig:boundaries}
\end{figure}
\begin{figure*}[t]
\centering
  \def\svgwidth{\textwidth}
\footnotesize
  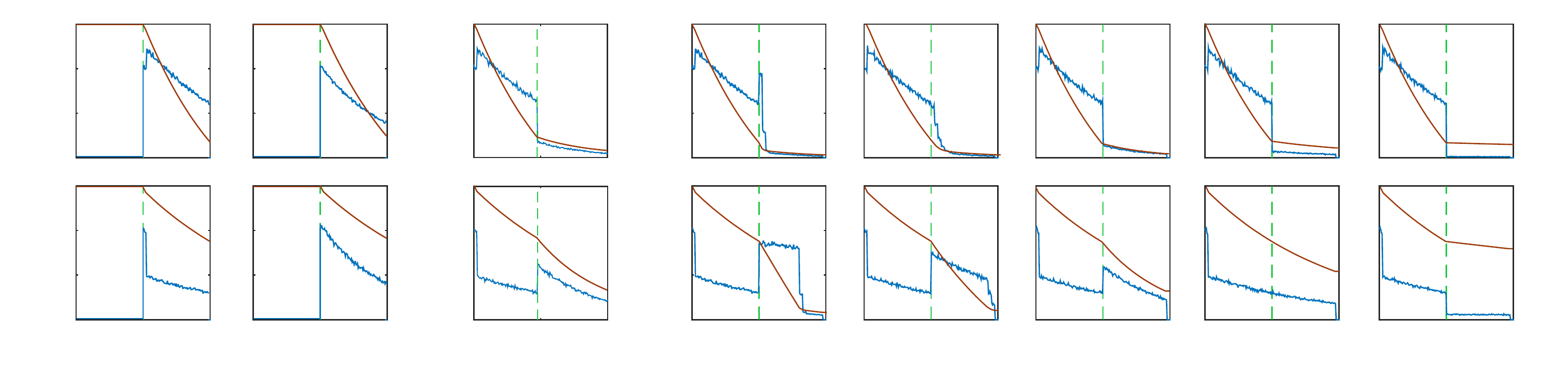
  \vspace{-.5cm}
\caption{\new{Probability of extinction $\sProb{t}$ (blue) and transmittance $\sTranmittanceCorr(t)$ (orange) as a function of $t$, for example cases of our three different boundary conditions. The vertical dotted line indicates the boundary.
The first medium has a negative correlation $\sCorr{1}=-0.5$ on the top row, and positive  $\sCorr{1}=0.5$ on the bottom.
(a) Vacuum to medium. 
When light enters the medium, it acts as a source term $\sEmission(\px,\diro)$, which depends on the angle of incidence $\sAngle$, since correlation might present some directionality (see \Fig{boundaries}).
(b) Medium to surface. As light is reflected on a surface boundary and changes direction, it acts as a directionally-resolved source $\sEmission(\px,\diro)$ which depends on the surface BSDF. 
%
%
c) Medium to medium, for a varying correlation $\sCorr{1,2}=[-0.9, 0.9]$ between the two media. 
For increasingly positive correlation $\sCorr{1,2}$ (high probability of the first medium shadowing the second), $p_{2}(t)$ becomes lower. 
For increasingly negative correlation $\sCorr{1,2}$ (low shadowing probability), $p_{2}(t)$ becomes higher near the boundary (then depends on $\sCorr{2}$). 
For uncorrelated media ($\sCorr{1,2}=0$), incoming photons can be modeled as sources at the entry boundary points, with $\sEmission(\px,\diro)$ dependent on the correlation of the second medium $\sCorr{2}$. 
For all cases modeled as light sources  $\sEmission(\px,\diro)$, $t$ is set to 0. Refer to the supplemental material for a more comprehensive set of examples. }}
\label{fig:freepaths_boundaries}
\end{figure*}
\subsection{Boundary Conditions}
\label{sec:boundary_conditions}
\new{The assumption that the medium is infinite and homogeneous ignores changes in correlation that occur at boundaries, such as photons entering a medium, the presence of surfaces inside, or the interface between two different media. 
}
%
%
%
\new{\Fig{boundaries} illustrates the different boundary conditions and their effects in light transport. Here, we describe them and show how to incorporate them to our model.
}

%
%

\new{
\paragraph{Vacuum to Medium (\Fig{boundaries}a): } This is the simplest case, where an \emph{uncorrelated} photon (from an uncorrelated medium or the vaccuum) enters a correlated medium. It can be modeled as a source $\sEmission_1(\px,\diro)$ at the entry boundary point $\px$, with $t=0$. 
\paragraph{Medium to Surface (\Fig{boundaries}b): }This case accounts for the interaction with surfaces such as a dielectric boundary, or an object placed inside the medium. Such surfaces are uncorrelated with respect to the medium. We can model this as a new source $\sEmission_2(\px,\diro)=\sRadICorr(\px,\diri)\sBCSDF(\px,\diri,\diro)$, with $\sRadICorr(\px,\diri)$ and $\sBCSDF(\px,\diri,\diro)$ the incoming radiance at $\px$ and the BSDF respectively, and setting $t=0$. 
%
\paragraph{Medium to Medium (\Fig{boundaries}c): } 
A photon crosses the interface between two different homogeneous media of different structure and correlation (this boundary condition therefore enables modeling heterogeneous media as well). The probability of extinction in the second medium $\sProbI{2}{t}$  depends not only on its correlation $\sCorr{2}$ and the correlation of the first medium $\sCorr{1}$, but also on the correlation \textit{between the two media} $\sCorr{1,2}$.  
}

\new{\Fig{freepaths_boundaries} shows results for all three boundary conditions; please refer to \Supp{app_boundaries} for a more exhaustive set of examples. }
\section{Rendering with the Extended GBE}
\label{sec:ggbe_rendering}
\new{
Up to this point, we have extended Larsen's original GBE formulation~\ParEq{larsengrte}, lifting the assumptions that made it unsuitable for rendering, presenting it also in integral form.
Our Extended GBE  \ParEqs{grteours}{integral_grte} supports an arbitrary mixture of scatterers (see \App{mixtures}), and accounts for the effect of different correlation between scatterers and sources. 
%
\newad{To use it for rendering, the required differential extinction probabilities can be tabulated~\cite{Larsen2011Generalized,Frank2010Generalized} by simulating via Monte Carlo an estimate of $\sProb{t}$, based on an explicit representation of the volume. This is a similar approach to our numerical results in \Supp{app_simulations}, and the validation curves computed by Meng et al.~\shortcite[Figure 6]{Meng2015Granular} to validate their uncorrelated radiative transfer-based approximation. Alternatively, and empirical $\sProb{t}$ can be used, fitting the observed transmittance in experimental setups, as is common in atmospheric sciences~\cite{Davis1999Horizontal}. In both cases, $\sExtinctionResolved(\px,t)$ is computed by inverting $\sProb{t}$ via \Eq{extinctionresolved}.}
%
%
}

\new{
In computer graphics, participating media are usually described in terms of their optical parameters. However, in our current formulation of the Extended GBE, there is no explicit connection with such parameters. 
In the following, we provide the missing connection: We formulate a model for correlated media based on the optical parameters commonly used in rendering, which is intuitive to use and easy to plug into our Extended GBE.
Then, we propose a simplified version of the model based on the assumption of positive correlation, which is easy to use and efficient to sample and evaluate. 
}

\subsection{Modeling Correlated Media from Optical Parameters}
\label{sec:local_optical_parameters}
\new{In a rendering context, the optical properties of a participating medium (e.g. extinction coefficients, scattering albedo, or phase function) are usually defined locally. Unfortunately, at the heart of our Extended GBE \ParEq{grteours} lies the differential extinction probability $\sExtinctionResolved(\px,t)$, whose \emph{memory effect} depends on the spatial correlation at neighboring points, and thus cannot be defined locally.
Our goal then is to model $\sExtinctionResolved(\px,t)$ and its derived quantities $\sProb{t}$ and $\sTranmittanceCorr(t)$, based on probability distributions of extinction $\sProbExtinction{\sExtinction}$. 
}
\revised{In the following, we assume both homogeneity in the neighborhood of $\px$ and isotropy, so we remove the spatial and angular dependence from the following derivations for clarity. }

\begin{figure}[t]
  \def\svgwidth{\columnwidth}
  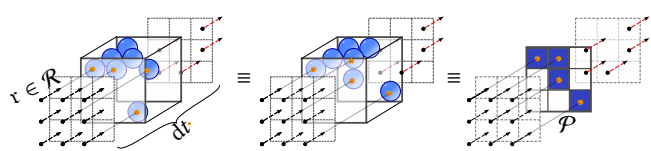
\caption{\new{Left and center: Examples of two differential volumes in a medium, each with different distribution of scatterers, but with a similar projection on the plane perpendicular to the direction of propagation (right). 
}}
\label{fig:kostinskis_differential}
\end{figure}

\new{Given a ray $\sRay$ in a medium, we can define its input radiance as $\sRad_i(\sRay)$, and its attenuation as $\sAttenuation(\sOpticalDepth_t(\sRay))$, \old{a probabilistic function of the optical depth $\sOpticalDepth_t(\sRay)$ at distance $t$}\revised{the ratio of input and output radiance of a single ray $\sRay$ defined as a probabilistic function, which depends on the ray's optical depth $\sOpticalDepth_t(\sRay)$}. 
Considering now a beam of light $\sBeamFront$ composed of several parallel rays $\sRay\in\sBeamFront$ (see \Fig{kostinskis_differential}), the total radiance $\sRad_o(t)$ traveling a distance $t$ in a correlated medium can be expressed as
%
\begin{align}
\sRad_o(t) & =  \int_\sBeamFront \sRad_i(\sRay) \,\sAttenuation(\sOpticalDepth_t(\sRay)) \,\diff{\sRay}.
\label{eq:light_transmitted_general_spatial}
\end{align}
In granular media~\cite{Moon2007Discrete}, where the correlation length is larger than a differential distance $\diff{t}$ (and usually larger than the granular particle's size), the probability of extinction $\sProb{t}$ depends on the distribution of scatterers along the direction of propagation of light, and needs to be taken into account explicitly. However, local correlation is assumed to be smaller than $\diff{t}$; this means that the exact position of the scatterers within the volume becomes irrelevant, and only their projection onto the plane $\mathcal{P}$ perpendicular to the propagation direction beam $\sBeamFront$ matters. 
%
%
We can then simplify the expression for the optical depth\footnote{``Optical depth'' is a standard term in physics, defined as the natural logarithm of the ratio of incident to transmitted radiant power through a material} to the homogeneous case where $\sOpticalDepth_t(\sRay)=\sExtinction(\sRay)\,t$\revised{, with $\sExtinction(\sRay)$ the density of scatterers found by an individual ray $\sRay$ when traversing the medium. }}
%

\new{However, explicitly integrating over all rays in $\sBeamFront$ is not practical. Instead, we would like to find a compact way of relating $\sRad_i(\sRay)$ to the extinction coefficient $\sExtinction(\sRay)$. 
We can remove its dependence on ray $\sRay$ by modeling $\sRad_i$ as a probability distribution $\sProbLight{\sRad_i}$ (e.g. by taking the histogram of $\sRad_i(\sRay)$), and explicitly relating it with the extinction coefficient $\sExtinction$ via a conditional probability distribution $\sProbExtinctionConditional{\sExtinction}{\sRad_i}$. We therefore transform \Eq{light_transmitted_general_spatial} into
\begin{align}
\sRad_o(t)& = \int_0^\infty \int_0^\infty \sProbLight{\sRad_i}\, \sProbExtinctionConditional{\sExtinction}{\sRad_i} \, \sRad_i\, \sAttenuation(\sExtinction \,t ) \,\diff{\sExtinction}\,\diff{\sRad_i}.
\label{eq:light_transmitted_general}
\end{align}
Defining $\hat{\sRad}_i=\int_\sBeamFront \sRad_i(\sRay)\,\diff{\sRay}=\int_0^\infty \sProbLight{\sRad_i}\, \sRad_i \,\diff{\sRad_i}$, and using $\sTranmittanceCorr(t)=\frac{\sRad_o(t)}{\hat{\sRad}_i}$  we get
\begin{align}
\sTranmittanceCorr(t) = \int_0^\infty \int_0^\infty \sProbLight{\sRad_i}\, \sProbExtinctionConditional{\sExtinction}{\sRad_i} \, \frac{\sRad_i}{\hat{\sRad}_i} \, \sAttenuation(\sExtinction \,t ) \, \diff{\sExtinction}\,\diff{\sRad_i},
\label{eq:transmittance_general}
\end{align}
}
\revisedSecond{which models transmittance $\sTranmittanceCorr(t)$ as a function of the correlation between the light and the distribution of local scatterers.}
\new{Finally, from \Eq{transmittance_general} we can compute the probability of extinction as $\sProb{t}=\left|\frac{\diff{\sTranmittanceCorr(t)}}{\diff{t}}\right|$, while $\sExtinctionResolved(t)$ can be obtained as $\sExtinctionResolved(t)=\sProb{t}/\sTranmittanceCorr(t)$ \ParEq{extinctionresolved}. }

\subsection{An intuitive local model for positively-correlated media}
\label{sec:rendering}

\new{\Eq{transmittance_general} is general and can model any type of correlation;
For the common case of positive correlation, 
we can set $\sAttenuation(\sExtinction \,t)=e^{-\sExtinction \,t}$ (see \cite{Kostinski2002} for details), and assume that $\sProbLight{\sRad_i}$ and $\sProbExtinction{\sExtinction}$ are independent, so that $\sProbExtinctionConditional{\sExtinction}{\sRad_i} = \sProbExtinction{\sExtinction}$. We can then rewrite \Eq{transmittance_general} as (see \Supp{app_simplification})
\begin{align}
\sTranmittanceCorr(t) &= \int_0^\infty \,e^{-\sExtinction\,t}\sProbExtinction{\sExtinction}\diff{\sExtinction}.
\label{eq:transmittance}
\end{align}
Using again the relationship in \Eq{extinctionresolved}, we obtain the differential extinction probability
\begin{align}
\sExtinctionResolved(t) &= \frac{\sProb{t}}{\sTranmittanceCorr(t)} = \frac{\int_0^\infty \,\sExtinction e^{-\sExtinction\,t}\sProbExtinction{\sExtinction}\diff{\sExtinction}}{\int_0^\infty \,e^{-\sExtinction\,t}\sProbExtinction{\sExtinction}\diff{\sExtinction}}.
\label{eq:diff_extinction_probability}
\end{align}
Note that this form of $\sProb{t}$ [numerator in \Eq{diff_extinction_probability}] is a generalization of the simple example in \Eq{prob_two_extinction} for a mixture of two different extinction coefficients. 
%
Assuming that the light distribution $\sProbLight{\sRad_i}$ from both sources $\sEmission_j$ and scatterers $S$ is uncorrelated with the scatterers distribution $\sProbExtinction{\sExtinction}$ 
, then $\sExtinctionResolved(t) = \sExtinctionResolvedI{\sRadI}(t) = \sExtinctionResolvedI{\sEmission_j}(t)$. } 


\newcommand{\angstructsize}{.19\textwidth}
\newcommand{\angstructspacesize}{.08cm}

\begin{figure}
\centering

\raisebox{-0.15cm}[0pt][0pt]{
\includegraphics[width=.24\textwidth]{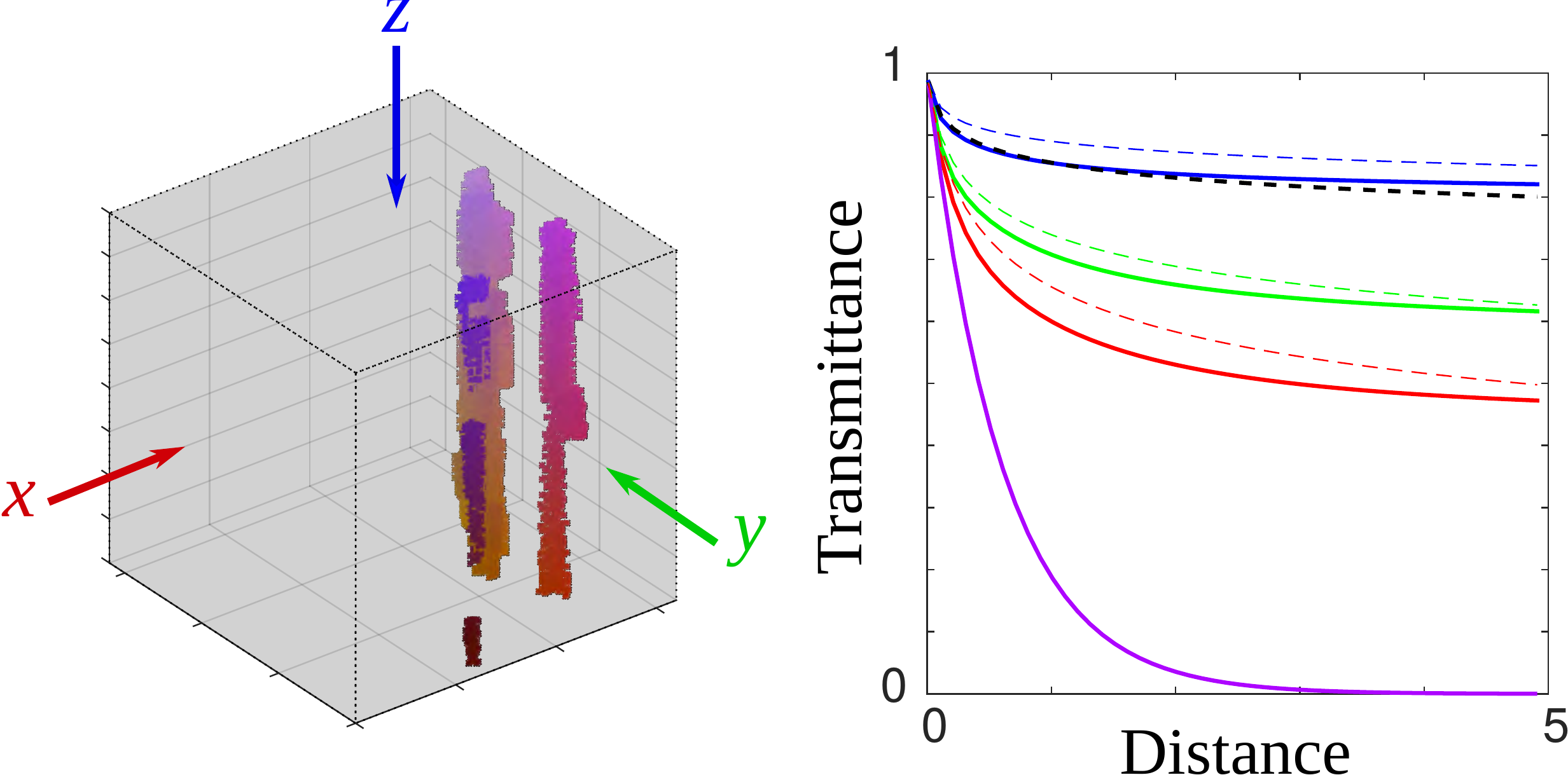} \hspace{\angstructspacesize}}
\includegraphics[width=\angstructsize]{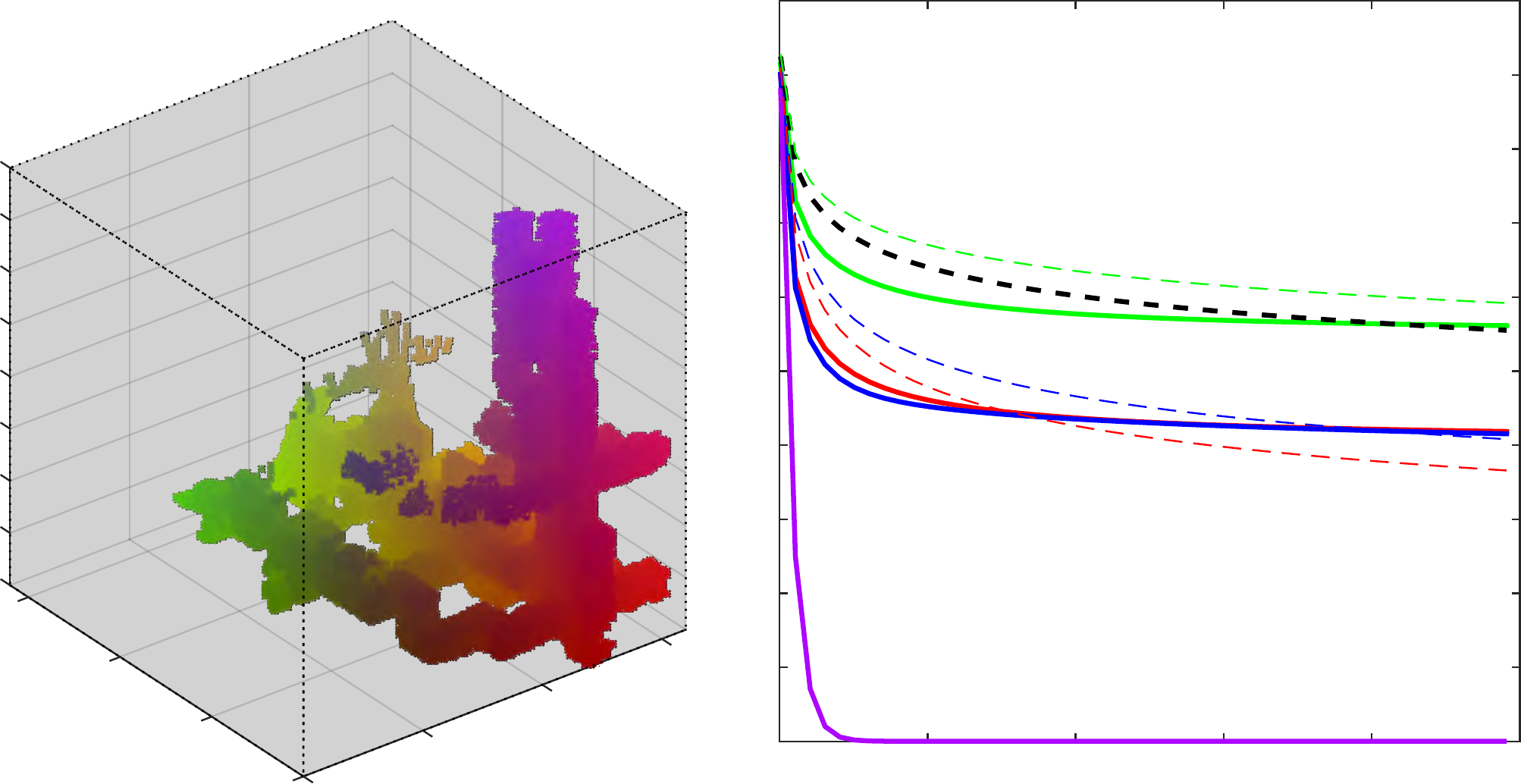} \hspace{\angstructspacesize} \\
\includegraphics[width=.24\textwidth]{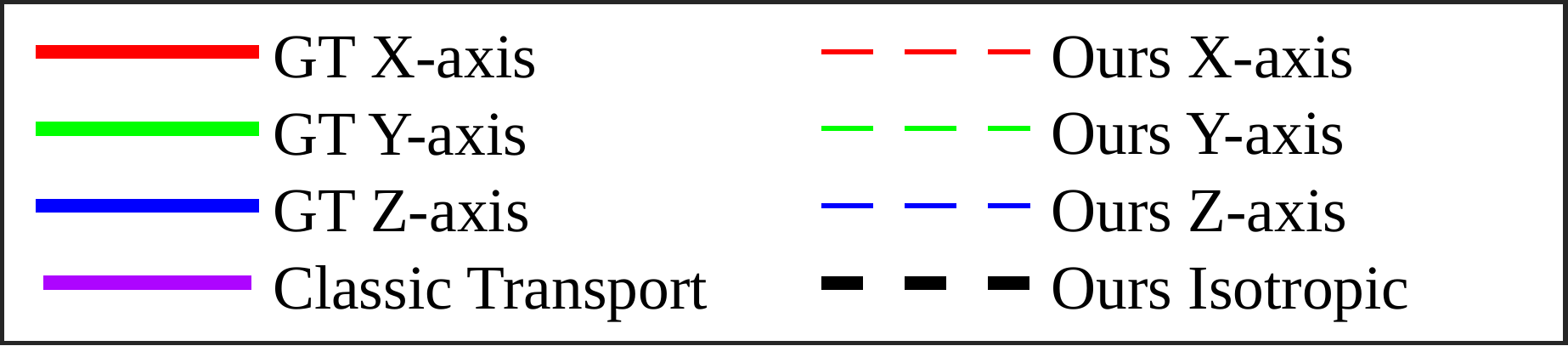} \hspace{\angstructspacesize} 
\includegraphics[width=\angstructsize]{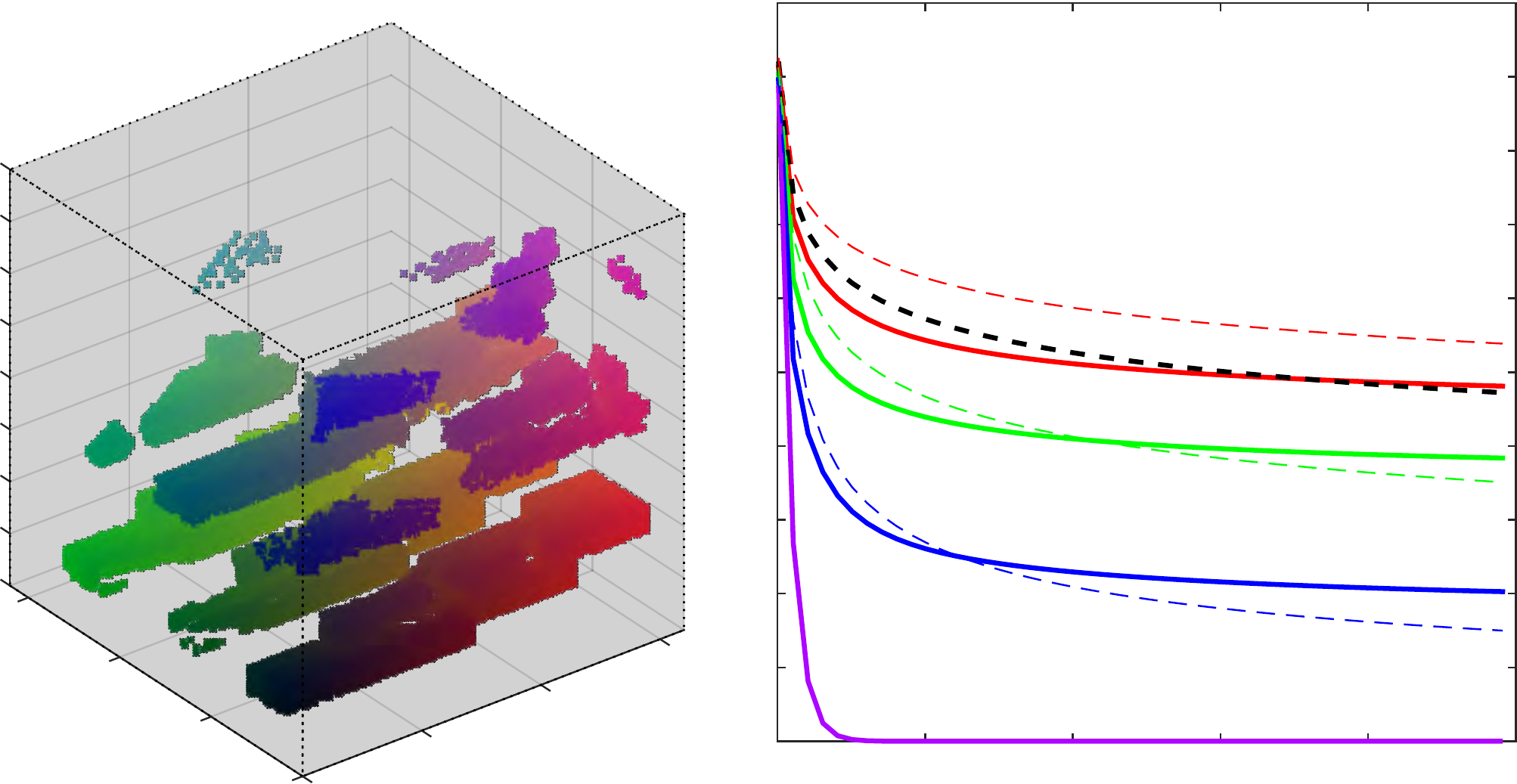} \hspace{\angstructspacesize}
 \caption{Transmittance in high-resolution volumes of locally-correlated media (procedurally generated after \cite{Lopezmoreno2014Gpu}). Beams of light travel through each volume, aligned in succession to the $x$, $y$, and $z$ axes. Ground-truth transmittance (red, green, and blue solid lines) has been computed by brute force regular tracking~\cite{Amanatides1987}, while our simulation (dotted lines) uses the gamma distribution proposed in Equation~\ref{eq:gamma}.
Classic transport governed by the RTE significantly overestimates extinction through the volume, resulting in a exponential decay (purple line). In contrast, our model matches ground-truth transmission much more closely. The black dotted line is the result of isotropic correlation, which is clearly also non-exponential. Please refer to \Fig{app_angular_structure} in the supplemental for more examples. }
%
\label{fig:angular_structure}
\end{figure}

\paragraph{Finding a good distribution $\sProbExtinction{\sExtinction}$}
\label{sec:rendering}
To be able to use \Eqs{transmittance}{diff_extinction_probability}, we need to find a good optical depth distribution $\sProbExtinction{\sExtinction}$ for the medium. 
Taking the average scatterers' cross section $\sCrossSection$, we can define
\begin{equation}
\sExtinction\,\sProbExtinction{\sExtinction}= \sConcentration\,\sProbConc{\sConcentration}\, \sCrossSection,
\label{eq:isotropic_structure}
\end{equation}
where $\sConcentration$ is the scatterers concentration and $\sProbConc{\sConcentration}$  its probability distribution.
%
To find a practical $\sProbConc{\sConcentration}$ we analyzed a wide range of high-resolution volumes exhibiting different correlation (see \Fig{angular_structure} for some examples). We observed that a gamma distribution fits $\sProbConc{\sConcentration}$ reasonably well, so that 
\begin{equation}
\sProbConc{\sConcentration}\approx \dGamma{\sConcentration}{\alpha}{\beta}=\frac{\beta^\alpha\sConcentration^{\alpha-1}e^{-\sConcentration\beta}}{\fGamma{\alpha}},
\label{eq:gamma}
\end{equation}
with  $\alpha=\Mean{\sConcentration}^2\cdot\Var{\sConcentration}^{-1}$, $\beta=\Mean{\sConcentration}\cdot\Var{\sConcentration}^{-1}$, and $\fGamma{\alpha}$ the gamma function. 
Moreover, previous research has shown that the gamma distribution is also very adequate for modeling the concentration of turbulent media such as clouds~\cite{Barker1996Parameterization}, or particulate media~\cite{Peltoniemi1992}.

\Eq{gamma} provides a compact and intuitive description of the statistical properties of $\sProbConc{\sConcentration}$ (and in turn of $\sProbExtinction{\sExtinction}$ in \Eq{isotropic_structure}), by only using its mean $\Mean{\sConcentration}$ and variance $\Var{\sConcentration}$ (intuitively, a higher variance indicates clusters of scatterers with gaps between them). 
%
In contrast, traditional (uncorrelated) media depend only on the mean concentration $\Mean{\sConcentration}$, and assume $\Var{\sConcentration}=0$. 
For simplicity, we have assumed that both $\sProbConc{\sConcentration}$ and $\sCrossSection$ are isotropic. \App{directional} shows how to add directional dependencies as $\sProbConcConditional{\sConcentration}{\diro}$ and $\sCrossSection(\diro)$.

\paragraph{Rendering} 
Using \Eqs{gamma}{transmittance}, and noting that the latter is related with the moment distribution function $\sMGFProb{t}$ of $\sProbConc{\sConcentration}$ as $\sTranmittanceCorr(t)=\sMGFProb{-\sCrossSection\,t}$~\cite{Davis2014Generalized},  we can compute the transmittance, probability of extinction, and differential probability of extinction as
\begin{align}
\label{eq:gamma_analytic_transmittance}
\sTranmittanceCorr(\sDistance) & = \left(1+\frac{\sCrossSection\cdot\sDistance}{\beta}\right)^{-\alpha}, \\
\label{eq:gamma_analytic_extinction_probability}
\sProb{\sDistance} & = \frac{\alpha\,\sCrossSection}{\beta}\,\left(\sCrossSection\frac{t}{\beta} + 1\right)^{-(1+\alpha)},  \\
\label{eq:gamma_analytic_differential_extinction_probability}
\sExtinctionResolved(\sDistance) & = \sCrossSection\frac{\alpha}{\beta}\, \left( 1 + \frac{\sCrossSection}{\beta}\,t\right)^{-1}.
\end{align}
%
\begin{figure}[t]
\centering
  \def\svgwidth{1.1\columnwidth}
  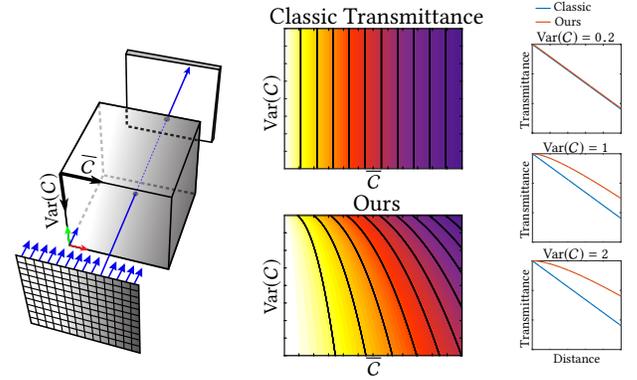
\caption{Comparison between traditional transmittance as predicted by the Beer-Lambert law, and our transmittance for correlated media. The scene consists of a cube embedding a participating medium, placed in front of a light source. The media has a constant cross section $\sCrossSection=1$, and increasing scatterers concentration $\Mean{\sConcentration}$ and correlation (i.e. density variance $\Var{\sConcentration}$) along the horizontal and vertical axes, respectively. 
Correlation does not affect transmittance in the classic model, which follows the Beer-Lambert law as shown in the log-scale plots on the right. In contrast, our model captures the slower-than-exponential decay as variance $\Var{\sConcentration}$ increases. We use  $\Mean{\sConcentration}=1$ for the plots. Figure after \cite{Novak2014}.}
\label{fig:gamma_vs_delta}
\end{figure}
\noindent 
In \Fig{angular_structure} we analyze the performance of our analytic expression of transmittance for correlated media in \Eq{gamma_analytic_transmittance}, against the exponential transmittance predicted by the Beer-Lambert law, and ground-truth transmittance computed by brute force regular tracking~\cite{Amanatides1987}. Our model is much closer to the ground truth than the result of classic light transport, which significantly overestimates extinction through the volume.
\Fig{gamma_vs_delta} explores our closed-form of transmittance: As variance increases, the slower-than-exponential behavior becomes more pronounced, as observed by Davis and Mineev-Weinstein~\shortcite{Davis2011} when analyzing the frequency of density fluctuations in correlated media. This effect is not captured by classic light transport. 

%
In a Monte Carlo renderer, we can compute a random walk by sampling transmittance using the probability defined in \Eq{gamma_analytic_extinction_probability}. However, as opposed to the classic exponential transmittance in Beer-Lambert law, $\sProb{\sDistance}$ is not proportional to $\sTranmittanceCorr(\sDistance)$, which may lead to increase variance of the estimate. To sample with a probability $\sProb{t}\propto\sTranmittanceCorr(\sDistance)$, assuming $\alpha > 1$ (i.e. $\Mean{\sConcentration}>\sqrt(\Var{\sConcentration}$) we can define $\sProb{\sDistance}$ as 
\begin{equation}
\sProb{\sDistance} = -\sCrossSection\frac{1-\alpha}{\beta} (1+\frac{\sCrossSection}{\beta}\sDistance)^{-\alpha} = -\sCrossSection\frac{1-\alpha}{\beta}\,\sTranmittanceCorr(\sDistance),
\label{eq:gamma_pdf_transmittance}
\end{equation}
which can be sampled using its inverse cdf
\begin{equation}
\sDistance(\xi) = -\frac{\beta}{\sCrossSection}\left(1-\sqrt[1-\alpha]{1-\xi}\right),
\label{eq:gamma_cdf_transmittance}
\end{equation}
with $\xi \in [0,1]$ a uniform random value. 
\new{
When the sampled distance $\sDistance$ is longer than the distance $\sDistance'$ to a boundary condition, the probability of an intersection at $\sDistance'$ becomes
\begin{equation}
\sProb{\sDistance'} = \left(1 + \frac{\sCrossSection}{\beta}\,\sDistance'\right)^{1-\alpha}.
\label{eq:gamma_invpdf_transmittance}
\end{equation}
%
}
%
\noindent We refer to \Supp{app_sampling_gamma_dist} for more detailed derivations, including the general case where $\alpha \leq 1$. 

\paragraph{Implementation}
\newad{While correlated media can be implemented as a volumetric definition in most renderers, there are a few details that need to be taken into account. The most important one is that the constants used when solving the classic RTE (e.g. $\sAlbedo$ or $\sExtinctionResolved(t)=\sExtinction$) are now defined as a function of $t$. As such, most of the optimizations typically done in the photon's random walk due to terms cancellation cannot be directly applied here. 
Additionally, the different correlations between scatterers and sources in \Eq{integral_grte} require keeping track on the previous vertex of the path when sampling a new one (via selecting either $\sProbI{\sEmission}{t}$ or $\sProbI{\sRadI}{t}$). This is also important when connecting with the light source via next-event estimation, where the source's transmittance and differential scattering probability need to be applied. }


\begin{figure}[t]
  \centering
  \def\svgwidth{1\columnwidth}
\footnotesize
  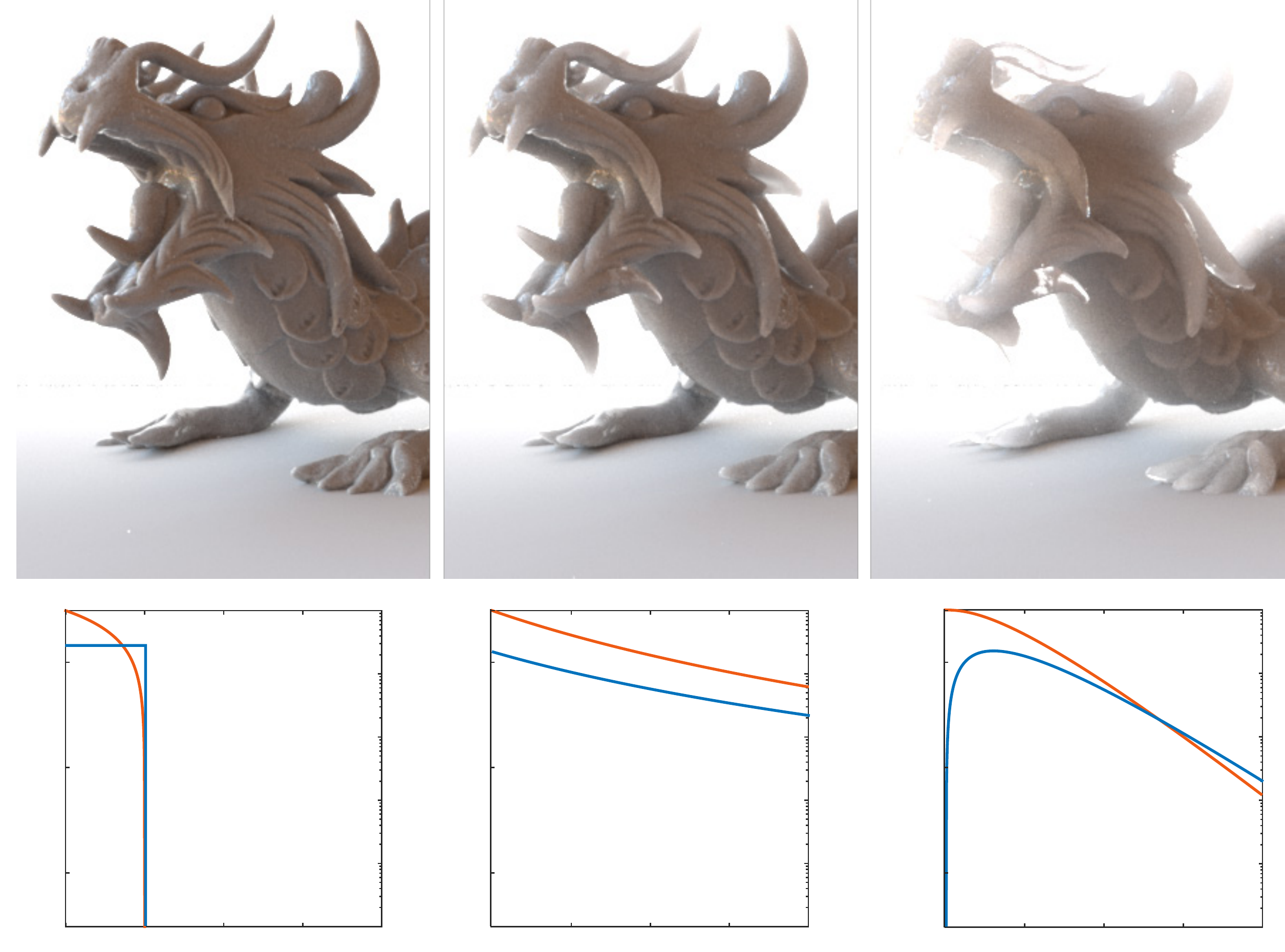
\caption{\new{Materials with different types of probability distributions of extinction $\sProb{t}$ (shown in the the bottom plots in blue, while transmittance $\sTranmittanceCorr(t)$ is shown in orange; both cases are in log-scale). From left to right: Negative correlation with linear extinction; a power-law $\sProb{t}$ resulting from our local model (\Sec{rendering}), with $\Var{\sConcentration(\px)}=1$; and an example of one empirical $\sProb{t}$ following a gamma distribution with $\Var{t}=.1$ (see \Supp{app_addpathlengths} for details). In all cases the mean extinction $\Mean{\mu}=$\SI{2}{\per\meter}, albedo $\sAlbedo=.8$, and isotropic phase function.} 
}
\label{fig:degrees_correlation}
\end{figure}

\begin{figure}[t]
\centering
\includegraphics[width=\columnwidth]{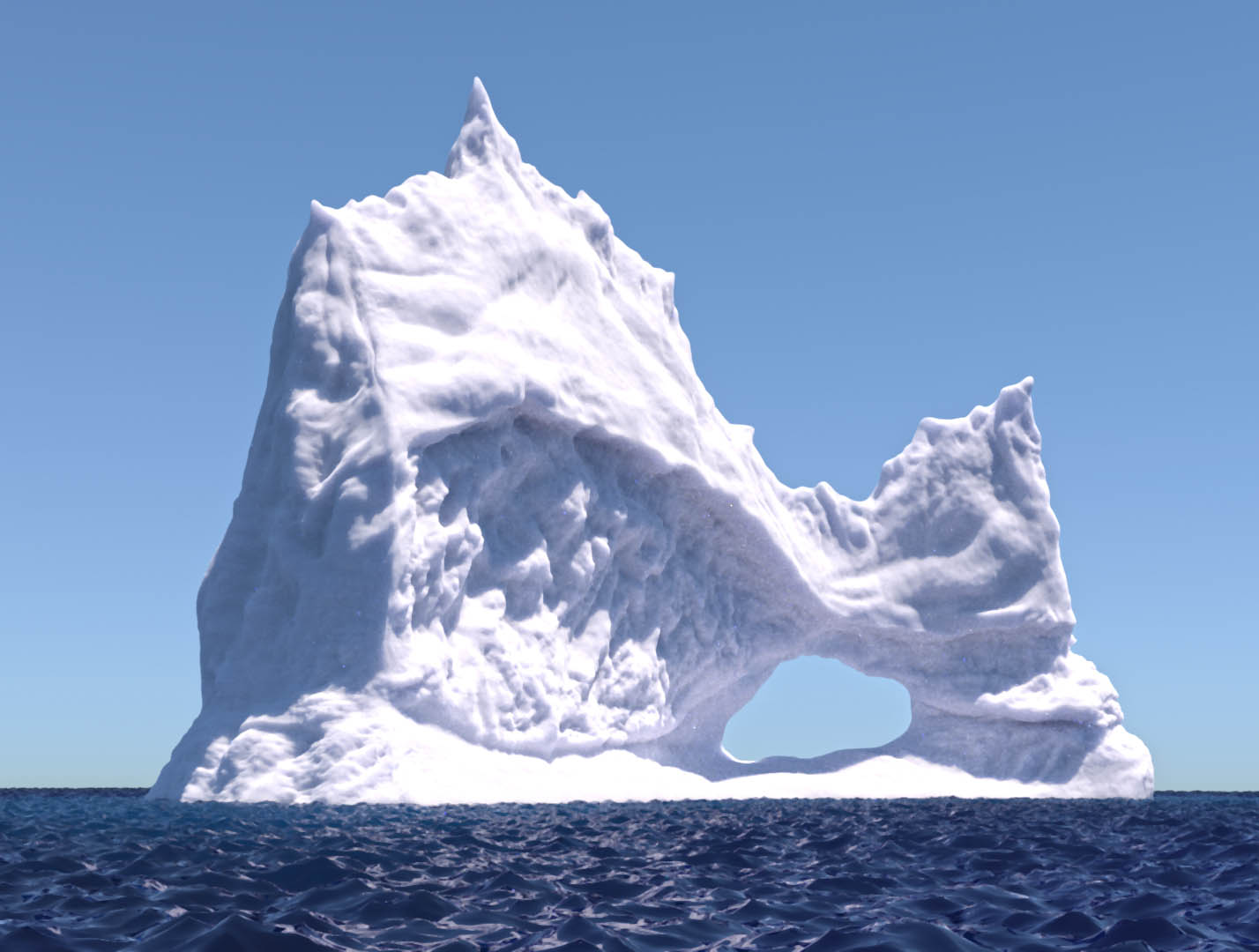}\\
\caption{\new{Rendering of an iceberg made of compacted snow (with snow's spectral cross section $\sCrossSection$ and albedo $\sAlbedo$ after Frisvad and colleagues~\shortcite{Frisvad2007}), using our model in \Sec{rendering}. }}
\label{fig:iceberg}
\end{figure}

\begin{figure*}[t]
  \centering
  \def\svgwidth{.99\textwidth}
  \input{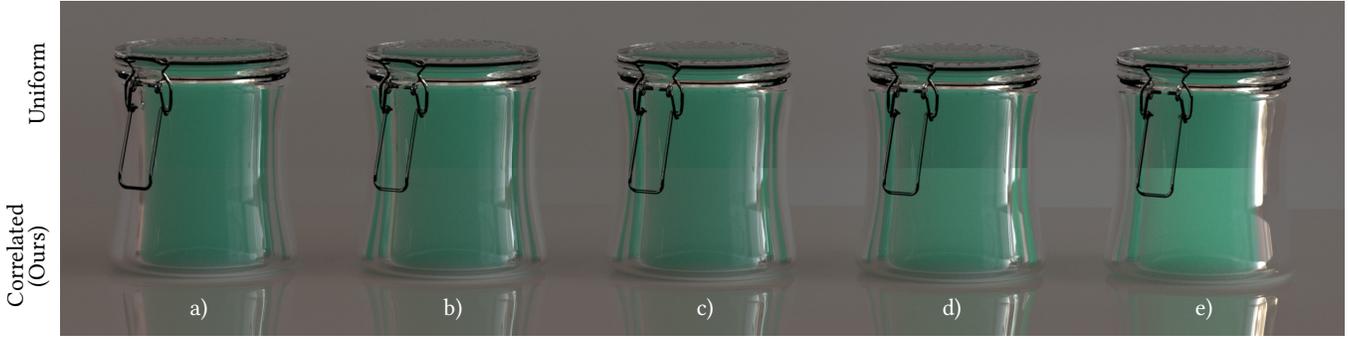}
    \caption{\new{Effect of local correlation in media with mean extinction of $\Mean{\sExtinction}=2$ and albedo $\sAlbedo = \{.1, .8, .6\}$, and increasing variance $\Var{\sConcentration(\px)}$: a) 0, b) 8, c) 12, d) 16, and e) 32. The top half shows the result of classic light transport, while the bottom half shows the result of our model. For $\Var{\sConcentration(\px)}=0$ the result is identical to classic light transport. }}
  \label{fig:jar_scattering}
\end{figure*}
\begin{table}[h]
\centering
\begin{tabular}{c c c c}
\hline
Figure & \# Samples & Uncorrelated & Correlated \\
\ref{fig:isotropic_dragons} & 4096 & 53 m & 45 m / 58 m\\
\ref{fig:degrees_correlation} & 4096 & 30 m & 33 m / 35 m \\
\ref{fig:iceberg} & 2048 & 185 m & 213 m \\
\ref{fig:jar_scattering} & 4096 & 26 m & 28 m\\
\ref{fig:cylinder} & 2048 & 5.6 m & 5.8 m \\
\ref{fig:anisotropic_dragons} & 8192 & 70 m &  82 m \\
\ref{fig:vs_rte} & 4096 & 17.71 m & 18.8 m \\ 
\hline
\end{tabular}
\caption{Computational cost for the images shown in the paper, for both uncorrelated (traditional model) and correlated media (ours). 
When different types of correlation are used, we show two measurements (positive / negative). 
}
\label{table:timings}
\end{table}
\begin{figure*}[t]
  \centering
  \def\svgwidth{1\textwidth}
  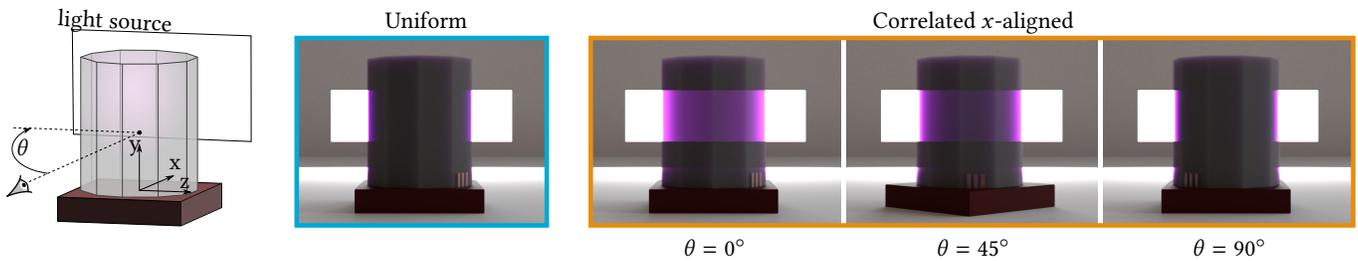
  \caption{Effect of directionally-dependent correlation on transmittance. The prism rotates around the y-axis. With uncorrelated media, appearance does not change with rotation. With a highly-correlated medium, appearance changes significantly as the prism rotates, according to the degree of alignment between the correlation and the view vector. The figure shows the case of x-axis-aligned correlation. \revised{Illustrative examples on the distribution of particles for each case are shown in \Fig{correlation}.}
  Please refer to the supplemental for the video. }
  \label{fig:cylinder}
\end{figure*}

\section{Results}
\label{sec:results}
In this section we show results using our new model for spatially-correlated participating media, including comparisons against the traditional RTE. We have implemented the integral form of our Extended GBE \ParEq{integral_grte} as a volumetric definition in Mitsuba~\cite{Mitsuba}.
\new{For materials with negative correlation we have used a linear transmittance decay (see \Supp{app_addpathlengths} for details); for positive correlation, we used our local model in \Sec{rendering}. }
Unless stated otherwise, we assume positively correlated media in our results.
All our tests were performed on an Intel Core i7-6700K at 4GHz with 16 GB of RAM.

The cost introduced by sampling and evaluating the correlated transmittance 
with respect to classical transmittance is negligible in comparison to the cost of tracing samples. Simulation parameters and timings are shown in \Tab{timings}; note that negatively correlated media tend to create longer paths, therefore increasing the total rendering cost. \revised{In terms of convergence, in some cases the pdf might not be proportional to the sampled transmittance [e.g. in \Eq{gamma_invpdf_transmittance}], which in turn might increase variance, we did not observe a strong effect when incorporating non-exponential transport. In \Supp{app_convergence} we analyze the convergence experimentally. }  

\Fig{isotropic_dragons} shows volumetric renderings of translucent dragons made of materials with the same density, but different correlation. The middle image shows positive correlation, following a gamma distribution with $\Var{\sConcentration}=40$. \new{On the right we show negative correlation, exhibiting linear transmittance.} In the three cases the media have scattering albedo $\sAlbedo=.8$, and mean extinction $\Mean{\sExtinction}=$~\SI{10}{\per\meter}. The net effect, due to the faster-than-exponential (negative correlation) and slower-than-exponential transmittance (positive correlation), is clearly visible in the final images.

\new{\Fig{degrees_correlation} highlights the versatility of our framework, with different scatterers correlation: negative correlation with linear transmittance decay, positive correlation according to our model, and an empirical distribution of $\sProb{t}$ (modeled as a gamma distribution, see \Supp{app_addpathlengths}). The mean extinction is in all cases \revised{$\Mean{\mu}= $\SI{2}{\per\meter}, with albedo $\sAlbedo=.8$}. Both the particles concentration and the cross section are isotropic. 
\Fig{iceberg} shows another non-exponential probability of extinction on granular compacted snow, using our model in \Sec{rendering}. Optical parameters of the snow have been computed after Frisvad et al.~\shortcite{Frisvad2007}.} 

In \Fig{jar_scattering} we analyze the effect of correlation with increasing variance [increasing $\Var{\sConcentration}$ in \Eq{gamma}]. The top half of the jars has been rendered with the classic RTE, and thus remain constant independent of the degree of correlation, as expected. The bottom half shows the result of our model; note that for $\Var{\sConcentration}=0$ the result converges with classic light transport.  

\Fig{cylinder} shows the effect of directional correlation. The scene is made up of a volumetric prism with very low scattering albedo, so the dominant effect is transmittance, and a strong rectangular area light placed behind it. The prism rotates around its $y$-axis. The first prism is made up of an uncorrelated medium, while the other three show a strong positive correlation along the $x$-axis, with $\Var{\sConcentration}=.5$; when the rotation angle is $\theta=0^\circ$, correlation is perfectly aligned with the x-axis [similar to the situation depicted in \Fig{correlation} (c)].  Both uncorrelated and correlated media have a mean particles concentration $\Mean{\sConcentration}=\{.8, 1.6, .7\}$ (RGB), and a mean cross section $\Mean{\sCrossSection}=1$. 
For the uncorrelated media, no changes occur in appearance as the prism rotates, as expected. For the x-aligned correlation, transmittance varies significantly as correlation progressively becomes unaligned with the view vector. 
We refer the reader to the supplemental video for the full animation, including other directions of correlation. 
\begin{figure*}[t]
  \centering
  \def\svgwidth{1\textwidth}
  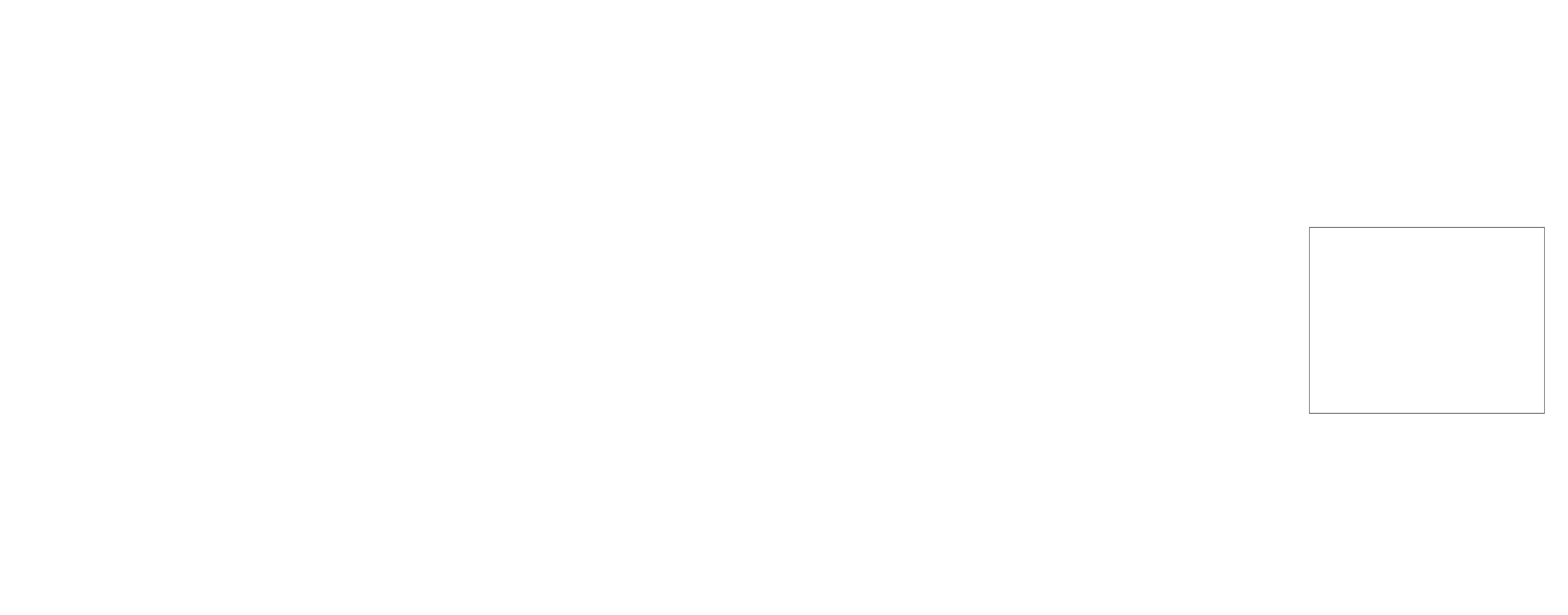
\caption{
Effect of directional correlation with varying albedos $\sScattering$. From left to right: uniformly distributed media, directionally correlated in the $x$ axis (aligned to the camera view), directionally correlated in $z$ axis, and with isotropic correlation. For all cases we keep $\Mean{\sConcentration}=10$, $\sCrossSection=1$, and $\Var{\sConcentration}=10$ in the main axis of correlation, while for the remaining directions is close to zero (so the mean free path is similar to the predicted by Beer-Lambert law). To the right we plot intensity values (green scanline shown in one of the dragons) for each medium, showing the differences between classic light transport and our approach.    }
   \label{fig:anisotropic_dragons}
\end{figure*}
\Fig{anisotropic_dragons} systematically analyzes the effect of directional correlation for varying albedos $\sAlbedo$, including uncorrelated media, isotropic correlated media, and directionally correlated media aligned with the $x$ and $z$ axes (being $y$ the up-vector). These four scenarios roughly correspond to the ones depicted in \Fig{correlation}. 

\begin{figure}[h]
  \centering
  \def\svgwidth{1\columnwidth}
  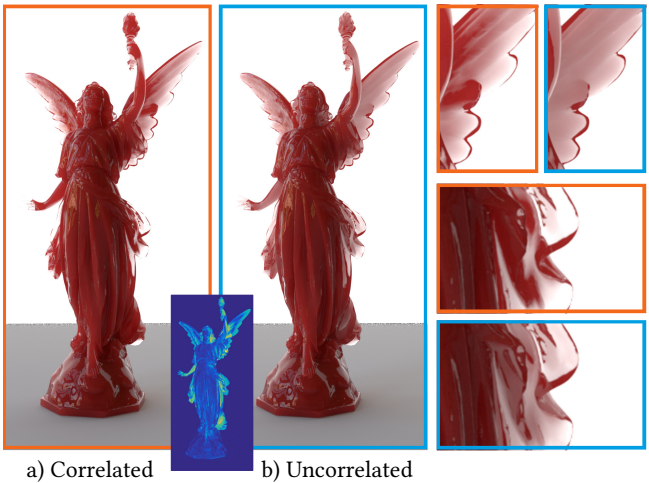
\caption{\new{Left: Render using our framework for correlated materials. Right: Render using the RTE, where the optical parameters of the material have been adjusted trying to match the appearance of the correlated case. Dissimilarities are evident, specially in thinner areas, since the extinction curves and diffusive behavior in both models are different (see false-color inset and zoomed-in areas). }}
   \label{fig:vs_rte}
\end{figure}
\new{
Last, in \Fig{vs_rte} we investigate if adjusting the optical parameters of an uncorrelated medium and using the classic RTE could produce the same results as our model for correlated media. 
In particular, we render the first statue with a correlated material (mean extinction $\Mean{\sConcentration}=20$, isotropic phase function, and scattering albedo $\sAlbedo={.8,.1,.1}$), and render an uncorrelated version adjusting $\Mean{\sConcentration}=45$, with the same phase function and scattering albedo.  
Although tweaking the parameters of the RTE can lead to an overall similar appearance, it cannot correctly reproduce the details due to the different extinction curves and diffusive behavior in both models (see also \cite{dEon2014diffusion}).
}

%

\section{Conclusions}
\label{sec:discussion}

We have introduced a novel framework to simulate light transport in spatially-correlated media, where the probability of extinction and transmittance no longer follow an exponential decay, as predicted by the Beer-Lambert law. 
\new{We have presented the Extended Generalized Boltzmann Equation, lifting the main limiting assumptions of the original GBE, and making it suitable for rendering applications. Our framework supports multiple sources, mixtures of particles, and directional correlation. } 
\new{In addition, we have proposed an intuitive model based on local optical properties for the most common case of positive correlation, providing a close-form solution for transmittance, without the need for costly numerical simulation or precomputations, allowing to model $\sExtinctionResolved(t)$ based on local definitions of $\sExtinction$. 
%
Interestingly, Davis and Xu~\shortcite{Davis2014Generalized} empirically proposed a similar expression to this model for transmittance in clouds.
%
However, the authors stated that an integro-differential counterpart of their formulation was yet unknown. Our \Eq{diff_extinction_probability} links this form of transmittance with the GBE, which is in turn an integro-differential equation. }
%

\new{
\paragraph{Limitations and future work} 
Our theoretical framework in \Sec{ourgbe} is general, and supports heterogeneous media through the medium-to-medium boundary condition.  
\old{In concurrent work, Camminady et al. proposed a solution for the simplest case, where the two media have identical structure $\sCorr{1}=\sCorr{2}=\sCorr{1,2}$, and therefore the probability of extinction $\sProb{t}$ only varies due to the different media density; however, finding an efficient, general solution remains a challenging problem.
This is because the differential probability of extinction $\sExtinctionResolved(\px)$ at point $\px$ affects $\sExtinctionResolved(\px+\diro\,\diff{t})$, according to the scatterers correlation at points $\px$ and $\px+\diro\,\diff{t}$, and to the cross-correlation between these two points. This means that we cannot model the probability of extinction in correlated heterogeneous media as the integral of the \emph{local} differential extinction probabilities along the ray, as in uncorrelated heterogeneous media. 
Revisiting numerical techniques for computing unbiased transmittance in heterogeneous media is thus an interesting topic of future work, since it is unclear how the underlying theory of virtual particles in existing methods could be adapted to correlated scatterers. }
\revised{However, practical implementation of such heterogeneities for continuous media is still challenging. This is because the differential probability of extinction $\sExtinctionResolved(\px)$ at point $\px$ affects $\sExtinctionResolved(\px+\diro\,\diff{t})$, according to the scatterers correlation at points $\px$ and $\px+\diro\,\diff{t}$, and to the cross-correlation between these two points. This means that the probability of extinction in correlated heterogeneous media cannot be modeled as the integral of the \emph{local} differential extinction probabilities along the ray, as in uncorrelated heterogeneous media. 
In concurrent work, Camminady et al.~\shortcite{Camminady2017} proposed a solution for the simplest case, where the two media have identical structure $\sCorr{1}=\sCorr{2}=\sCorr{1,2}$, and therefore the probability of extinction $\sProb{t}$ only varies due to the different media density; however, finding an efficient, general solution remains a challenging problem.
Revisiting numerical techniques for computing unbiased transmittance in heterogeneous media is thus an interesting topic of future work, since it is unclear how the underlying theory of virtual particles in existing methods~\cite{Woodcock1965,Coleman1968,Novak2014,Szirmay2017,Kutz2017} could be adapted to correlated scatterers. }
\revised{From a physical point of view, it would also be interesting to introduce in our \old{generalized RTE} Extended GBE~\ParEq{grteours} support for refractive media~\cite{gutierrez2006atmo,Ament2014}, as well as vector or bispectral scattering~\cite{Jarabo2018bidirectional}.}
%

Other open problems include extending our local model \ParEq{transmittance} to the case of negatively correlated media, thus removing precomputation or the definition of an empirical $\sProb{t}$, or finding a model for the continuous transition between correlated and particulate media. } 
\revised{While for perfect negative correlation we can model light-particle interactions as a Bernoulli process (see \Supp{app_negative}), for other degrees of negative correlation this process is not obvious: We hypothesize that such cases could be modeled as a mixture of Poissonian and Bernoulli processes, although an analytical model for negative correlation remains an open challenge that deserves a more in-depth exploration.}
\new{Finally, for our final model we have chosen a gamma distribution for $\sProbExtinction{\sExtinction}$; however other probability distributions might work better depending on the scenario. \toSuppMaterial{For completeness, we analyze additional distribution functions in the supplemental material, including their mathematical expressions to be used for rendering.} \revised{Moreover, our directionally-resolved model for the variance of the distribution might be too smooth for materials with high-frequency details: in such cases, a mixture of ellipsoids (similar to the approach of Zhao et al.~\shortcite{Zhao2016}) could result in a more accurate fit. }}

\begin{sloppypar}Our definition of locally-correlated light transport may be suitable for filtering volumetric appearances, avoiding costly optimization procedures~\cite{Zhao2016}, or ad-hoc shadowing functions~\cite{Schroder2011Volumetric}. 
%
Accelerating rendering of particulate media~\cite{Meng2015Granular,Muller2016Efficient} is another area that could benefit from our locally-correlated model. Introducing our compact representation into the shell transport functions proposed by Moon et al.~\shortcite{Moon2007Discrete} and M\"{u}ller et al.~\shortcite{Muller2016Efficient} could significantly decrease the storage cost of these representations. 
Last, similarity theory~\cite{Wyman1989,Zhao2014} is an important tool for accelerating light transport within the RTE. Redefining this theory within locally-correlated radiative transport is another interesting avenue of work, specially given the additional degrees of freedom introduced by the non-exponential probability of extinction $\sProb{t}$.
\end{sloppypar}

\appendix

\section*{Acknowledgments}
We thank Miguel Angel Otaduy, Carlos Castillo and Jorge Lopez-Moreno for comments and discussions on early stages of the project and the dataset in \Fig{angular_structure}; 
Julio Marco, Adolfo Mu\~{n}oz, and Ib\'{o}n Guill\'{e}n for discussions and proof-reading; Pilar Romeo for help with the figures; and the reviewers for the in-depth reviews. 
This project has been funded by the European Research Council (ERC) under the EU's Horizon 2020 research and innovation programme (project CHAMELEON, grant No 682080), DARPA (project REVEAL), and the Spanish Ministerio de Econom\'ia y Competitividad (projects TIN2016-78753-P and TIN2014-61696-EXP).

\bibliographystyle{ACM-Reference-Format}
\bibliography{bibliography}

\renewcommand{\thefigure}{S.\arabic{figure}}
\renewcommand{\theequation}{S.\arabic{equation}}
\renewcommand{\thesection}{S.\arabic{section}}

\renewcommand{\suppimages}{images}
\renewcommand{\Supp}[1]{\Sec{#1}}
\renewcommand{\Supps}[2]{\Secs{#1}{#2}}


\section{Mixtures of scatterers}
\label{app:mixtures}
\new{For media made up of a mixture of scatterers $\sParticles$, we compute the differential extinction probabilities $\sExtinctionResolved(\px,t)$ (for both scatterer-to-scatterer and source-to-scatterer transport) as 
\begin{equation}
\sExtinctionResolved(\px,t) = \sum_{\sParticle\in\sParticles}{w_\sParticle}\sExtinctionResolved_k(\px,t),
\label{eq:extinction_mixture}
\end{equation}
where the weights $w_\sParticle$ represent the probability of having a scatterer of type $\sParticle\in\sParticles$ ($\sum_{\sParticle\in\sParticles}{w_\sParticle}=1$), and $\sExtinctionResolved_k(\px,t)$ is the differential extinction probability of each type. 
For the phase function and scattering albedo, we have
\begin{gather}
\sAlbedo(\px,t) = \sum_{\sParticle\in\sParticles} \frac{w_\sParticle\,\sExtinctionResolved_k(\px,t)}{\sum_{\sParticle\in\sParticles} w_\sParticle\,\sExtinctionResolved_k(\px,t)}\,\sAlbedoI{k}(\px,t), \label{eq:albedo_mixture}\\
\sPF(\px,\diri,\diro,t) = \sum_{\sParticle\in\sParticles} \frac{w_\sParticle\,\sExtinctionResolved_k(\px,t)\,\sAlbedoI{k}(\px,t)}{\sum_{\sParticle\in\sParticles} w_\sParticle\,\sExtinctionResolved_k(\px,t)\,\sAlbedoI{k}(\px,t)}\,\sPFI{k}(\px,\diri,\diro,t), \label{eq:pf_mixture}
\end{gather}
Last, using \EqsRange{extinction_mixture}{pf_mixture}, we can compute the scattering operator for a mixture of scatterers as
\begin{equation}
\sScatteringOperator(\px,\diri,\diro,t) = \sum_{\sParticle\in\sParticles}{w_\sParticle}\sScatteringOperator_{\sParticle}(\px,\diri,\diro,t).
\label{eq:freepath_mixture}
\end{equation}
%
%
%
%
%
}

\section{Modeling directional correlation}
\label{app:directional}
Similar to the anisotropy on the cross section described by Jakob et al.~\shortcite{Jakob2010Radiative}, the scatterers correlation might have also an important directional effect, as illustrated in \Fig{correlation} and observed by Vasques and Larsen~\shortcite{Vasques2014anisotropic}. By considering the directional dependency on both $\sProbConc{\sConcentration}$ and $\sCrossSection$, we transform \Eq{isotropic_structure} into:
\begin{align}
\sExtinction \, \sProbExtinctionConditional{\sExtinction}{\diro} & = \sConcentration \, \sProbConcConditional{\sConcentration}{\diro} \, \sCrossSection(\diro),
\label{eq:anisotropic_structure}
\end{align}
where $\sProbConcConditional{\sConcentration}{\diro}$ and $\Mean{\sCrossSection}$ are the probability distribution of the concentration and the mean cross section along $\diro$ respectively. 

To model $\sProbConcConditional{\sConcentration}{\diro}$, we noted that its only varying parameter is its variance, which we redefine as a directional function $\Var{\sConcentration;\diro} \in \Sphere$. Following the same approach as the SGGX model \cite{Heitz2015SGGX}, we model $\Var{\sConcentration;\diro}$ as a zero-mean ellipsoid, using the matrix $\sVarMatrix$  defining the eigenspace of the variance of the projected concentration in $\Sphere$ (see~\cite{Heitz2015SGGX} for details). We thus obtain:
\begin{align}
\Var{\sConcentration;\diro} = \sqrt{\diro^T \, \sVarMatrix \, \diro}.
\label{eq:variance}
\end{align}
For each direction $\diro$, we first obtain the projected variance, and then define the corresponding gamma distribution $\dGamma{\sConcentration}{\alpha(\diro)}{\beta(\diro)}$ with $\alpha(\diro)$ and $\beta(\diro)$ computed from $\Mean{\sConcentration}$ and $\Var{\sConcentration; \diro}$. 
%
This has several benefits over other directional distributions: it is compact and efficient to evaluate; it supports anisotropy on the main axes; it is symmetric, smooth and non-negative in the full domain $\Sphere$; and it is intuitive to characterize. 
\section{The Generalized Boltzmann Equation}
\label{sec:app_gbe}

Here we include for completeness the derivations of the Generalized Boltzmann Equation (GBE) by Larsen and Vasques~\shortcite{Larsen2007Generalized,Larsen2011Generalized}. 

Let us define $N(\px,\diro,t)\vV\diff{\Sphere}\diff{t}$ [\si{\per\cubic\meter\per\steradian\per \meter}] as the number of particles in $\vV\diff{\Sphere}\diff{t}$ over $\px$ and $\diro$ that have traveled a distance $t$ since its last interaction (scattering or emission). %
By considering the net flux of particles $\sFlux(\px,\diro,t)$ as the number of particles moving a distance $\diff{t}$ in a differential time $\diff{\sTime}$ we get
\begin{align}
\sFlux(\px,\diro,t) & =\frac{\diff{t}}{\diff{\sTime}} N(\px,\diro,t) & [\si{\per\square\meter\per\steradian\per\second \per\meter}] \nonumber \\ 
& =\sParticlesSpeed \, N(\px,\diro,t),  &
\label{eq:app_flux}
\end{align}
where $\sParticlesSpeed=\frac{\diff{t}}{\diff{\sTime}}$ [\si{\meter\per\second}] is the speed of the particles. 

By using the classic conservation equation that relates the sources of gain and loss of particles with the rate of change of particles, we get (we use Arvo's notation~\shortcite{Arvo1993}):
\begin{align}
\frac{\diff{}}{\diff{\sTime}} N(\px,\diro,t) & = \underbrace{\left(\sArvoSource(\px,\diro,t) + \sArvoInscattering(\px,\diro,t) \right)}_\text{gains} \nonumber \\
& - \underbrace{\left(\sArvoStream(\px,\diro,t) + \sArvoExtinction(\px,\diro,t) \right)}_\text{losses},
\label{eq:app_gbe_transport}
\end{align}
with $\sArvoSource(\px,\diro,t)$ and $\sArvoInscattering(\px,\diro,t)$ the gains due to particles emission (source) and inscattering respectively, and $\sArvoStream(\px,\diro,t)$ and $\sArvoExtinction(\px,\diro,t)$ the losses due to particles leaking (streaming) and extinction due to absorption and outscattering. 

In the classic steady-state Bolzmann Equation (and therefore the RTE), it holds that the particles and in equilibrium, and therefore $\frac{\diff{}}{\diff{\sTime}} N(\px,\diro,t) = 0$. However, the introduction on the $t$ dependence on $N(\px,\diro,t)$ introduces a non-zero particles rate over $(\px,\diro,t)$. 
By using the relationship in \Eq{app_flux}, we can compute the rate of change in the number of particles in $\vV\diff{\Sphere}\diff{t}$ around $\px,\diro,t$ as
\begin{align}
\frac{\diff{}}{\diff{\sTime}} N(\px,\diro,t)\vV\diff{\Sphere}\diff{t} & = \frac{\diff{}}{\sParticlesSpeed\diff{\sTime}} \sParticlesSpeed\, N(\px,\diro,t)\vV\diff{\Sphere}\diff{t} \\
& = \frac{\diff{}}{\diff{t}} \sFlux(\px,\diro,t) \vV\diff{\Sphere}\diff{t}. \nonumber
\end{align}
Using a similar relationship, we can compute net rate of particles leaking out of $\vV$ around $\px$ in direction $\diro$ after traveling a distance $t$ as 
\begin{align}
\sArvoStream(\px,\diro,t) = \diro\cdot\Diff{\sFlux(\px,\diro,t)} \vV\diff{\Sphere}\diff{t}.
\label{eq:app_gbe_stream}
\end{align}

Now, let us define $\sExtinctionResolved(t)$ [\si{\per\meter}] as the differential probability of extinction, and $\sExtinctionResolved(t)\diff{t}$ as the probability of a particle to interact at distance $\diff{t}$ after having traveled a distance $t$ since its last interaction (emission or scattering). With these definitions, we can compute the rate of collision (extinction) as
\begin{align}
\sArvoExtinction(\px,\diro,t) &= \frac{1}{\diff{\sTime}} \sExtinctionResolved(t)\diff{t}\, N(\px,\diro,t)\vV\diff{\Sphere}\diff{t} \nonumber \\
& = \frac{\diff{t}}{\diff{\sTime}}\sExtinctionResolved(t)\,N(\px,\diro,t)\vV\diff{\Sphere}\diff{t} \nonumber \\
& = \sExtinctionResolved(t)\,\sFlux(\px,\diro,t)\,\vV\diff{\Sphere}\diff{t}.
\label{eq:app_gbe_extinction}
\end{align}

The treatment of inscattering and source terms is slightly more complex, given that they set the \emph{memory} of the particles to $t=0$. Assuming that scattering and absorbers have the same distribution, and therefore we can formulate the differential probability of scattering as $\sScatteringResolved(t)=\sAlbedo \, \sExtinctionResolved(t)$ [\si{\per\meter}], with $\sAlbedo$ [\si{\unitless}] the probability of scattering of a particle that has suffered collision (scattering albedo). From \Eq{app_gbe_extinction}, we can compute the rate of particles colliding at $\px$ from direction $\diri$ as 
\begin{align}
\sArvoExtinction(\px,\diri) & = \int_0^\infty \sArvoExtinction(\px,\diri,t)\diff{t}\\
& = \left[\int_0^\infty \sExtinctionResolved(t)\,\sFlux(\px,\diri,t) \diff{t}\right] \,\vV\diff{\Sphere}. \nonumber
\end{align}
Then, by multiplying $\sArvoExtinction(\px,\diri)$ by the phase function $\sPF(\diri,\diro)$ [\si{\per\steradian}] and the scattering albedo $\sAlbedo$, and integrating over the sphere $\Sphere$ we get
\begin{align}
\left[\sAlbedo \int_\Sphere \sPF(\diri,\diro) \sArvoExtinction(\px,\diri) \diff{\diri}\right] \vV\diff{\Sphere}.
\label{eq:app_gbe_scattered}
\end{align}
Since as particles emerge from a scattering event they reset their value $t$ to $t=0$, then the path length spectrum of inscattering is a delta function $\delta(t)$. Multiplying \Eq{app_gbe_scattered} by $\delta(t)\diff{t}$ we get $\sArvoInscattering(\px,\diro,t)$ as
\begin{align}
\sArvoInscattering(\px,\diro,t)  & = \delta(t) \sAlbedo \left[\int_\Sphere \sPF(\diri,\diro) \sArvoExtinction(\px,\diri) \diff{\diri}\right] \vV\diff{\Sphere} \diff{t}.
\label{eq:app_gbe_inscattering}
\end{align}

Similarly to scattering, emission also requires to set particles to $t=0$. Following the same reasoning as before, we define the source term $\sArvoSource(\px,\diro,t)$ as:
\begin{align}
\sArvoSource(\px,\diro,t)  & = \delta(t) \sEmissionFlux(\px,\diro) \vV\diff{\Sphere} \diff{t},
\label{eq:app_gbe_source}
\end{align}
where $\sEmissionFlux(\px,\diro) \vV\diff{\Sphere}$ [\si{\per\cubic\meter\per\steradian\per\second}] is the rate at which particles are emitted by an internal source in $\px$ in direction $\diro$. 

Substituting \Eqsss{app_gbe_stream}{app_gbe_extinction}{app_gbe_inscattering}{app_gbe_source} into \Eq{app_gbe_transport}, and dividing both sides of the equation by $\vV\diff{\Sphere} \diff{t}$ we get the final GBE for generic particles transport proposed by Larsen and Vasques~\shortcite[Eq. (2.3)]{Larsen2011Generalized}
\begin{multline}
\frac{d}{dt}\sFlux(\px,\diro,t) + \diro\cdot\Diff{\sFlux(\px,\diro,t)} + \sExtinctionResolved(t)\sFlux(\px,\diro,t) = \\
\delta(t)\,\sAlbedo\int_0^\infty\sExtinctionResolved(s)\int_\Sphere \sFlux(\px,\diri,s)\sPF(\diri,\diro)\diff{\diri}\diff{s} + \delta(t)\,\sEmissionFlux(\px,\diro),
\label{eq:app_delta_larsengrte_flux}
\end{multline}

\Eq{app_delta_larsengrte_flux} defines models transport for general particles as a function of their flux $\sFlux(\px,\diro,t)$. Since we are interested on light, we want to express such equation in terms of radiance. We can then set $\sParticlesSpeed=\sLightSpeed$, with $\sLightSpeed$ the speed of light, and assuming monoenergetic photons with wavelength $\lambda$ [\si{\per\hertz}], then we define the radiance at $\px$ from direction $\diro$, that has traveled a distance $t$ since its last interaction as
\begin{align}
\sRad(\px,\diro,t) = \frac{\sPlank\sLightSpeed}{\lambda}\,N(\px,\diro,t) = \frac{\sPlank}{\lambda}\,\sFlux(\px,\diro,t), && \left[\si[per-mode=fraction]{\watt \per\square\meter\per\steradian\per \meter}\right]
\end{align}
with $\sPlank$ is Plank's constant.
Note that the $t$-resolved radiance $\sRad(\px,\diro,t)$ relates with the classic radiance as:
\begin{align}
\sRad(\px,\diro) = \int_0^\infty \sRad(\px,\diro,t)\diff{t}. && \left[\si[per-mode=fraction]{\watt \per\square\meter\per\steradian}\right]
\end{align}
Similarly, the source term for light $\sEmission(\px,\diro)$ is defined in terms of radiant power, and related with $\sEmissionFlux(\px,\diro)$ as
\begin{align}
\sEmission(\px,\diro) = \frac{\sPlank}{\lambda}\,\sEmissionFlux(\px,\diro). && \left[\si[per-mode=fraction]{\watt \per\cubic\meter\per\steradian}\right]
\end{align}
Therefore, by multiplying \Eq{app_delta_larsengrte_flux} by $\sPlank\lambda^{-1}$ we get the GBE in terms of radiance as
\begin{multline}
\frac{d}{dt}\sRad(\px,\diro,t) + \diro\cdot\Diff{\sRad(\px,\diro,t)} + \sExtinctionResolved(t)\sRad(\px,\diro,t) = \\
\delta(t)\,\sAlbedo\int_0^\infty\sExtinctionResolved(s)\int_\Sphere \sRad(\px,\diri,s)\sPF(\diri,\diro)\diff{\diri}\diff{s} + \delta(t)\,\sEmission(\px,\diro).
\label{eq:app_delta_larsengrte}
\end{multline}
Finally, we can obtain the equivalent delta-less form presented in \Eq{larsengrte}: We first set \Eq{app_delta_larsengrte} for $t>0$ as
\begin{equation}
\frac{d}{dt}\sRad(\px,\diro,t) + \diro\cdot\Diff{\sRad(\px,\diro,t)} + \sExtinctionResolved(t)\sRad(\px,\diro,t) = 0.
\label{eq:app_larsengrte}
\end{equation}
Then, to define the initial value for $t=0$ of the ODE defined by \Eq{app_larsengrte} we operate \Eq{app_delta_larsengrte} with $\lim_{\epsilon\to0}\int_{-\epsilon}^{\epsilon}(\cdot)\diff{t}$, and using $\sRad(\px,\diro,t)=0$ for $t<0$ we define 
\begin{equation}
\sRad(\px,\diro,0) = \lim{t\to0^+}=\sRad(\px,\diro,0^+)
\end{equation}
to obtain
\begin{equation}
\sRad(\px,\diro,0) = \int_0^\infty\sScatteringResolved(t)\int_\Sphere \sRad(\px,\diri,t)\sPF(\diri,\diro)\diff{\diri}\diff{t} + \sEmission(\px,\diro),
\end{equation}
which is the second line in \Eq{larsengrte}.

\section{The RTE as a special case of the GBE}
\label{sec:app_gbe2rte}

Here we will see that the classic RTE is a special case of Larsen's Generalized Boltzmann Equation (GBE)~\cite{Larsen2007Generalized,Larsen2011Generalized}, in which the differential extinction probability $\sExtinctionResolved(t)$ is independent of $t$, and therefore a constant defined by the extinction coefficient $\sExtinctionResolved(t)=\sExtinction$. 

%

%
\revised{Let us use the equivalent delta-based form of \Eq{larsengrte} shown in \Eq{app_delta_larsengrte}. }
In the classic RTE, the differential probability of extinction is a constant $\sExtinctionResolved(t)=\sExtinction$, so that \Eq{app_delta_larsengrte} becomes
\begin{multline}
\frac{d}{dt}\sRad(\px,\diro,t) + \diro\cdot\Diff{\sRad(\px,\diro,t)} + \sExtinction\sRad(\px,\diro,t) = \\
\delta(t)\,\sScattering\int_0^\infty\int_\Sphere \sRad(\px,\diri,s)\sPF(\diri,\diro)\diff{\diri}\diff{s} + \delta(t)\,\sEmission(\px,\diro) = \\
\delta(t)\,\sScattering\int_\Sphere \sRad(\px,\diri)\sPF(\diri,\diro)\diff{\diri} + \delta(t)\,\sEmission(\px,\diro),
\label{eq:app_delta_larsengrte1}
\end{multline}
where $\sScattering = \sAlbedo\,\sExtinction$, $\sRad(\px,\diri)=\int_0^\infty \sRad(\px,\diri,s) \diff{s}$. Then, by operating \Eq{app_delta_larsengrte1} by $\int_{-\epsilon}^\infty (\cdot) \diff{t}$ (with $\epsilon \approx 0$; note that we cannot use $\epsilon=0$ because otherwise the integral of the delta function $\delta(t)$ would be undefined) we get
\begin{multline}
\sRad(\px,\diro,-\epsilon) + \sRad(\px,\diri,\infty) + \diro\cdot\Diff{\sRad(\px,\diro)} + \sExtinction\sRad(\px,\diro) = \\
\sScattering\int_\Sphere \sRad(\px,\diri)\sPF(\diri,\diro)\diff{\diri} + \sEmission(\px,\diro).
\end{multline}
Finally, by using $\sRad(\px,\diri,-\epsilon)=\sRad(\px,\diri,\infty)=0$ we get
\begin{multline}
\diro\cdot\Diff{\sRad(\px,\diro)} + \sExtinction\sRad(\px,\diro) = \\
\sScattering\int_\Sphere \sRad(\px,\diri)\sPF(\diri,\diro)\diff{\diri} + \sEmission(\px,\diro),
\end{multline}
which is the RTE \ParEq{rte}.

\section{From our Extended GBE to Larsen's GBE}
\label{sec:app_ggbe2gbe}
Here we demonstrate that our Extended GBE~\Sec{ourgbe} is a generalization of Larsen's GBE~\ParEq{larsengrte}, and how the latter can be obtained from ours. 

Our Extended GBE is defined as
\begin{align}
\label{eq:app_grteours}
\frac{d}{dt}\sRad(\px,\diro,t) & + \diro\cdot\Diff{\sRad(\px,\diro,t)} + \sExtinctionResolvedI{\sRadI}(\px,t)\,\sRadICorr(\px,\diro,t) \nonumber \\ & + \sum_j \sExtinctionResolvedI{\sEmission_j}(\px,t)\,\sRadQiCorr{j}(\px,\diro,t) = 0, \\
\label{eq:app_grteours_initinsc}
\sRadICorr(\px,\diro,0) & = \int_0^\infty\int_\Sphere\Big(\sScatteringOperator_{\sRadI}(\px,\diri,\diro,t)\,\sRadICorr(\px,\diri,t)  \\ &+ \sum_j \sScatteringOperator_{\sEmission_j}(\px,\diri,\diro,t)\,\label{eq:app_grteours_initsource}
\sRadQiCorr{j}(\px,\diri,t)\Big) \,\diff{\diri}\diff{t}, \nonumber \\
\sRadQiCorr{j}(\px,\diro,0) & = \sEmission_j(\px,\diro),
\end{align}
%
where 
\begin{equation}
\label{eq:app_rad}
\sRad(\px,\diro,t) = \sRadICorr(\px,\diro,t) + \sum_j \sRadQiCorr{j}(\px,\diro,t),
\end{equation} 
the differential extinction probabilities for the scattered photons and the (unscattered) photons emitted by light source $\sEmission_j$ are respectively $\sExtinctionResolvedI{\sRadI}(\px,t)$ and $\sExtinctionResolvedI{\sEmission_j}(\px,t)$, the scattering operator for scattered photons is $\sScatteringOperator_{\sRadI}(\px,\diri,\diro,t) = \sAlbedoI{\sRadI}(\px,t)\,\sExtinctionResolvedI{\sRadI}(\px,t)\sPFI{\sRadI}(\px,\diri,\diro,t)$ , and $\sScatteringOperator_{\sEmission_j}(\px,\diri,\diro,t)$ is the scattering operator for photons emitted by light source $\sEmission_j$. 

\Eq{app_grteours} does not impose any assumption on the correlation between scatterers and sources. If they were somehow positively correlated, so that the scatterers and emitters would have the exact same correlation with respect to extincting particles (which could be scatterers or not), then
\begin{align*}
\forall j, \quad &\sExtinctionResolvedI{\sRadI}(\px,t)=\sExtinctionResolvedI{\sEmission_j}(\px,t) = \sExtinctionResolved(\px,t) \\  &\text{and} \\ \forall j, \quad &\sScatteringOperator_{\sRadI}(\px,\diri,\diro,t) = \sScatteringOperator_{\sEmission_j}(\px,\diri,\diro,t) = \sScatteringOperator(\px,\diri,\diro,t).
\end{align*}
This allows us to transform \Eq{app_grteours} into
\begin{align}
\label{eq:app_grteours2larsen}
\frac{d}{dt}\sRad(\px,\diro,t) & + \diro\cdot\Diff{\sRad(\px,\diro,t)} \nonumber \\ + &\sExtinctionResolved(\px,t)\Big(\sRadICorr(\px,\diro,t) + \sum_j \sRadQiCorr{j}(\px,\diro,t)\Big) = \nonumber \\
\frac{d}{dt}\sRad(\px,\diro,t) & + \diro\cdot\Diff{\sRad(\px,\diro,t)} + \sExtinctionResolved(\px,t)\sRad(\px,\diro,t) = 0,
\end{align}
while \Eqs{app_grteours_initinsc}{app_grteours_initsource} become
\begin{align}
\label{eq:app_grteours2larsen_initinsc}
\sRadICorr(\px,\diro,0) & = \int_0^\infty\int_\Sphere \sScatteringOperator(\px,\diri,\diro,t)\,\Big(\sRadICorr(\px,\diri,t)  \\ & + \sum_j\sRadQiCorr{j}(\px,\diri,t)\Big) \,\diff{\diri}\diff{t} \nonumber \\
& = \int_0^\infty\int_\Sphere \sScatteringOperator(\px,\diri,\diro,t)\sRad(\px,\diro,t)\,\diff{\diri}\diff{t}, \nonumber \\
\label{eq:app_grteours2larsen_source}
\sRadQiCorr{j}(\px,\diro,0) & = \sEmission_j(\px,\diro).
\end{align}

From \Eqss{app_rad}{app_grteours2larsen_initinsc}{app_grteours2larsen_source} we can simplify the initial value of \Eq{app_grteours2larsen} as:
\begin{equation}
\label{eq:app_grteours2larsen_init}
\sRad(\px,\diro,0) = \int_0^\infty\int_\Sphere \sRad(\px,\diri,t)\sScatteringOperator(\px,\diri,\diro,t)\diff{\diri}\diff{t} + \sEmission(\px,\diro).
\end{equation}
Finally, by removing the spatial dependence on $\sExtinctionResolved(t)$ and the $t$-dependence on albedo and phase function from \Eqs{app_grteours2larsen}{app_grteours2larsen_init} we get Larsen's GBE~\ParEq{larsengrte}.

\section{Integral form of the Extended GBE}
\label{sec:app_integralggbe}
In this section we compute the integro-differential form of our Extended GBE, modeled in differential form in \EqsRange{app_grteours}{app_grteours_initsource}. Let us first expand \Eq{app_grteours} by using \Eq{app_rad} as
\begin{align}
& \frac{d}{dt}\sRadICorr(\px,\diro,t) + \diro\cdot\Diff{\sRadICorr(\px,\diro,t)} + \sExtinctionResolvedI{\sRadI}(\px,t)\,\sRadICorr(\px,\diro,t) \nonumber \\ + & \sum_j \Big(\frac{d}{dt}\sRadQiCorr{j}(\px,\diro,t) + \diro\cdot\Diff{\sRadQiCorr{j}(\px,\diro,t)} + \sExtinctionResolvedI{\sEmission_j}(\px,t)\,\sRadQiCorr{j}(\px,\diro,t)\Big) \nonumber \\ = & \,0.
\end{align}
This expression is a sum of multiple independent differential equations on $\sRadICorr(\px,\diro,t)$ and $\sRadQiCorr{j}(\px,\diro,t)$ with $j\in[1,\infty)$. Since they are independent on each other, we can solve them individually, and them put them back together. Let us first start with the simpler case of $\sRadQiCorr{j}$, by setting $\sRadICorr(\px,\diro,t) = 0$ and $\sRadQiCorr{k}=0$ for all $k\neq j$, and getting
\begin{align}
& \frac{d}{dt}\sRadQiCorr{j}(\px,\diro,t) + \diro\cdot\Diff{\sRadQiCorr{j}(\px,\diro,t)} + \sExtinctionResolvedI{\sEmission_j}(\px,t)\,\sRadQiCorr{j}(\px,\diro,t) = 0, \nonumber \\
& \sRadQiCorr{j}(\px,\diro,0) = \sEmission_j(\px,\diro).
\end{align}
By solving this partial differential equation we get
\begin{align}
\sRadQiCorr{j}(\px,\diro,t) & = \sRadQiCorr{j}(\px_{t},\diro,0) \, e^{-\int_0^t\sExtinctionResolvedI{\sEmission_j}(\px,s)\diff{s}}\nonumber \\ & =\sEmission_j(\px,\diro)\sTranmittanceCorrI{\sEmission_j}(\px,\px_{t}),
\end{align}
where $\px_{t}=\px-\diro\,t$ and $\sTranmittanceCorrI{\sEmission_j}(\px,\px_{t})=e^{-\int_0^t\sExtinctionResolvedI{\sEmission_j}(\px,s)\diff{s}}$. Then we apply the definite integral on $t$ in the interval $[0,\infty)$ to remove the $t$ dependence as
\begin{align}
\label{eq:app_source_integral}
\sRadQiCorr{j}(\px,\diro) & = \int_0^\infty \sRadQiCorr{j}(\px,\diro,t) \diff{t} \nonumber \\ 
& = \int_0^\infty \sEmission_j(\px,\diro)\,\sTranmittanceCorrI{\sEmission_j}(\px,\px_{t}) \,\diff{t}.
\end{align}

Now let's consider the case of $\sRadICorr(\px,\diro,t)$ by setting $\sRadQiCorr{k}=0$ for all $k$ 
\begin{align}
\label{eq:app_inscattering_ode}
 \frac{d}{dt}\sRadICorr(\px,\diro,t) & + \diro\cdot\Diff{\sRadICorr(\px,\diro,t)} + \sExtinctionResolvedI{\sRadI}(\px,t)\,\sRadICorr(\px,\diro,t) = 0, \nonumber \\
 \sRadICorr(\px,\diro,0) & = \int_0^\infty\int_\Sphere\Big(\sScatteringOperator_{\sRadI}(\px,\diri,\diro,t)\,\sRadICorr(\px,\diri,t)  \\ &+ \sum_j \sScatteringOperator_{\sEmission_j}(\px,\diri,\diro,t)\,\sRadQiCorr{j}(\px,\diri,t)\Big) \,\diff{\diri}\diff{t}. \nonumber 
\end{align}
Again, by solving \Eq{app_inscattering_ode}, and applying $\sRadICorr(\px,\diro,0)=\sRadI(\px,\diro)$ we get
\begin{align}
\sRadICorr(\px,\diro,t) & = \sRadICorr(\px,\diro,0)\, e^{-\int_0^t\sExtinctionResolvedI{\sRadI}(\px,s)\diff{s}}\nonumber \\ 
& = \sRadI(\px,\diro)\,\sTranmittanceCorrI{\sRadICorr}(\px,\px_{t}).
\end{align}
which by applying again $\int_{0}^\infty (\cdot) \diff{t}$ gives
\begin{align}
\label{eq:app_inscatterin_integral}
\sRadICorr{j}(\px,\diro) & = \int_0^\infty \sRadICorr{j}(\px,\diro,t) \diff{t} \nonumber \\ 
& = \int_0^\infty \sRadI(\px,\diro)\,\sTranmittanceCorrI{\sRadICorr}(\px,\px_{t}) \,\diff{t}.
\end{align}

Finally, from \Eqs{app_source_integral}{app_inscatterin_integral} we compute the total radiance $\sRad(\px,\diro)$ via \Eq{app_rad} as
\begin{align}
\sRad(\px,\diro) = \int_0^\infty & \sTranmittanceCorrI{\sRadI}(\px,\px_t)\, \sRadI(\px_t,\diro) \\ 
+ & \sum_j \sTranmittanceCorrI{\sEmission_{j}}(\px,\px_t)\,\sEmission_{j}(\px_t,\diro) \,\diff{t}. \nonumber
\end{align}


\section{Simplifying Equation (15)} 
\label{sec:app_simplification}

In this section we include the derivations taking from \Eq{transmittance_general} to \Eq{transmittance} in \Sec{local_optical_parameters} of the main text. \Eq{transmittance_general} computes the transmittance of an incoming beam as
\begin{align}
\sTranmittanceCorr(t) = \int_0^\infty \int_0^\infty \sProbLight{\sRad_i}\, \sProbExtinctionConditional{\sExtinction}{\sRad_i} \, \frac{\sRad_i}{\hat{\sRad}_i} \, \sAttenuation(\sExtinction \,t ) \, \diff{\sExtinction}\,\diff{\sRad_i}.
\label{eq:app_transmittance_general}
\end{align}
where $\sProbLight{\sRad_i}$ is a probability distribution describing the incoming radiance $\sRad_i$, $\sProbExtinctionConditional{\sExtinction}{\sRad_i}$ is the conditional probability distribution describing the distribution of particles as a function of the incoming radiance $\sRad_i$, and ${\hat{\sRad}_i}=\int_0^\infty \sProbLight{\sRad_i}\, \sRad_i \,\diff{\sRad_i}$ is the total incoming radiance. 

The first assumption we make is that the spatial distributions of incoming light and scatterers are decorrelated. This means that $\sProbLight{\sRad_i}$ and $\sProbExtinction{\sExtinction}$ are independent, so that $\sProbExtinctionConditional{\sExtinction}{\sRad_i} = \sProbExtinction{\sExtinction}$. This transforms \Eq{app_transmittance_general} into
\begin{align}
\sTranmittanceCorr(t) & = \int_0^\infty \int_0^\infty \sProbLight{\sRad_i}\, \sProbExtinction{\sExtinction} \, \frac{\sRad_i}{\hat{\sRad}_i} \, \sAttenuation(\sExtinction \,t ) \, \diff{\sExtinction}\,\diff{\sRad_i} \nonumber \\
& = \int_0^\infty \sProbExtinction{\sExtinction} \sAttenuation(\sExtinction \,t ) \, \int_0^\infty  \, \sProbLight{\sRad_i}\, \frac{\sRad_i}{\hat{\sRad}_i} \, \diff{\sRad_i}\, \diff{\sExtinction} \nonumber \\
& = \int_0^\infty \sProbExtinction{\sExtinction} \sAttenuation(\sExtinction \,t ) \frac{\hat{\sRad}_i}{\hat{\sRad}_i} \diff{\sExtinction} \nonumber \\
& = \int_0^\infty \sProbExtinction{\sExtinction} \sAttenuation(\sExtinction \,t ) \diff{\sExtinction}.
\end{align}

Finally, $\sAttenuation(\sExtinction \,t )$ is the attenuation function, that describes the probability of extinction of an \emph{individual ray}. Note that we have used a generic attenuation function $\sAttenuation(\sOpticalDepth_t(\sRay))$; if the particles distribution is random (although correlated) then the extinction at each differential ray of the beam is Poissonian, holding $\sAttenuation(\sOpticalDepth_t(\sRay))=e^{-\sExtinction \,t}$. In other cases, in particular in ordered media presenting negative correlation, this attenuation does not hold and extinction becomes a Bernouilli stochastic proccess, which in the limit reduces to a deterministic linear attenuation. By keeping the exponential attenuation, we transform \Eq{app_transmittance_general} into
\begin{align}
\sTranmittanceCorr(t) = \int_0^\infty \sProbExtinction{\sExtinction} e^{-\sExtinction \,t } \diff{\sExtinction}.
\label{eq:app_transmittance_poisson}
\end{align}


\section{The RTE as a special case of our local model}
\label{sec:app_gamma2rte}
In this section we show how the exponential transmittance predicted by the Beer-Lambert law is a particular case of our model in \Sec{local_optical_parameters}, in particular how \Eq{transmittance_general}, its simplified form \ParEq{transmittance}, and the final gamma-based transmittance \ParEq{gamma_analytic_transmittance} converge to $\sTranmittanceCorr(\sDistance) = e^{-\Mean{\sExtinction}\,\sDistance}$, with $\Mean{\sExtinction}$ the mean extinction in the differential volume $\vV$.

\subsection{\Eq{transmittance_general} to exponential transmittance}
Starting from \Eq{app_transmittance_general} [\Eq{transmittance_general} in the paper], let us first define the scatterers distribution by setting the probability distribution of extinction $\sProbExtinction{\sExtinction}$. In the classic RTE the assumption is that particles are uniformly distributed in a differential volume, so that the extinction probability is always the same $\Mean{\sExtinction}$. Mathematically, this is equivalent to setting
\begin{equation}
\sProbExtinction{\sExtinction} = \delta(\Mean{\sExtinction}-\sExtinction),
\label{eq:app_deltadistribution}
\end{equation}
where $\delta(s)$ is the Dirac delta function. With that, we can transform \Eq{app_transmittance_general} into
\begin{align}
\sTranmittanceCorr(t) & = \int_0^\infty \int_0^\infty \sProbLight{\sRad_i}\, \delta(\Mean{\sExtinction}-\sExtinction) \, \frac{\sRad_i}{\hat{\sRad}_i} \, \sAttenuation(\sExtinction \,t ) \, \diff{\sExtinction}\,\diff{\sRad_i} \nonumber \\
& = \int_0^\infty \sProbLight{\sRad_i}\, \frac{\sRad_i}{\hat{\sRad}_i} \, \int_0^\infty  \, \delta(\Mean{\sExtinction}-\sExtinction) \sAttenuation(\sExtinction \,t ) \, \diff{\sExtinction} \, \diff{\sRad_i} \nonumber \\
& = \int_0^\infty \frac{\sRad_i}{\hat{\sRad}_i} \diff{\sRad_i} \sAttenuation(\Mean{\sExtinction} \,t )  \nonumber \\
& = \frac{\hat{\sRad}_i}{\hat{\sRad}_i}  \sAttenuation(\sExtinction \,t )   \nonumber \\
& = \sAttenuation(\sExtinction \, t). 
\label{eq:app_transgeneral_to_delta}
\end{align}

Finally, we need to define the attenuation process of extinction defined by $\sAttenuation(\sExtinction \, t)$. Since we are assuming that particles are randomly distributed, then we can safely assume that $\sAttenuation(\sExtinction \, t)$ is a Poissonian proccess (see \Sec{app_simplification}), where $\sAttenuation(\sExtinction \, \sRay)=e^{-\sExtinction \,t}$ holds. By substitution, we therefore transform \Eq{app_transgeneral_to_delta} into the exponential transmittance $\sTranmittanceCorr(\sDistance) = e^{-\Mean{\sExtinction}\,\sDistance}$. Finally, by applying that $\sExtinctionResolved(t)=\sProb{t}/\sTranmittanceCorr(t)$ we can verify that 
\begin{align}
\sExtinctionResolved(t) & = \frac{\sProb{t}}{\sTranmittanceCorr(t)} = \left|\frac{\diff{\sTranmittanceCorr(t)}}{\diff{t}}\right| \frac{1}{\sTranmittanceCorr(t)} = \frac{\Mean{\sExtinction}\,e^{-\Mean{\sExtinction}\,\sDistance}}{e^{-\Mean{\sExtinction}\,\sDistance}} = \Mean{\sExtinction},
\end{align}
which is the $t$-independent classic form of the differential extinction probability, and which as shown in \Sec{app_gbe2rte} reduces the GBE to the classic RTE. 

\subsection{\Eq{transmittance} to exponential transmittance}
\label{sec:app_trans_to_exptrans}
Following the same procedure as in the previous section it is easy to verify that by defining a uniform distribution of particles with mean extinction $\Mean{\sExtinction}$ via \Eq{app_deltadistribution} we reduce \Eq{transmittance} to the exponential decay as:
\begin{align}
\sTranmittanceCorr(t) & = \int_0^\infty \sProbExtinction{\sExtinction} e^{-\sExtinction \,t } \diff{\sExtinction} \nonumber \\
& = \int_0^\infty \delta(\Mean{\sExtinction}-\sExtinction) e^{-\sExtinction \,t } \diff{\sExtinction} \nonumber \\
& = e^{-\Mean{\sExtinction}\,\sDistance}. 
\end{align}

\subsection{\Eq{gamma_analytic_transmittance} to exponential transmittance}
Finally, we will show that practical gamma-based form of transmittance \ParEq{gamma_analytic_transmittance}, defined as
\begin{align}
\sTranmittanceCorr(\sDistance) & = \int_0^\infty \dGamma{\sConcentration}{\alpha}{\beta} e^{-\sExtinction \sCrossSection \,t } \diff{\sExtinction} \nonumber \\ & = \left(1+\frac{\Mean{\sCrossSection}\cdot\sDistance}{\beta}\right)^{-\alpha},
\label{eq:app_gamma_analytic_transmittance1}
\end{align}
with $\dGamma{\sConcentration}{\alpha}{\beta}$ the gamma distribution, $\alpha=\Mean{\sConcentration}^2\cdot\Var{\sConcentration}^{-1}$, $\beta=\Mean{\sConcentration}\cdot\Var{\sConcentration}^{-1}$, with $\Mean{\sConcentration}$ and $\Var{\sConcentration}$ the mean and variance of particles concentration $\sConcentration$ respectively, and $\sCrossSection$ the mean cross section.
By plugging the definition of $\alpha$ and $\beta$ in \Eq{app_gamma_analytic_transmittance1} we get
\begin{equation}
\sTranmittanceCorr(\sDistance) = \left(1+\sCrossSection\,\sDistance\frac{\Var{\sConcentration}}{\Mean{\sConcentration}}\right)^{-\frac{\Mean{\sConcentration}^2}{\Var{\sConcentration}}}.
\label{eq:app_gamma_analytic_transmittance2}
\end{equation}
Then, by applying the limit to \Eq{app_gamma_analytic_transmittance2} we get
\begin{equation}
\lim_{\Var{\sConcentration}\to0}\left(1+\sCrossSection\,\sDistance\frac{\Var{\sConcentration}}{\Mean{\sConcentration}}\right)^{-\frac{\Mean{\sConcentration}^2}{\Var{\sConcentration}}}=e^{-\Mean{\sConcentration} \, \sCrossSection \,\sDistance}.
\label{eq:app_gamma_analytic_transmittance2}
\end{equation}
Finally, by using $\Mean{\sExtinction}=\Mean{\sConcentration}\,\sCrossSection$ we get $\sTranmittanceCorr(\sDistance)= e^{-\Mean{\sExtinction}\,\sDistance}$. A complementary way of formulating this proof is by noticing that $\lim_{\Var{\sConcentration}\to0}\dGamma{\sConcentration}{\alpha}{\beta} = \delta(\Mean{\sConcentration}-\sConcentration)$, which results into a very similar derivation to \Sec{app_trans_to_exptrans}.


\section{Derivation of Sampling Procedures for Equation (20)} 
\label{sec:app_sampling_gamma_dist}
In \Sec{rendering} of the main text we define transmittance for distance $\sDistance$ in correlated media as \ParEq{gamma_analytic_transmittance}
\begin{equation}
\sTranmittanceCorr(\sDistance) = \left(1+\frac{\Mean{\sCrossSection}\cdot\sDistance}{\beta}\right)^{-\alpha},
\label{eq:app_gamma_analytic_transmittance}
\end{equation}
with $\alpha=\Mean{\sConcentration}^2\cdot\Var{\sConcentration}^{-1}$, $\beta=\Mean{\sConcentration}\cdot\Var{\sConcentration}^{-1}$, with $\Mean{\sConcentration}$ and $\Var{\sConcentration}$ the mean and variance of particles concentration $\sConcentration$ respectively, and $\sCrossSection$ the mean cross section.

\paragraph{General case $\alpha\in(0,\infty)$}

\noindent In order to sample a distance $t$ with respect to \Eq{gamma_analytic_transmittance} we need to define the probability function of sampling distance $t$ as $\sProb{\sDistance}$. We can compute it using the physical definition of transmittance as:
\begin{equation}
\sTranmittanceCorr(\sDistance) = \int_\sDistance^\infty \sProb{\sDistance'} \diff{\sDistance'},
\label{eq:app_transmittance}
\end{equation}
from which follows \ParEq{gamma_analytic_extinction_probability}
\begin{align}
\sProb{\sDistance} & = \left| \frac{\diff{\sTranmittanceCorr(t)}}{\diff{t}}\right| \nonumber \\
& = \alpha\,\sCrossSection\,\frac{\left(\sCrossSection\frac{t}{\beta} + 1\right)^{-(1+\alpha)}}{\beta},
\label{eq:badpdf}
\end{align}
which has as CDF 
\begin{equation}
P(\sDistance) = \sTranmittanceCorr(0)-\sTranmittanceCorr(t)  = 1-\sTranmittanceCorr(t).
\label{eq:badcdf}
\end{equation}
We sample $\sTranmittanceCorr(\sDistance)$ by using the inverse of \Eq{badcdf} as
\begin{equation}
\sDistance(\xi) = -\frac{\beta}{\sCrossSection}\left(1-\sqrt[-\alpha]{1-\xi}\right). 
\label{eq:inverse_badcdf}
\end{equation}

\paragraph{Sampling \Eq{gamma_analytic_transmittance} for $\alpha\in(0,1)$}
Unfortunately, \Eq{badpdf} is not proportional to \Eq{app_transmittance}, which is desirable for minimizing variance in Monte Carlo integration. In order to compute such sampling probability we impose $\sProb{t}\propto\sTranmittanceCorr(\sDistance)$ as $\sProb{\sDistance} = C \, \sTranmittanceCorr(\sDistance)$, 
where $C$ is a constant that ensures that $\int_0^\infty\sProb{\sDistance'}\diff{\sDistance'}=1$. We can thus write
\begin{align}
C & = \frac{1}{\int_0^\infty \sTranmittanceCorr(\sDistance')\diff{\sDistance'}}.
\label{eq:app_constant_pdf}
\end{align}
Solving the integral in the denominator we get
\begin{align}
\int_0^\infty \left(1+\frac{\sCrossSection\cdot\sDistance'}{\beta}\right)^{-\alpha}\diff{\sDistance'} & = \left.\frac{(\beta+\sCrossSection\,\sDistance')}{\sCrossSection(1-\alpha)}\left(1+\frac{\sCrossSection\cdot\sDistance'}{\beta}\right)^{-\alpha} \right|_0^\infty \\
& = -\frac{\beta}{\sCrossSection\,(1-\alpha)} + \lim_{\sDistance'\to\infty}\frac{(\beta+\sCrossSection\,\sDistance')}{\sCrossSection\,(1-\alpha)}\left(1+\frac{\sCrossSection\cdot\sDistance'}{\beta}\right)^{-\alpha}, \nonumber
\end{align}
which is convergent for $\alpha>1$ to 
\begin{align}
\int_0^\infty \sTranmittanceCorr(\sDistance')\diff{\sDistance'} = -\frac{\beta}{\sCrossSection(1-\alpha)}
\label{eq:app_inttransmittance}
\end{align}

Finally, by using \Eqs{app_constant_pdf}{app_inttransmittance} we can compute the sampling probability as \ParEq{gamma_pdf_transmittance}
\begin{align}
\sProb{\sDistance} = -\sCrossSection\frac{1-\alpha}{\beta} (1+\frac{\sCrossSection}{\beta}\sDistance)^{-\alpha} = -\sCrossSection\frac{1-\alpha}{\beta}\,\sTranmittanceCorr(\sDistance),
\label{eq:app_gamma_pdf_transmittance}
\end{align}
which has CDF
\begin{align}
P(\sDistance) = 1-(1+\frac{\sCrossSection}{\beta}\sDistance)^{1-\alpha}.
\label{eq:app_goodcdf}
\end{align}
Finally, we sample \Eq{app_gamma_pdf_transmittance} by inverting \Eq{app_goodcdf} as \ParEq{gamma_cdf_transmittance}
\begin{equation}
\sDistance(\xi) = P(\sDistance)^{-1} = -\frac{\beta}{\sCrossSection}\left(1-\sqrt[1-\alpha]{1-\xi}\right). 
\label{eq:inverse_goodcdf}
\end{equation}


\section{Additional Probability Distributions of Extinction}
\label{sec:app_addpathlengths}
Here we list additional probability distributions of extinction $\sProb{t}$ used in the results of the paper. The first one (\Sec{app_negative}) a perfect negative correlation results into a Bernoulli process (rather than a Poisson process), leading to linear transmittance; the second one (\Sec{app_gammapl}) models $\sProb{t}$ as a gamma probability distribution. Note that the later is different from our local model in \Sec{rendering}. This second $\sProb{t}$ is important, given that a gamma distribution is in general in good agreement with measured (or computed via Monte Carlo simulations) probability distributions of extinction in particulate materials (see e.g. \cite[Figure 6]{Meng2015Granular}). In the following we list the close-forms of transmittance $\sTranmittanceCorr(t)$, probability distribution of extinction $\sProb{t}$, and differential probability of extinction $\sExtinctionResolved(t)$.

\subsection{Perfect negative correlation}
\label{sec:app_negative}
We define this probability distribution of extinction via the mean extinction coefficient $\Mean{\sExtinction}$, as
\begin{align}
\sTranmittanceCorr(t) & =\max(0, 1-\Mean{\sExtinction}\,t), \\
\sProb{t} & = 
\begin{cases}
\sExtinction & \text{for } t < \frac{1}{\sExtinction}\\
0 & \text{elsewhere}
\end{cases} \label{eq:app_constant_transmittance}\\
\sExtinctionResolved(t) & = 
\begin{cases}
\frac{\sExtinction}{1-\sExtinction\,t} & \text{for } t < \frac{1}{\sExtinction}\\
0 & \text{elsewhere}
\end{cases}.
\end{align}
We can sample \Eq{app_constant_transmittance} by using
\begin{equation}
t(\xi) = \frac{\xi}{\sExtinction},
\end{equation}
with $\xi\in(0,1)$ a uniform random number. 

\subsection{Gamma probability distribution of extinction}
\label{sec:app_gammapl}
In this case, the gamma distribution defines the probability distribution of extinction $\sProb{t}=\dGamma{t}{k}{\theta}$, parametrized by the parameters $k=\sMFP^2/\Var{t}$ and $\theta=\Var{t}/\sMFP$, where $\sMFP$ is the mean free path, and $\Var{t}$ the variance of the distribution. 
Note that to avoid confusion with \Eq{gamma} used as $\sProbConc{\sConcentration}$, we used the alternative parametrization of the gamma distribution, where $k=\alpha$ and $\theta=\beta^{-1}$. This distribution leads to
\begin{align}
\sTranmittanceCorr(t) & = 1-\frac{\fGammaIncomplete{k, \theta^{-1}\,t}}{\fGamma{k}}, \\
\sProb{t} & = \dGamma{t}{k}{\theta}, \label{eq:app_gamma_transmittance}\\
\sExtinctionResolved(t) & = \frac{\dGamma{t}{k}{\theta}}{1-\frac{\fGammaIncomplete{k, \theta^{-1}\,t}}{\fGamma{k}}},
\end{align}
where $\fGammaIncomplete{s,x}=\int_0^x t^{s-1}e^{-t} \diff{t}$ is the incomplete gamma function, and $\fGamma{x}$ is the gamma function. 
In order to sample \Eq{app_gamma_transmittance} we do not have a closed form, and need to use numerical methods. In our case, we used the rejection method by Marsaglia and Tsang~\shortcite{Marsaglia2000}, which can sample the full space of $k$, and with cost approximately constant with $k$. 

\section{Details on Figure 2}
\label{sec:app_figure2}
In order to validate the existence of non-exponential transmittance, in addition to findings from other fields such as neutron transport or atmospheric sciences, we performed a simple experiment where we capture the transmittance of different correlated (non-exponential) and uncorrelated (exponential) media. For capture, we use a setup inspired in Meng et al.~\cite[Figure 3]{Meng2015Granular}, where we filled a glass-made vase with the material. The vase was placed on top of a mobile flash for lighting, and captured using a Nikon D200 placed over the vase. 

We capture a set of HDR images of increasing thickness for each material. Each image was captured by multi-bracketing 36 RAW images, with fixed aperture set at $4.9$, ISO-1600, and exposition time ranging from \SI{1/6400}{\second} to \SI{1/2}{\second}. To get rid of the effect of the container we also captured an HDR image of the empty glass. 

Finally, to assess whether the exponential transmittance holds or not on the captured materials, we fit them to an exponential function. As shown in Figure 2, for correlated materials this fitness is not very accurate, while diluted milk shows a very good fit, as expected. 

\newcommand{\sizecaptured}{.09\textwidth}


\section{Rendering convergence}
\label{sec:app_convergence}
\begin{figure*}
\includegraphics[width=.6\textwidth]{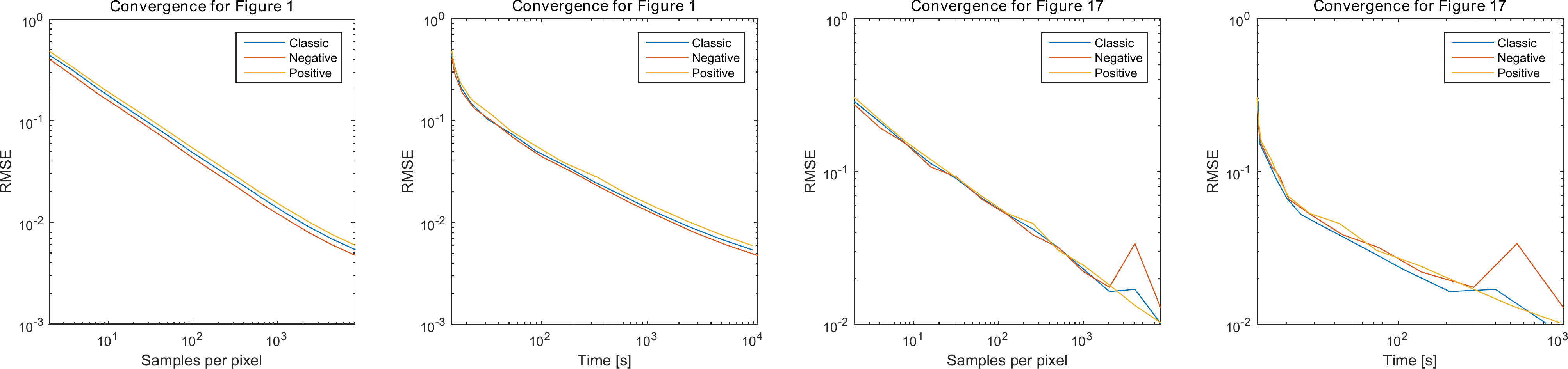}
\caption{Convergence plots for the scenes in \Figs{isotropic_dragons}{anisotropic_dragons}, relating the RMSE with the number of samples and rendering time, respectively. Each scene is rendered with different types of media: a classic exponential medium, a negatively correlated medium, and a positively correlated medium modeled with our model in \Sec{rendering}.}
\label{fig:app_convergence}
\end{figure*}
In order to evaluate whether our new model increases variance, we evaluate the convergence of our model with respect to classic exponential transport. We perform this evaluation in the scenes shown in \Figs{isotropic_dragons}{anisotropic_dragons}, for three different types of media: classic exponential, and non-exponential with positive and negative correlation. We used volumetric path tracing for rendering. \Fig{app_convergence} shows how the error of both media converge in a similar rate with the number of samples (as expected), but that the increment in variance is marginal, and in fact only observable in the case of positive correlation, where the increased transmittance might result in an increase of variance. A similar behavior is observed with the converge with respect to rendering time, since ray-geometry intersections dominate.


\section{Monte Carlo Numerical Simulations}
\label{sec:app_simulations}

In order to gain understanding on the problem, illustrate the results, and compare our solutions against a ground truth, we computed a number of simulations on procedurally generated explicit media. This allowed us to investigate on the differences on the probability distribution of extinction $\sProb{t}$ between scatterers and sources (\Sec{ourgbe}), as well to study the effect of boundary conditions (\Sec{boundary_conditions}). Here we explain the details of such simulations, including the definitions of the media, and include the full set of results of our simulations. 

\subsection{Modeling correlated scatterers}
Since we are interested on the average behavior of $\sProb{t}$, we procedurally generated different types of media in 2D. We opted for a two-dimensional problem since it is simpler but valid to our problem (extinction is a 1D problem), as has been shown in many previous works in transport related fields. 
Each media was formed by a number of circular scatterers with same (very small) radius $r$. 
For each realization of the 2D volume, we build a randomized procedural media. These procedural definitions were different for the case of positive and negative media. 
\begin{itemize}
\item \textbf{Negatively Correlated Media}: Based on previous work on transport on Lorentz gases~\cite{Dumas1996}\footnote{A Lorentz gas is a periodic array of scatterers forming a lattice.}, we generate a perfect negative media by deterministically defining the position of the particles in an array. We introduced the constraint of having each particle in the middle of an hexagon, where the closest neighbor particles where at the vertices of that hexagon. That ensured that the closest particles where all at the same distance. We then slightly displaced each vertex position to ensure that none of them masked any other particle along the propagation direction of the ray. Finally, we stochastically perturbed the position of the particles based on the desired degree of correlation $\sCorr{}$: We decided whether a particle should be perturbed with probability $p_p = 1-|\sCorr{}|$, and perturbed its position $\px$ as $\px=\px_0+\dir\,s$, where $\px_0$ is the particle original position, $\dir$ is a unit vector uniformly sampled in the circle of directions, and $s=-\log(\xi)\,(1-|\sCorr{}|)^2$ with $\xi$ a uniform random number. 

\item \textbf{Positively Correlated Media}: Here we follow the approach of Shaw et al.~\shortcite{Shaw2002} and Larsen and Clark~\shortcite{Larsen2014link}: We  select the position of a first particle $\px_0$ by uniformly random sampling the unit square. Then, we begin a random walk from this initial position, so that the position $\px_i$ of a particle $i>0$ is computed as $\px_i=\px_{i-1}+\dir\,s$, where $\dir$ is a unit vector uniformly sampled in the circle of directions, and $s=-\log(\xi)\,(1-\sCorr{})^2$ with $\xi$ a uniform random number and $\sCorr{}\in(0,1)$ the degree of positive correlation.
\end{itemize}

In both cases, we use a periodic boundary condition following previous work~\cite{Shaw2002}. Note that we did not impose a minimum distance of particles (that could be another form of negative correlation by using a dart-throwing sampling approach; we did so to avoid introducing some form of correlation when each of the approaches converge to the uncorrelated behavior (i.e. $\sCorr{}\to 0$); however, given the small radii of the simulated particles ($10^{-5}$, distributed in a unit squared medium) we found that they were unlikely to intersect each other.   

In the following, we show numerical solutions for source-to-scatterer and scatterer-to-scatterer probability distributions of extinction and transmittance (\Sec{app_pathlengths}), as well as simulations on the medium-to-medium boundary conditions (see \Sec{boundary_conditions}) for a variety of different media correlations.

\subsection{Source-to-Scatterer and Scatterer-to-Scatterer Extinction}
\label{sec:app_pathlengths}

\begin{figure*}[t]
  \def\svgwidth{.8\textwidth}
\footnotesize
  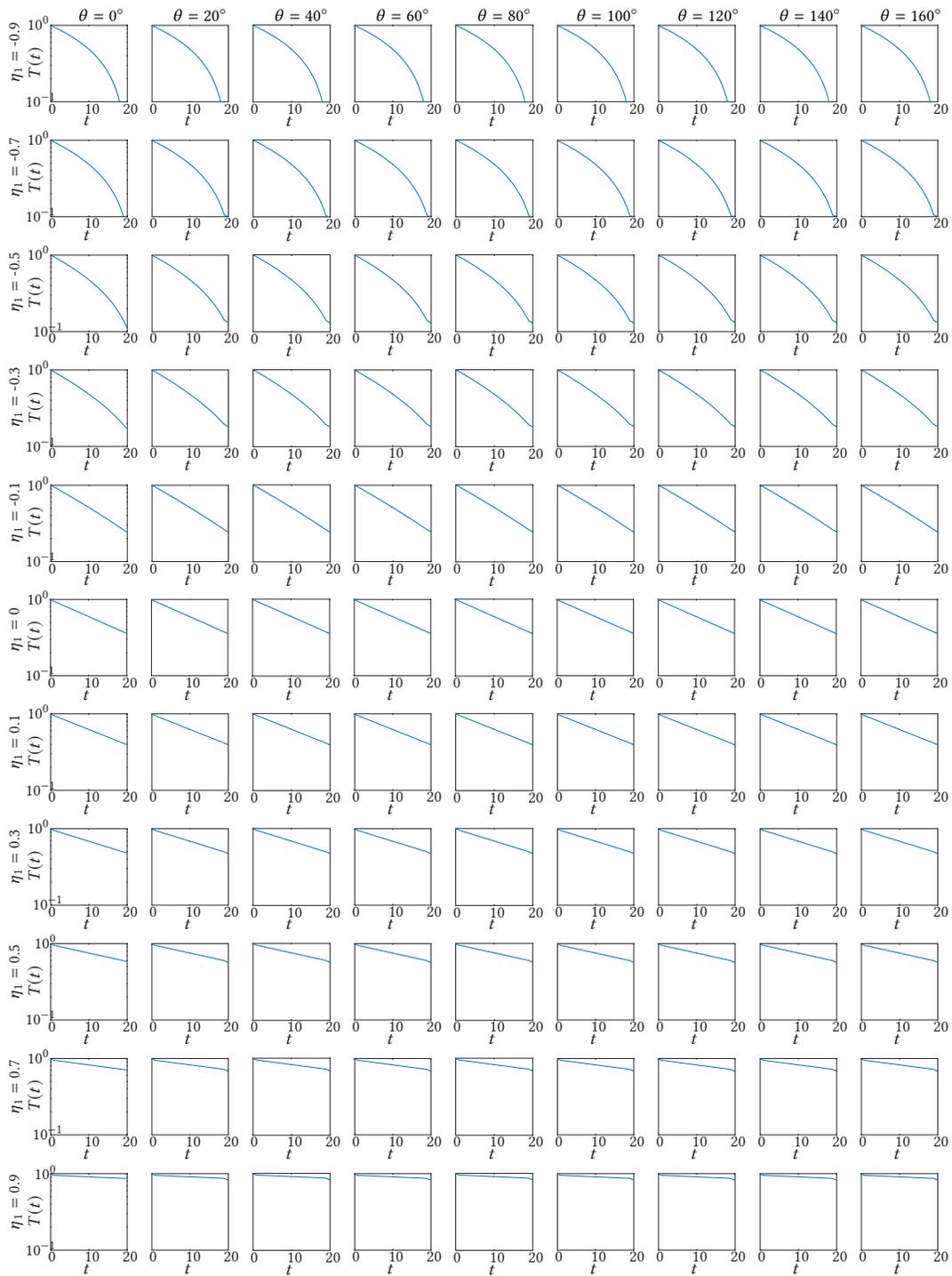
\caption{Monte Carlo simulations of the transmittance $\sTranmittanceCorrI{\sEmission}(t)$ from rays with origin at sources, for infinite media with correlation varying from $\sCorr{1}\in[-1,0.9]$, in logarithmic scale, for different angles $\theta$. 
}
\label{fig:app_freepaths_sources}
\end{figure*}

\Figs{app_freepaths_sources}{app_freepaths_scatterers} show a series transmittances $\sTranmittanceCorr(t)$ for a source term at the boundary of the medium, and for scatterer-to-scatterer transport, respectively. Each of them has been computed for a different level of correlation $\sCorr{1}=\in[-1,0.9]$. We have simulated each of them by averaging 2000 iterations each iteration with a different randomly generated medium, and 1000 samples per iteration.
The samples from the source $\sEmission(\px,\diro)$ where traced from the boundary of the medium, for a given direction $\sAngle$. In contrast, the samples for the scatterer-to-scatterer extinction were traced from the scatterers, by randomly selecting the scatterer of origin, and with a random direction. 
%

\begin{figure*}[t]
  \def\svgwidth{\textwidth}
  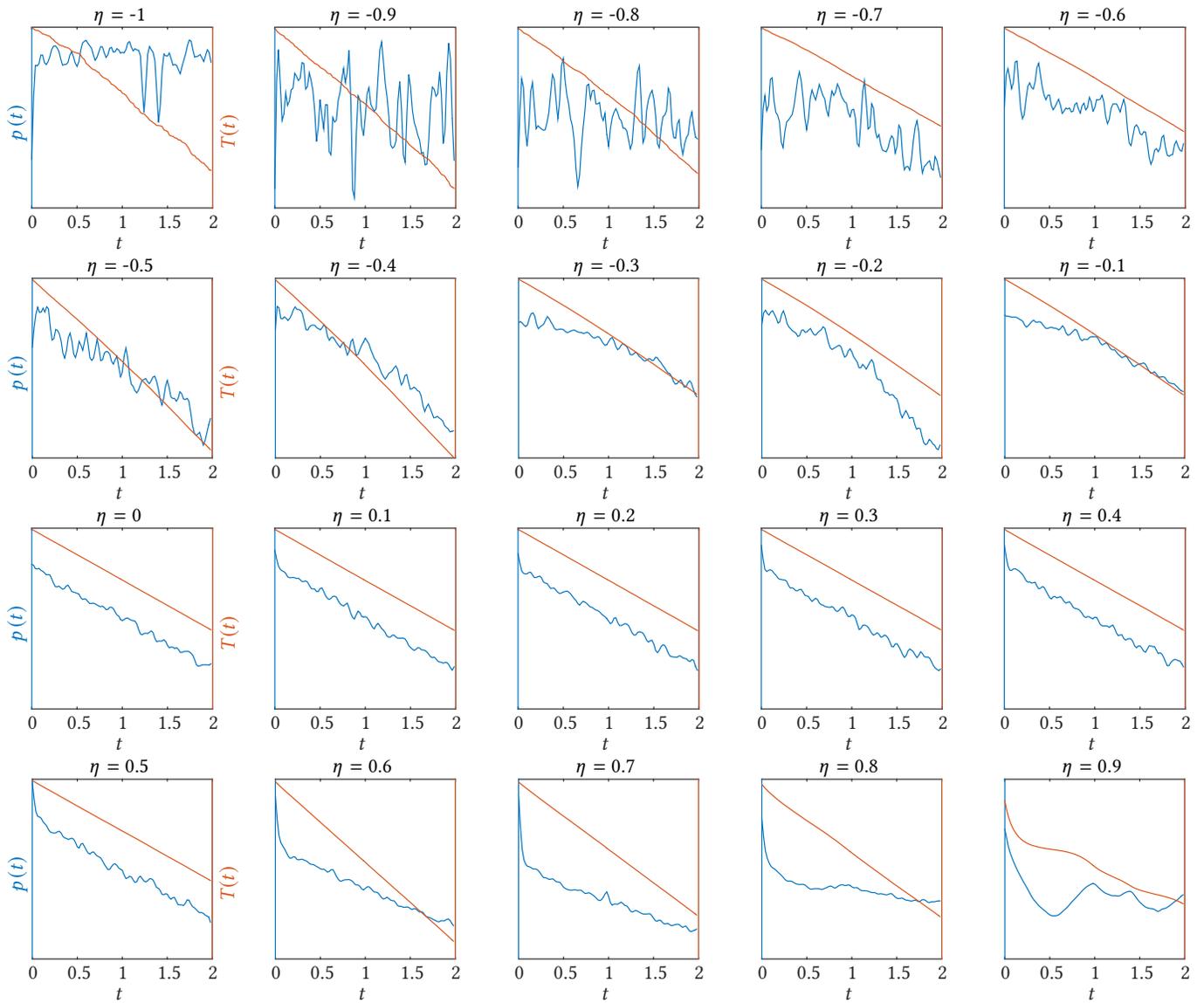
\caption{Monte Carlo simulations of the probability distribution of extinction $\sProbI{\sRadI}{t}$ (blue) and transmittance $\sTranmittanceCorrI{\sRadI}(t)$ (orange) from rays with origin at scatterers, for infinite media with correlation varying from $\sCorr{1}\in[-1,0.9]$, in logarithmic scale.
}
\label{fig:app_freepaths_scatterers}
\end{figure*}

\subsection{Boundary Conditions}
\label{sec:app_boundaries}

\begin{figure*}[t]
  \def\svgwidth{\textwidth}
\footnotesize
  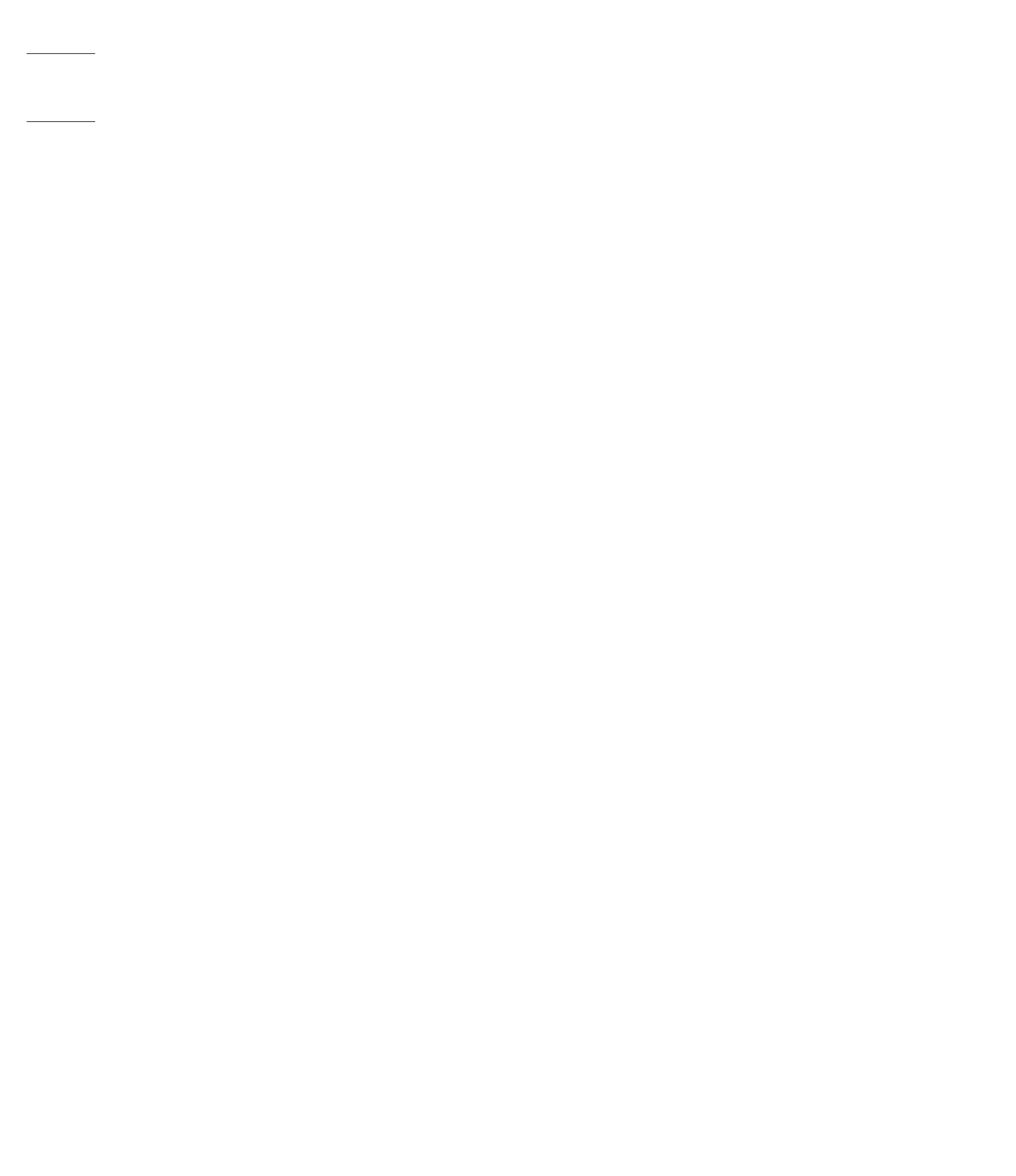
\caption{Monte Carlo simulations for the medium-to-medium boundary (marked as a green dashed line), showing the probability distribution of extinction $\sProb{t}$ (blue), and transmittance $\sTranmittanceCorrI{\sEmission}(t)$ (orange), for original media with correlation $\sCorr{1}\in[-0.9,0]$, and second media defined so that the correlation between both media $\sCorr{1,2}\in[-0.9,0.9]$ infinite media with correlation varying from $\sCorr{1}\in[-1,0.9]$. 
}
\label{fig:app_freepaths_boundaries_negative}
\end{figure*}

\Figs{app_freepaths_boundaries_negative}{app_freepaths_boundaries_positive} shows a wider range of results for the media-to-media boundary, complementing those in \Fig{freepaths_boundaries}, for $\sCorr{1}\in[-1,0.9]$. We follow the same procedure as in \Fig{app_freepaths_sources}, with a change of media at distance $t=20$. 

\begin{figure*}[t]
  \def\svgwidth{\textwidth}
\footnotesize
  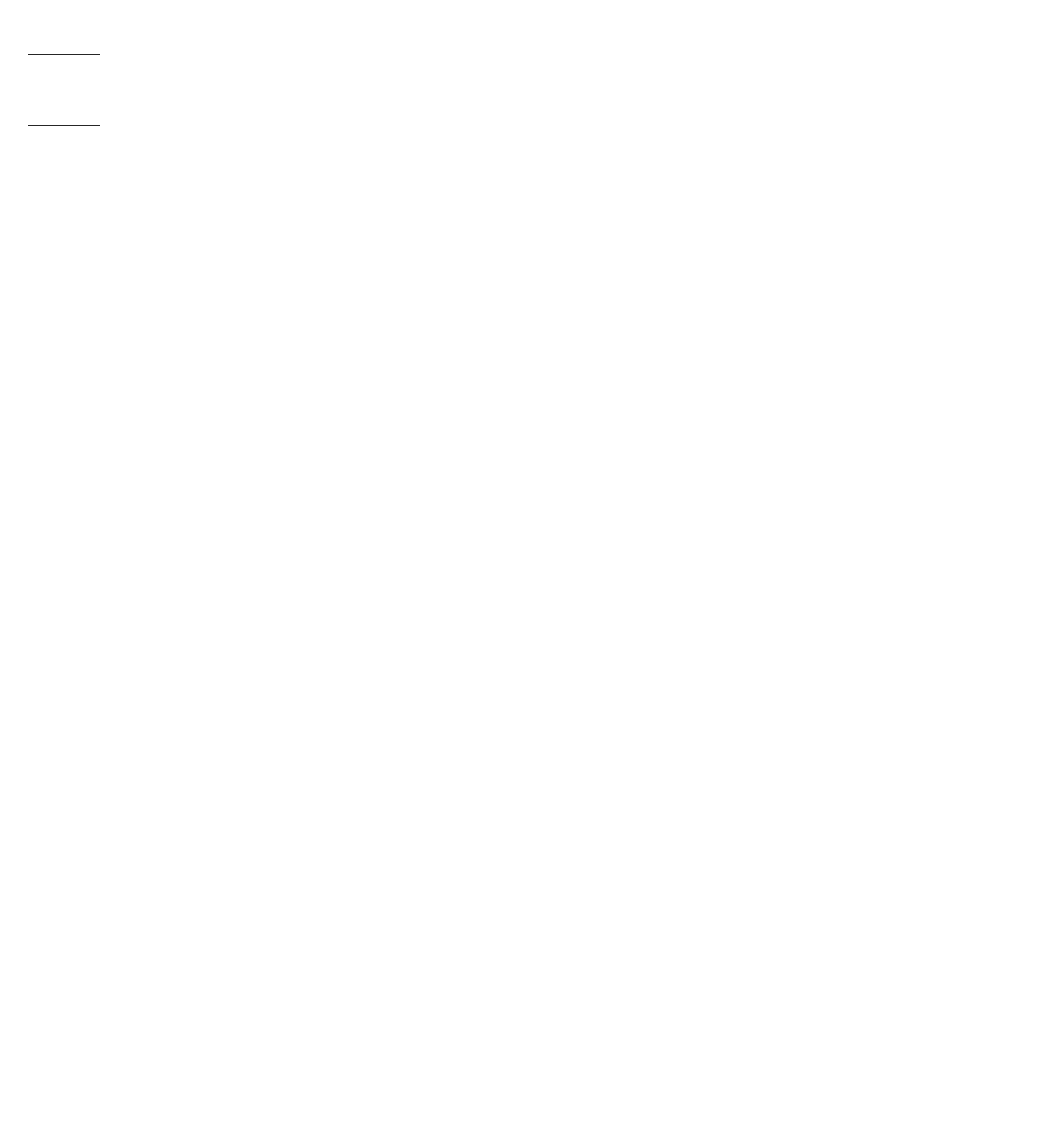
\caption{Monte Carlo simulations for the medium-to-medium boundary (marked as a green dashed line), showing the probability distribution of extinction $\sProb{t}$ (blue), and transmittance $\sTranmittanceCorrI{\sEmission}(t)$ (orange), for original media with correlation $\sCorr{1}\in[0,0.9]$, and second media defined so that the correlation between both media $\sCorr{1,2}\in[-0.9,0.9]$ infinite media with correlation varying from $\sCorr{1}\in[-1,0.9]$. 
}
\label{fig:app_freepaths_boundaries_positive}
\end{figure*}

\newcommand{\angstructsizesupp}{.3\textwidth}
\newcommand{\angstructspacesizesupp}{.1cm}

\begin{figure*}
\centering

\includegraphics[width=\angstructsizesupp]{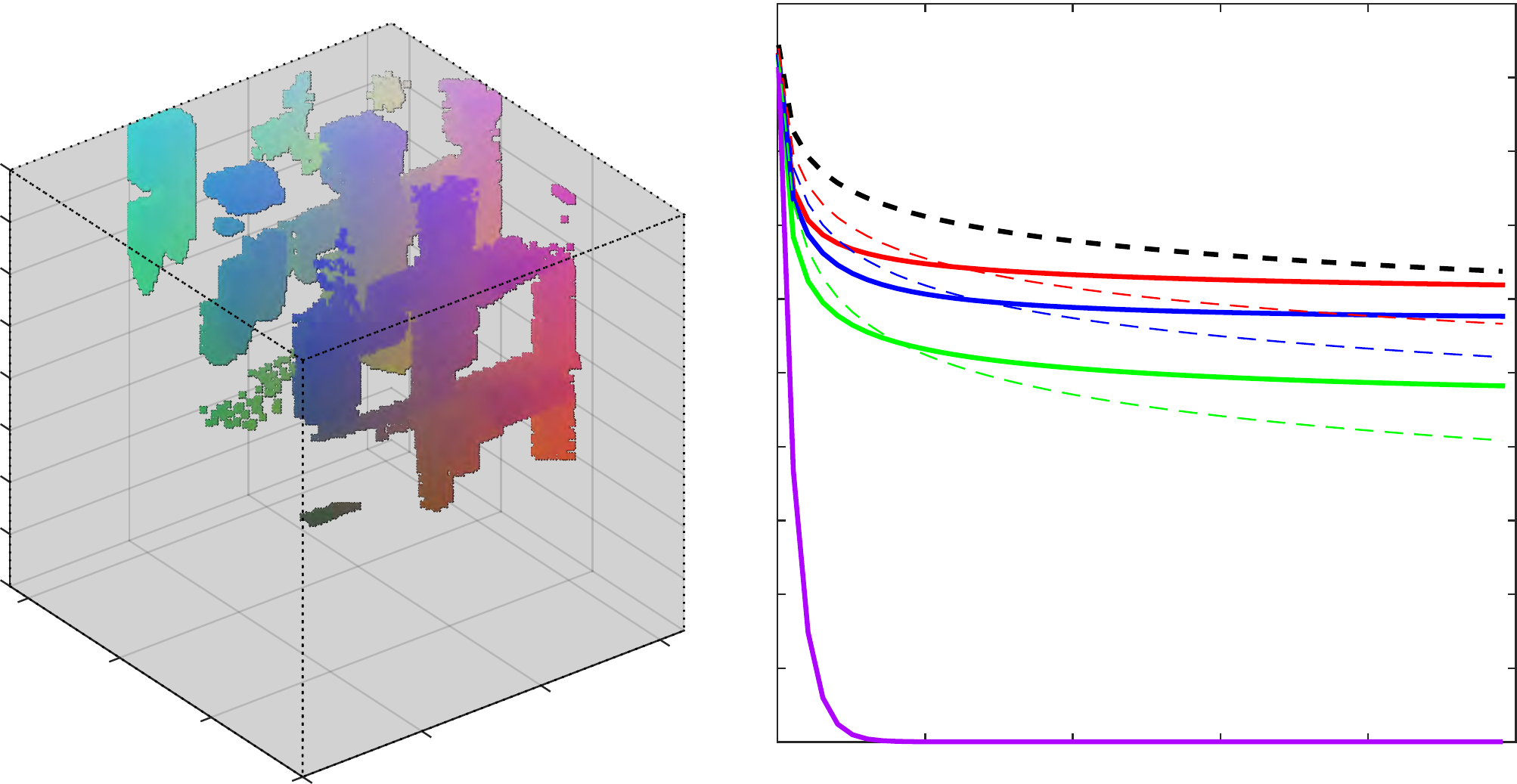} \hspace{\angstructspacesizesupp}
\includegraphics[width=\angstructsizesupp]{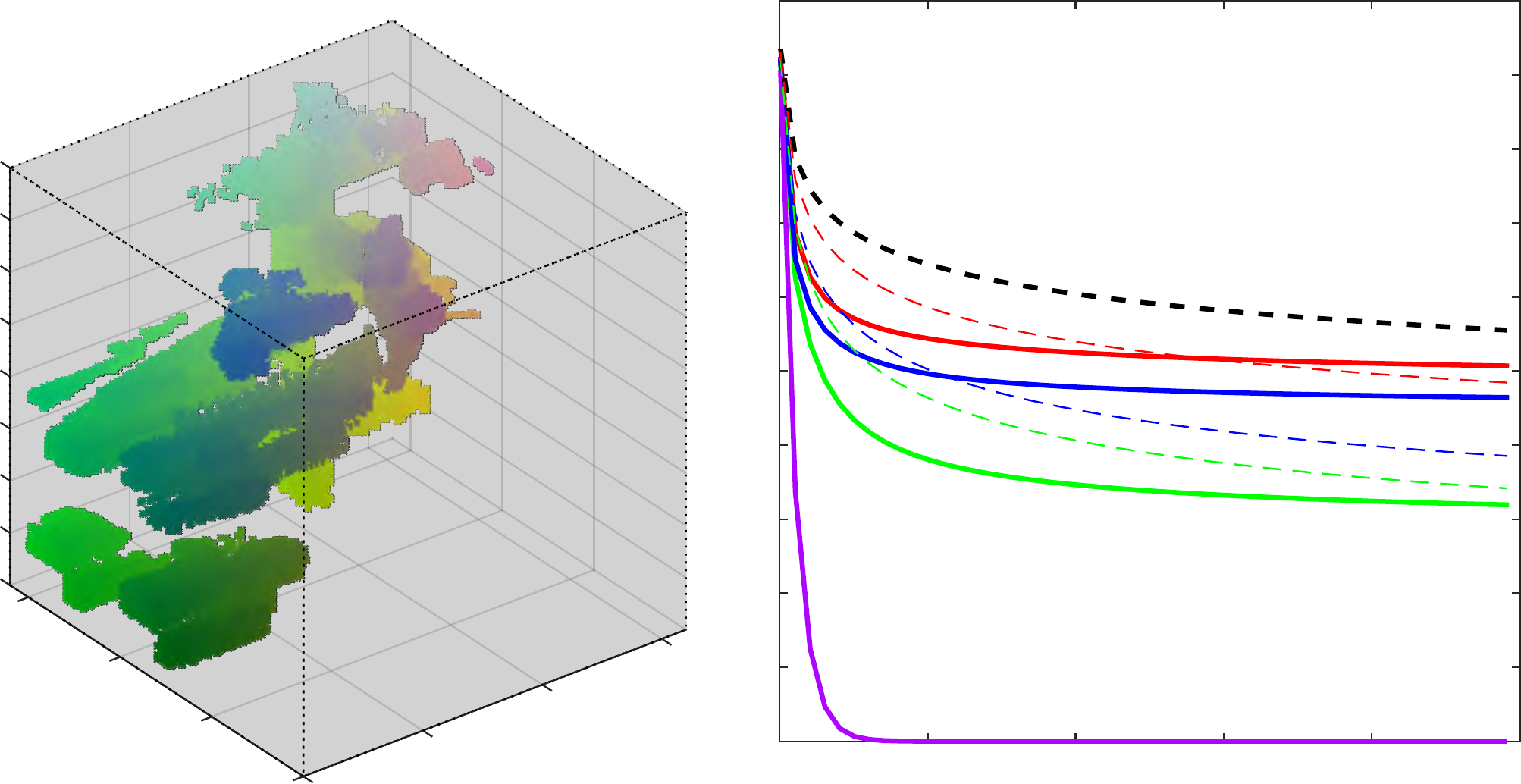} \hspace{\angstructspacesizesupp}
\includegraphics[width=\angstructsizesupp]{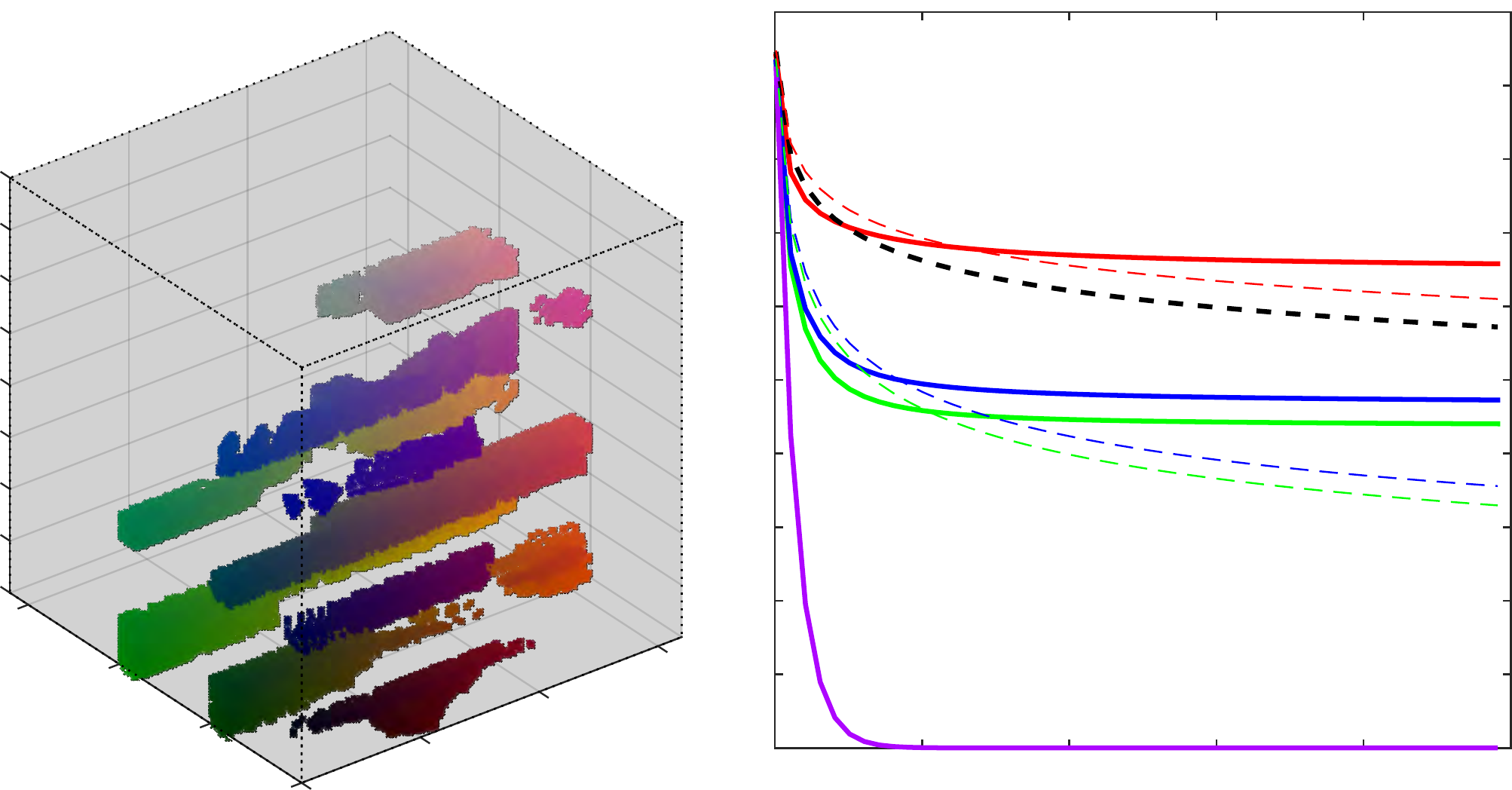}  \\ 
\includegraphics[width=\angstructsizesupp]{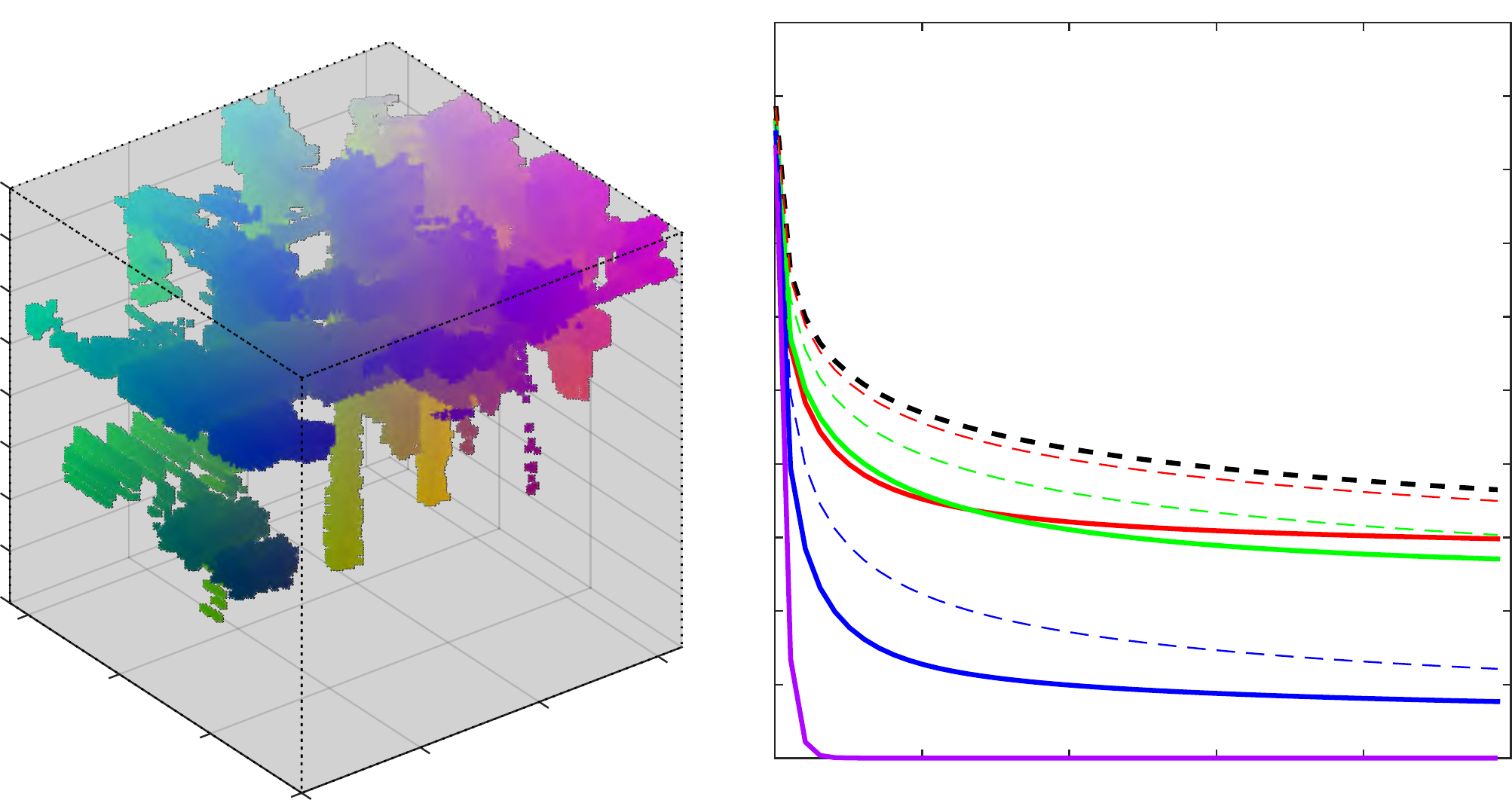} \hspace{\angstructspacesizesupp}
\includegraphics[width=\angstructsizesupp]{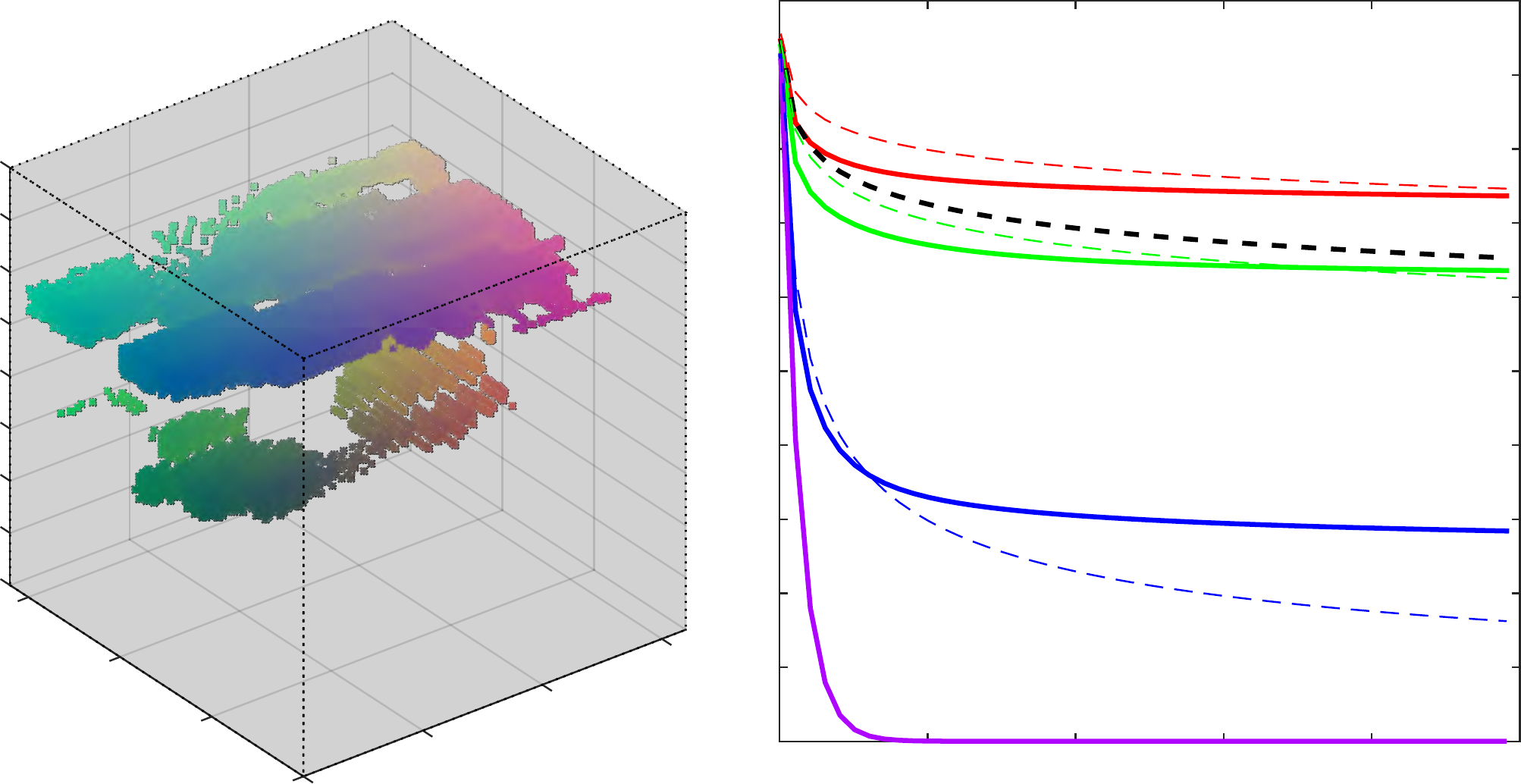} \hspace{\angstructspacesizesupp}
\includegraphics[width=\angstructsizesupp]{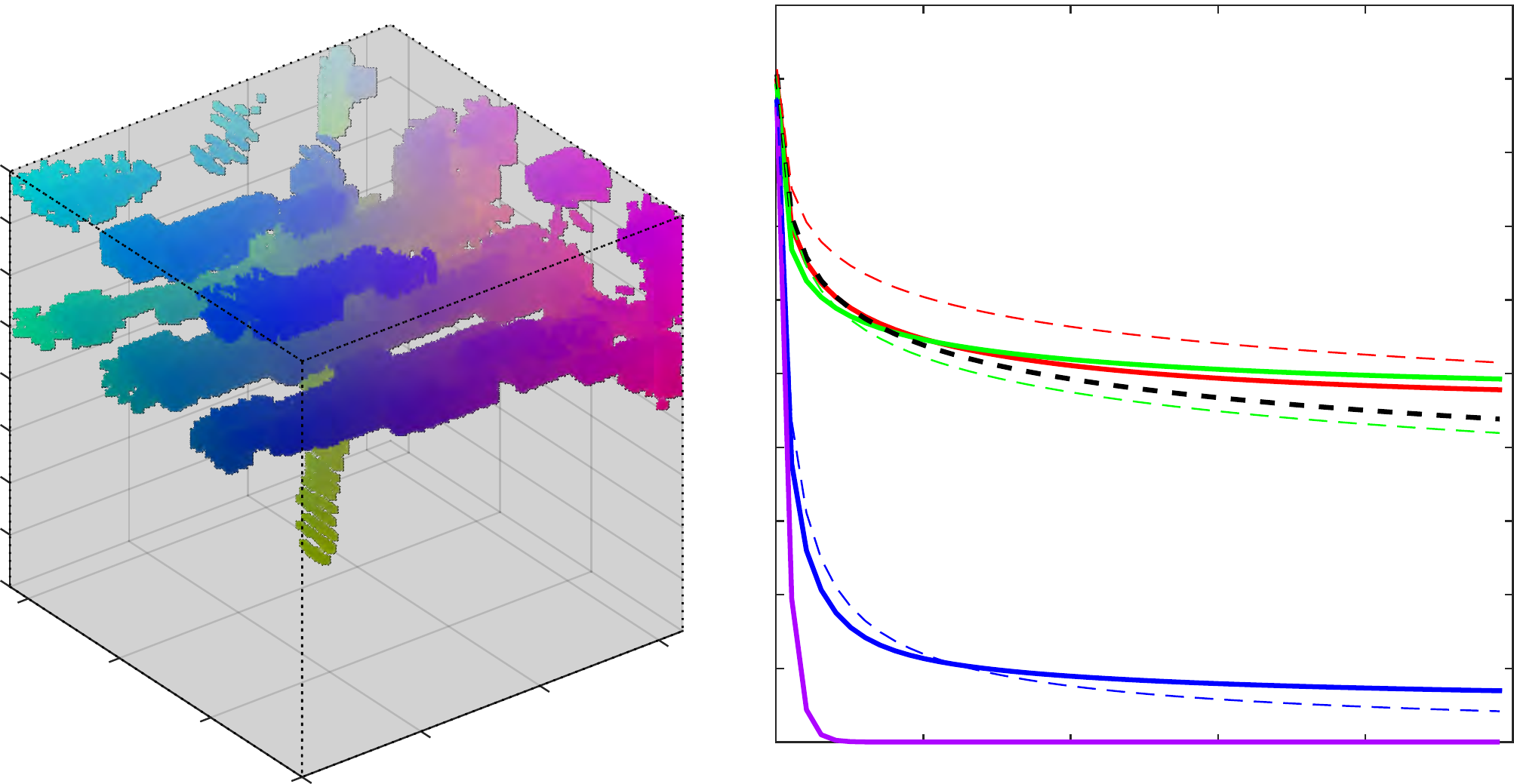} \\
\includegraphics[width=\angstructsizesupp]{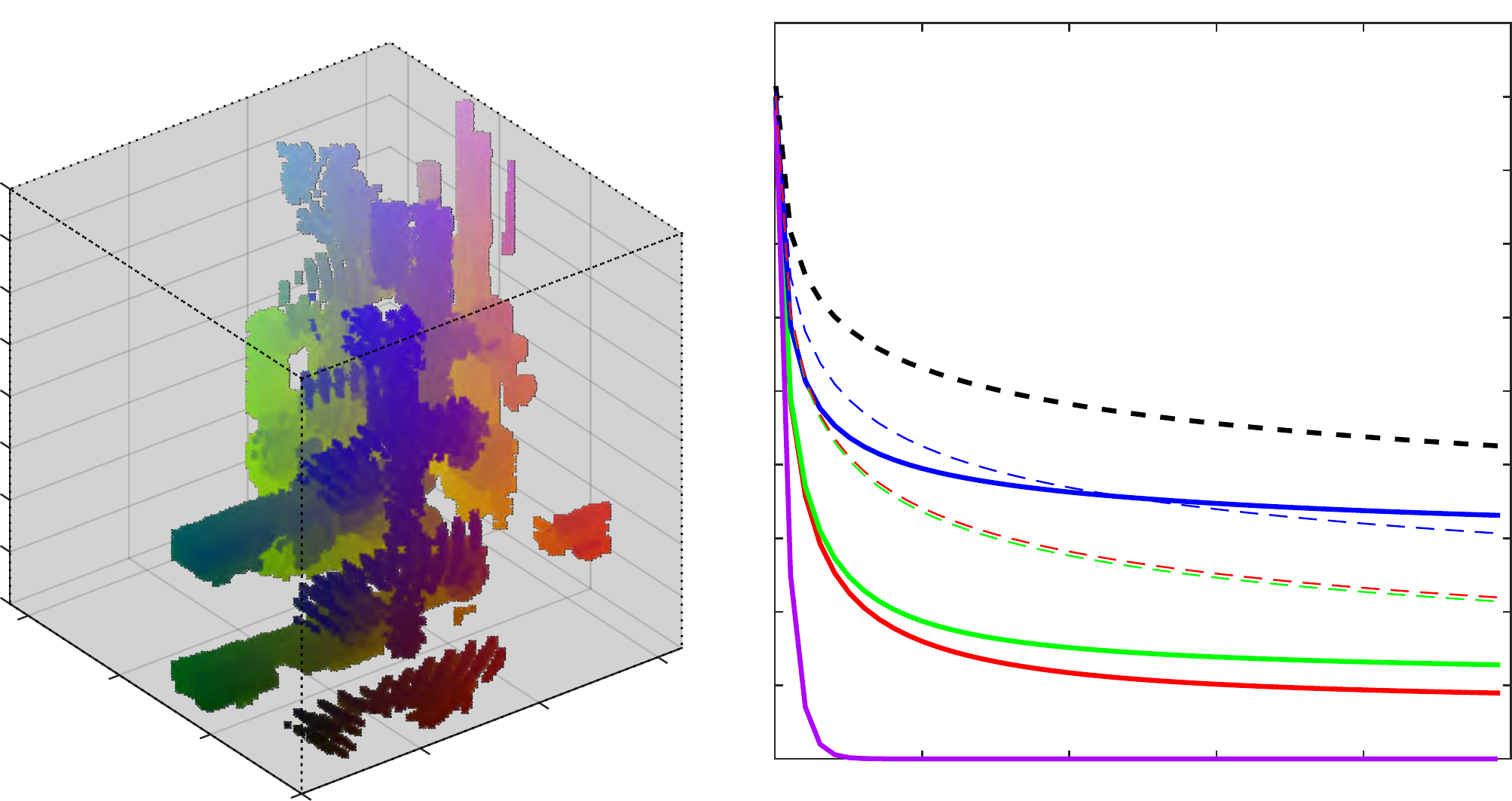} \hspace{\angstructspacesizesupp}
\includegraphics[width=\angstructsizesupp]{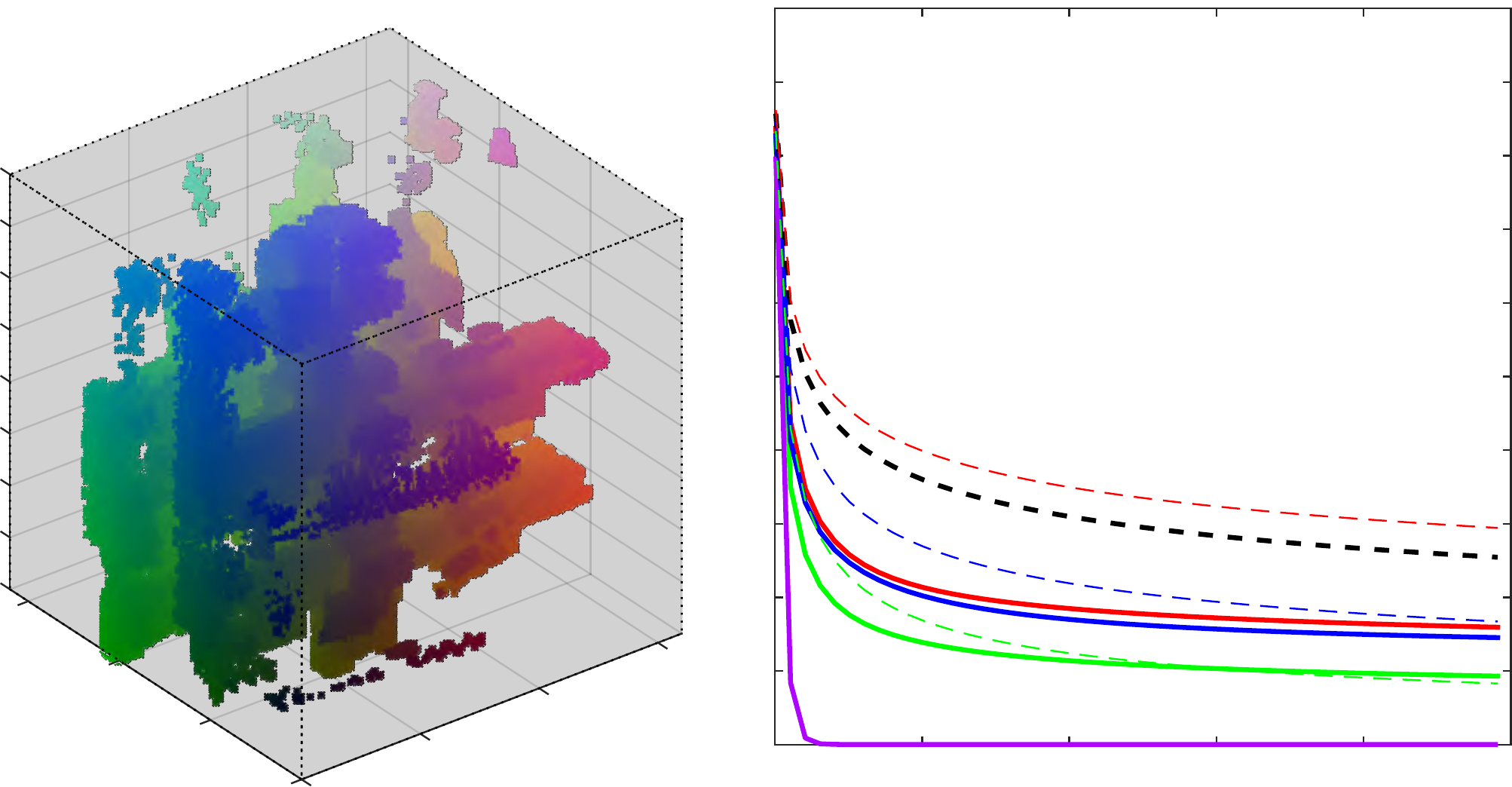} \hspace{\angstructspacesizesupp}
\includegraphics[width=\angstructsizesupp]{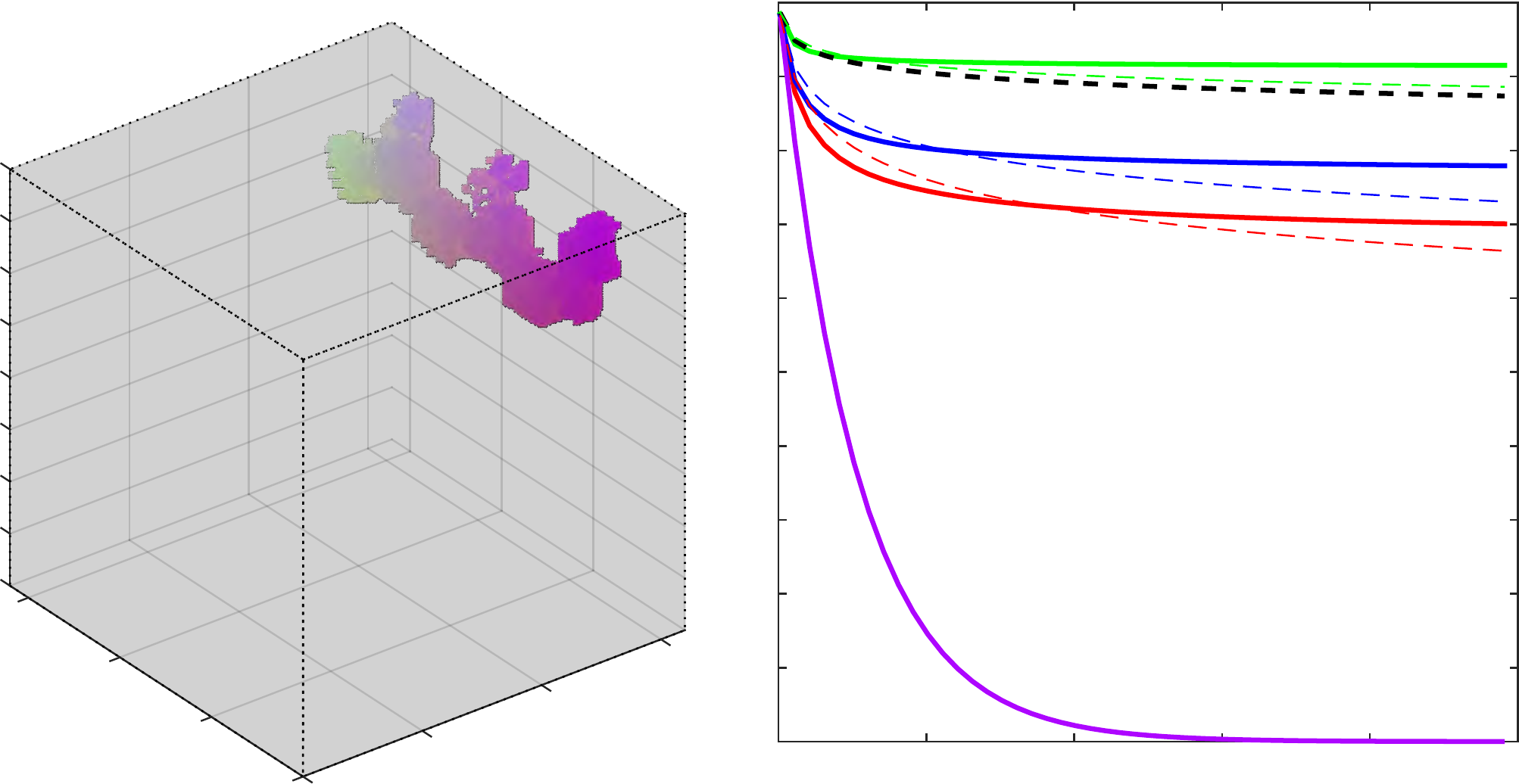} \\
\includegraphics[width=\angstructsizesupp]{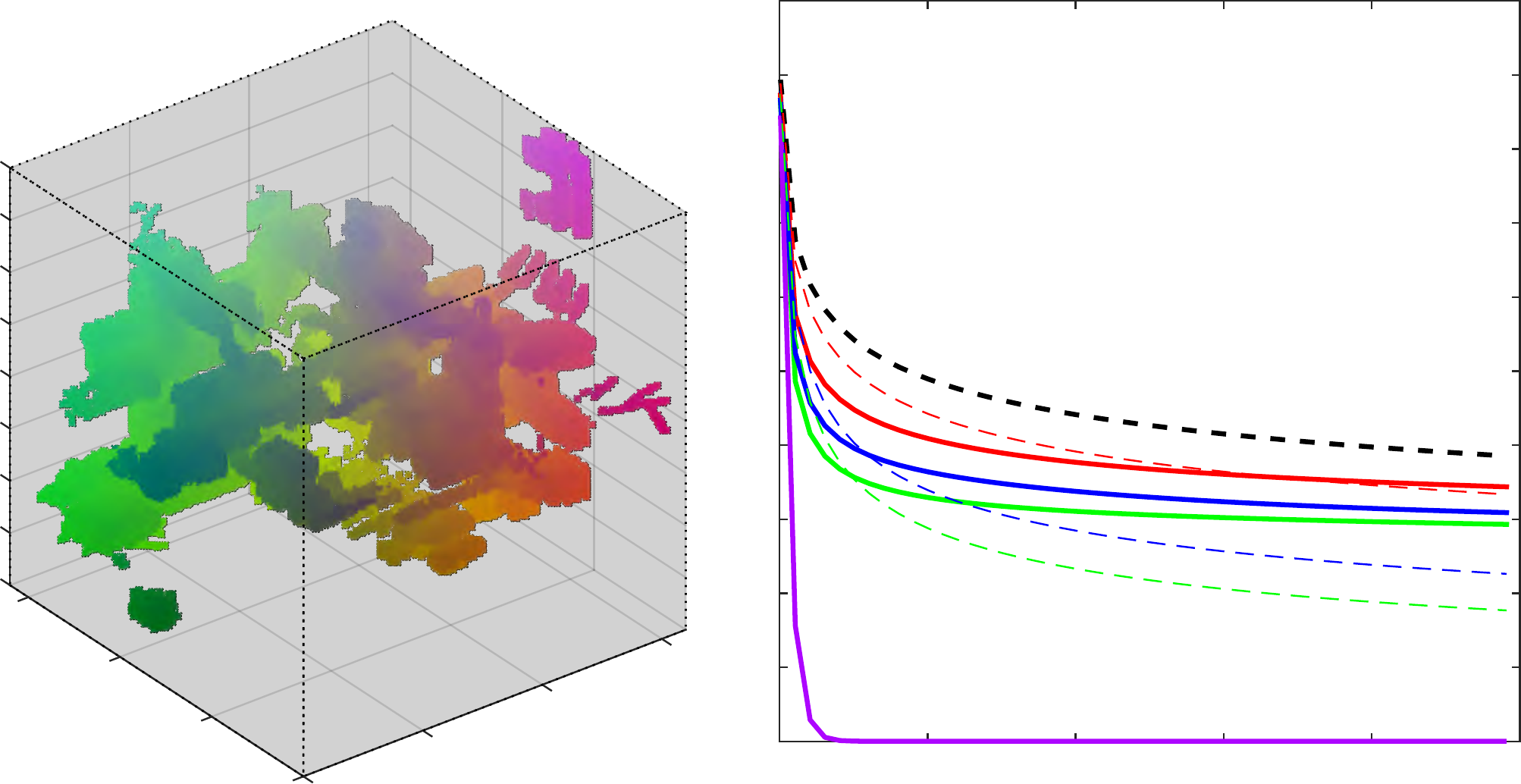} \hspace{\angstructspacesizesupp} 
\includegraphics[width=\angstructsizesupp]{\suppimages/volumes/vol05_000834_compact.pdf} \hspace{\angstructspacesizesupp}
\includegraphics[width=\angstructsizesupp]{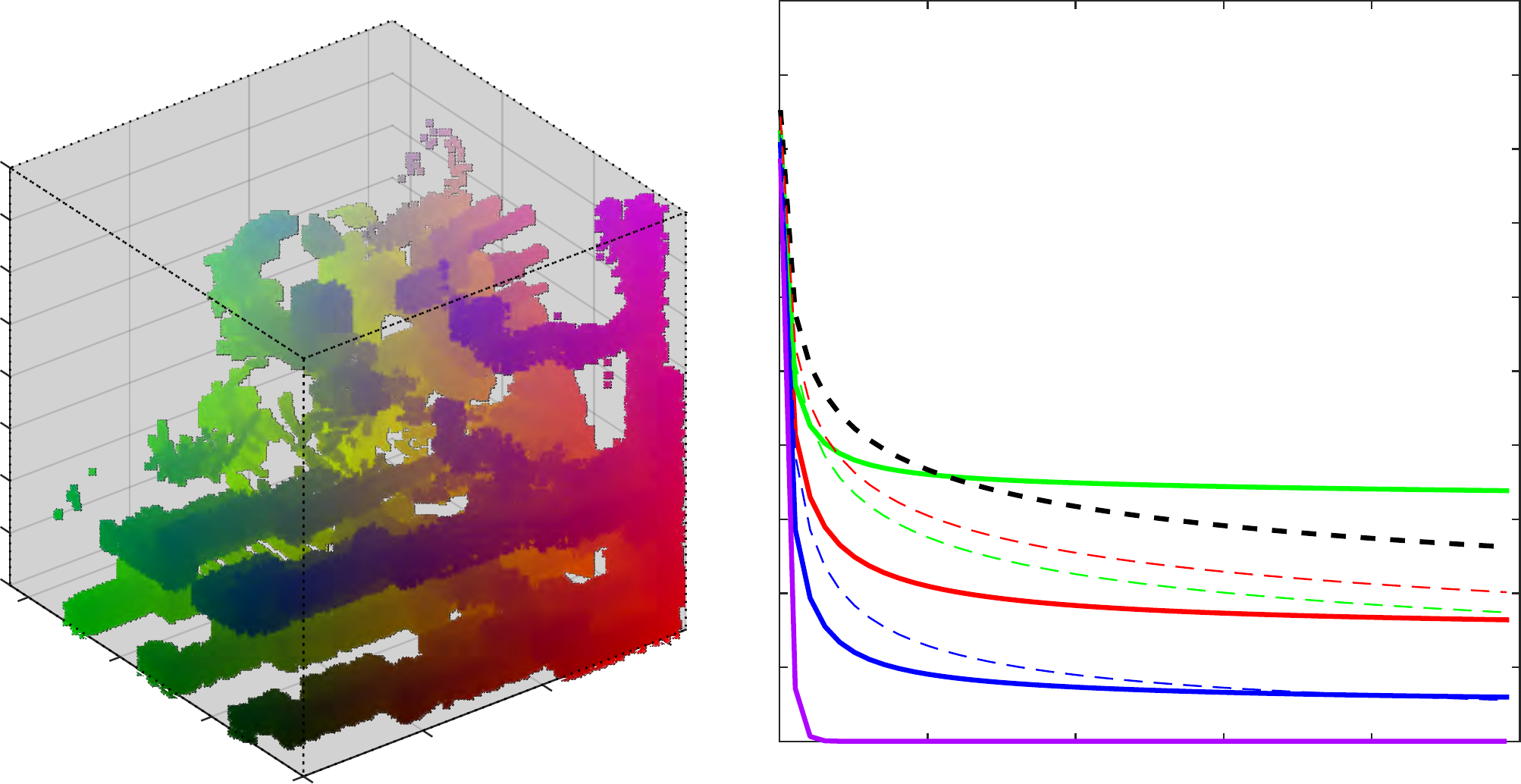} \\
\vspace{.5cm}\includegraphics[width=.20\textwidth]{\suppimages/volumes/vol_legend.pdf}

 \caption{Additional examples of transmittance in high-resolution volumes of locally-correlated media (procedurally generated after \cite{Lopezmoreno2014Gpu}). Beams of light travel through each volume, aligned in succession to the $x$, $y$, and $z$ axes. Ground truth transmittance (red, green, and blue solid lines) has been computed by brute force regular tracking~\cite{Amanatides1987}, while our simulation (dotted lines) uses the gamma distribution proposed in Equation~\ref{eq:gamma}.
Classic transport governed by the RTE significantly overestimates extinction through the volume, resulting in a exponential decay (purple line). In contrast, our model matches ground-truth transmission much more closely. The black dotted line is the result of isotropic correlation, which is clearly also non-exponential. }
%
\label{fig:app_angular_structure}
\end{figure*}

\end{document}